\documentclass[11pt]{article}

\usepackage{tikz}
\usepackage{amsmath}
\usepackage{amssymb}
\usepackage{xspace}
\usepackage[draft]{hyperref}
\usepackage{stmaryrd}
\usepackage{multirow}
\usepackage{verbatim}
\usepackage{algorithm}
\usepackage[noend]{algpseudocode}
\usepackage{pifont}
\usepackage{pgfplots}
\pdfmapfile{+ltlfonts.map}
\usepackage{ltlfonts}

\usepackage{makecell}

\usepackage{subfig}

\usepackage{graphicx}

\usepackage{wasysym}

\usepackage{enumerate}
\usepackage{enumitem}
\usepackage{caption}
\usepackage{diagbox}
\usepackage{amssymb}

\usepackage{pifont}
\usepackage{adjustbox}

\usepackage[margin=0.7in]{geometry}

\newcommand{\mycomment}[1]{}
\newcommand{\system}{\textsc{Phoenix}\xspace}
\newcommand{\systemtitle}{PHOENIX\xspace}

\newcommand{\rlfReport}{\ensuremath{\mathsf{rlfReport}}\xspace}

\newcommand{\securityModeCommand}{\ensuremath{\mathsf{securityModeCommand}}\xspace}
\newcommand{\securityModeComplete}{\ensuremath{\mathsf{securityModeComplete}}\xspace}
\newcommand{\ueInformationRequest}{\ensuremath{\mathsf{ueInformationRequest}}\xspace}

\newcommand{\rrcConnectionRequest}{\ensuremath{\mathsf{rrcConnectionRequest}}\xspace}
\newcommand{\attachRequest}{\ensuremath{\mathsf{attachRequest}}\xspace}

\newcommand{\ueCapabilityEnquiry}{\ensuremath{\mathsf{ueCapabilityEnquiry}}\xspace}

\newcommand{\pltl}{PLTL\xspace}

\newcommand{\signatureSynthesizer}{signature synthesizer\xspace}
\newcommand{\monitor}{monitor\xspace}

\newtheorem{mydef}{Definition}

\newtheorem{example}{Example}

\newcommand{\bigO}{\mathcal{O}}

\newcommand{\yesterday}{\ensuremath{\LTLcircleminus}\xspace}

\newcommand{\historically}{\ensuremath{\LTLsquareminus}\xspace}
\newcommand{\once}{\ensuremath{\LTLdiamondminus}\xspace}
\newcommand{\since}{\ensuremath{\,\mathrm{\cal S}\,}\xspace}

\algnewcommand{\true}{\textbf{true}}
\algnewcommand{\false}{\textbf{false}}

\newcommand{\fml}[1]{\ensuremath{\mathcal{#1}}}

\newcommand{\lang}{\ensuremath{\mathcal{L}}\xspace}

\newcommand{\calc}[1]{\ensuremath{\mathcal{#1}}\xspace}

\newcommand{\xmark}{\ding{55}}%

\begin{document}
%

\title{\systemtitle: Device-Centric Cellular Network Protocol Monitoring using Runtime Verification\footnote{
This is the extended version of the NDSS 2021 paper with the same title.
}}

\author{Mitziu Echeverria$^*$, Zeeshan Ahmed$^*$, Bincheng Wang$^*$, M. Fareed Arif$^*$, \\Syed Rafiul Hussain$^\dagger$, Omar Chowdhury$^*$\\ \\
$^*$The University of Iowa, $^\dagger$Pennsylvania State University\\ \\
Email: $^*$\{mitziu-echeverria, zeeshan-ahmed, bincheng-wang,\\ muhammad-arif, omar-chowdhury\}@uiowa.edu,\\$^\dagger$\{hussain1\}@psu.edu}

\date{}



\maketitle

\begin{abstract}
	End-user-devices in the current cellular ecosystem are prone to many different
	vulnerabilities across different generations and protocol layers.
	Fixing these vulnerabilities retrospectively can be
	expensive, challenging, or just infeasible. A pragmatic approach for dealing with
	such a diverse set of vulnerabilities would be to identify attack attempts at runtime on the device side,
	and thwart them with mitigating and corrective actions. Towards this goal, in the paper we propose a general
	and extendable approach called \system for identifying n-day cellular network
	control-plane vulnerabilities as well as dangerous practices of network operators from
	the device vantage point. \system monitors the device-side cellular network traffic
	for performing signature-based unexpected behavior detection through lightweight
	runtime verification techniques. Signatures in \system can be manually-crafted by a
	cellular network security expert or can be automatically \emph{synthesized} using
	an optional component of \system, which reduces the signature synthesis problem to
	the \emph{language learning from the informant} problem. Based on the corrective
	actions that are available to \system when an undesired behavior is detected,
	different instantiations of \system are possible: a full-fledged defense
	when deployed inside a baseband processor; a user warning system when deployed
	as a mobile application; a probe for identifying attacks in the wild. One such
	instantiation of \system was able to identify all 15 representative
	n-day vulnerabilities and unsafe practices of 4G LTE networks considered in our
	evaluation with a high packet processing speed ($\sim$68000 packets/second) while
	inducing only a moderate amount of energy overhead ($\sim$4mW). 
\end{abstract}

\section{Introduction}
Along with global-scale communication, cellular networks facilitate
a wide range of critical applications and services including
earthquake and tsunami warning system (ETWS), telemedicine, and smart-grid electricity distribution.
Unfortunately, cellular networks, including the most recent generation, have been often plagued with debilitating
attacks due to design weaknesses~\cite{lteinspector, TORPEDO, 5g_reasoner, 5Gformal_authentication_basin}
and deployment slip-ups \cite{privacy_ndss16, kim_ltefuzz_sp19, lte_redirection,how_not_to_break_crypto}.
Implications of these attacks range from intercepting and eavesdropping messages, tracking users' locations,
and disrupting cellular services, which in turn may severely affect the security and privacy of both
individual users and primary operations of a nation's critical infrastructures. To make matters worse,
vulnerabilities discovered in this ecosystem take a long time to generate and distribute patches as they not only
require collaboration between different stakeholders (e.g., standards body, network operator,
baseband processor manufacturer) but also
incur high operational costs. To make matters worse, different patches could potentially lead to
unforeseen errors if their integration is not accounted for.

In addition to it, although a majority of the existing work focus on discovering new attacks through analysis of the \emph{control-plane} protocol
specification or deployment \cite{lteinspector, TORPEDO, privacy_ndss16, kim_ltefuzz_sp19, 5Gformal_authentication_basin,
5g_reasoner, lte_redirection, how_not_to_break_crypto}, only a handful of efforts have focused on proposing defense mechanisms or
any apparatus to detect attack occurrences~\cite{imsi_catcher_catchers, FBSRADAR, mobile_self_defense, FBSleuth, wisec_root}.
Unfortunately, these proposed mechanisms are far from being widely adopted since they suffer from one of the following
limitations: \textbf{(i)} Requires modifications to an already deployed cellular network protocol \cite{wisec_root} which require
network operator cooperation;
\textbf{(ii)} Focuses on identifying particular attacks and hence are not easily extensible \cite{imsi_catcher_catchers, FBSRADAR,
mobile_self_defense, FBSleuth}; and
\textbf{(iii)} Fails to handle realistic scenarios (e.g., roaming) \cite{wisec_root}.

A pragmatic approach for protecting users and their devices from such a wide-variety
of vulnerabilities and dubious practices of the operators
(referred to as \textbf{\emph{undesired behavior}}\footnote{
	In our context, \emph{not} all undesired behavior are necessarily exploitable
	attacks. We also call some not-necessarily-malicious behavior (e.g., the use of null encryption by real network operators)
	undesired behavior if they can be detrimental to a user's privacy and security. In our exposition, we use
	attack, vulnerability, and undesired behavior, interchangeably.
	}
at the abstract in this paper) is to deploy a device-centric defense.
Such a defense, similar to an intrusion prevention system in principle, will monitor the network traffic at
runtime to identify undesired behavior and then take different corrective actions to possibly thwart it
(e.g., dropping a packet). In this paper, we focus on the core problem of
developing a general, lightweight, and extendable mechanism \system that can empower cellular devices
to detect various undesired behavior. 
To limit the scope of the paper, we focus on monitoring the control-plane traffic for undesired behavior,
although \system is generalizable to data-plane traffic. Monitoring control-plane traffic is vital as
flaws in control-plane procedures, such as registration and mutual authentication, are
entry points for most attacks in both control- and data-plane procedures.

\system{}'s undesired behavior detection approach can induce different instantiations depending on the corrective
actions that are available to it. When deployed inside a baseband processor, \system can be used as a full-fledged
device-centric defense, akin to the pragmatic approach discussed above, that intercepts each message before getting
processed by the message handler and take corrective actions (e.g., drop the message, terminate the session) when it
identifies the message as part of an attack sequence. Alternatively, if \system is deployed as a mobile application that
can obtain a copy of the protocol message from the baseband processor, then one can envision building a warning system, which
notifies device owners when it detects that a protocol packet is part of an undesired behavior. Finally, \system can be
deployed and distributed as part of cellular network probes or honeypots that log protocol sessions with undesired behavior.

\paragraph{Approach.} In this paper, we follow a \emph{behavioral signature-based}
attack (or, generally undesired behavior) detection approach. It is enabled by the observation that a
substantial number of cellular network undesired behavior, which is detectable from the device's point-of-view,
often can be viewed as protocol state-machine bugs. Signatures of such undesired behavior can be constructed by
considering the relative temporal ordering of events (e.g., receiving an unprotected message
after mutual authentication).

Based on this above insight, we design a lightweight, generic, and in-device runtime
undesired behavior detection system
dubbed \system for cellular devices. In its core, \system{}'s detection has
two main components: (1) a pre-populated signature database for undesired behavior;
(2) a monitoring component that efficiently \emph{monitors} the device's cellular network
traffic for those behavioral signatures and takes corresponding corrective measures based
on its deployment (e.g., drop a message, log a message, warn the user).
Such a detection system is highly efficient and deployable as it
neither induces any extra communication overhead nor calls for any
changes in the cellular protocol. \system works with only a local view of the network,
yet is effective without provider-side support in identifying a wide array of undesired behavioral signatures.

For capturing behavioral signatures, we consider the following three different signature representations
that induce different tradeoffs in terms of space and runtime overhead, explainability, and detection accuracy:
(1) Deterministic Finite Automata (DFA);
(2) Mealy machine (MM) \cite{mealy1955method};
(3) propositional, past linear temporal (\pltl) \cite{ltl} formulas.
Cellular network security experts can add behavioral signatures in these representations
to \system{}'s  database. In case an expert is not
familiar with one of the above signature representations, they can get help/confirmation
from an \textbf{\emph{optional}} automatic signature synthesis component we propose.
We show that for all the above representations the automatic signature synthesis problem
can be viewed as an instance of the \emph{language learning from the informant} problem.
For DFA and MM representations, we rely on existing automata learning algorithms, whereas
for PLTL, we propose a new algorithm, an extension of
prior work \cite{learning_ltl}. For runtime monitoring of these signature representations
in \system, we use standard algorithms \cite{havelund2004}.

We consider two different instantiations for \system.
First, we implemented \system as an Android application and instantiated with the following monitors:
DFA-based, 
MM-based, 
and PLTL-based. 
In \system app, for capturing in-device cellular traffic,
we enhanced the MobileInsight Android \cite{mobile_insight} application
to efficiently parse messages and invoke the relevant monitors.
Second, we implemented \system inside srsUE, distributed as part of
the open-source protocol stack
srsLTE \cite{gomez2016srslte}, powered by
the PLTL-based monitor---the most efficient in our evaluation,
to mimic \system{}'s deployment inside the baseband processor.


We evaluated \system{}'s Android app instantiation based on both testbed generated
and real-world network traffic in 3 COTS devices. In our evaluation with 15
existing cellular network attacks for 4G LTE,
we observed that in general all of the approaches were able to identify the existing
attacks with a high degree of success. Among the different monitors, however,
DFA on average produced a higher number of false positives
(21.5\%) and false negatives (17.1\%) whereas MM and PLTL
turn out to be more reliable; producing a significantly less number of false positives
($\sim\!\!$ 0.03\%) and false negatives ($\sim\!\!$ 0.01\%).
In addition, we observed that all monitors can handle a high number of
control-plane packets (i.e., 3.5K-369K packets/second).
We measured the power consumption induced by different monitors
and observed that on average, they all consume a moderate amount of energy ($\sim\!\!\!$ 2-6 mW).
Interestingly, we discover that \system, when powered by the PLTL-based monitor,
	produces no false warnings on real networks and in fact, it helped us discover
	unsafe network operator practices in three major U.S. cellular network providers.
Finally, we evaluated \system instantiation as part of  srsUE \cite{gomez2016srslte} with testbed
generated traffic and observed that it only incurs a small memory overhead (i.e., 159.25 KB).

\paragraph*{Contributions.}
In summary, the paper makes the following  contributions:
\begin{itemize}\setlength{\itemsep}{0em}
	\item 
	We design an in-device, behavioral-signature based
	cellular network control-plane undesired behavior detection system called \system.
	We explore the design space of developing such a vulnerability detection system and consider
	different trade-offs.
	\item We implement \system as an Android app, which during our evaluation with
	3 COTS cellular devices in our testbed has been found to be effective
	in identifying 15 existing 4G LTE attacks while incurring a small overhead.

	\item We implement \system by extending srsUE \cite{gomez2016srslte}---mimicking  a full-fledged
	defense, and show its effectiveness at preventing
	attacks. 
	\item We finally show how one could automatically synthesize behavioral signatures \system expects by
	posing it as a learning from an informant problem \cite{informant_learning} and solve it with different
	techniques from automata learning and syntax-guided synthesis.
\end{itemize}

\section{Preliminaries}
\label{sec:preliminary}

In this section, we briefly overview the background material necessary to understand
our technical discussions.

\begin{figure}[t]
	\centering
	\includegraphics[width=.8\columnwidth]{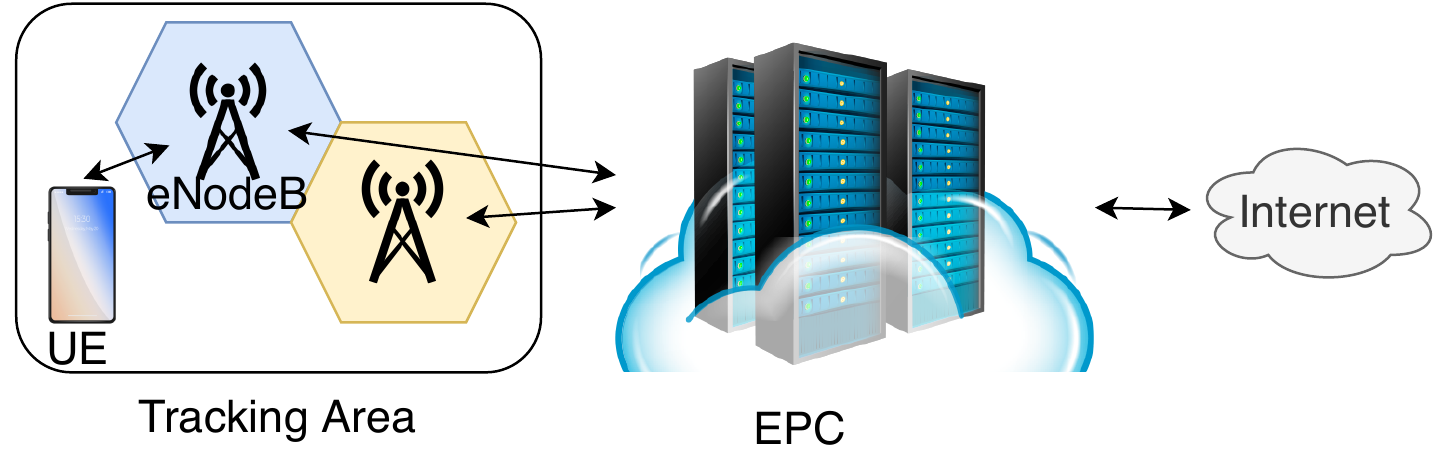}
	\caption{4G LTE Network Architecture.}
	\label{fig:lte_architecture}
\end{figure}

\textbf{LTE Architecture.}
The LTE network ecosystem can be broken down into 3 main components
(See Figure~\ref{fig:lte_architecture}): User Equipment (\textbf{UE}),
Evolved Packet Core (\textbf{EPC}) and the Radio Access Network (\textbf{E-UTRAN}).
The UE is a cellular device equipped with a SIM card. Each SIM card contains a
unique and permanent identifier known as the International Mobile Subscriber Identity (IMSI).
Also, each  device comes with a unique and device-specific identifier called
International Mobile Equipment Entity (IMEI). As both the IMSI and IMEI are unique
and permanent, their exposure can be detrimental to a user's privacy and security.
In LTE, the coverage area of a network can be broken down into hexagon cells where
each cell is powered by a base station (\textbf{eNodeB}). The network created by
the base stations powering up the coverage area and the UE is referred to as E-UTRAN.
The Evolved Packet Core (EPC) is the core network providing service to users.
The EPC can be seen as an amalgamation of services running together and continuously communicating with one another.

\textbf{LTE Protocols.}
The LTE network protocol consists of multiple layers, however,
this paper focuses only on the \emph{Network Layer}. This layer
consists of 3 protocols: NAS (Non-access Stratum), RRC (Radio Resource Control),
and IP (Internal Protocol). In this paper, we only explore NAS and RRC.
The NAS protocol is the logical channel between the UE and the EPC.
This protocol is in charge of highly critical procedures such as the
attach procedure which provides mutual authentication between the EPC and the UE.
The RRC protocol can be seen as the backbone of multiple protocols, including NAS.
In addition, RRC is the main channel between the UE and the eNodeB.

\textbf{Past-Time Propositional Linear Temporal Logic (\pltl).}
\label{sec:pltl_syntax_and_semantics}
\pltl extends propositional logic with past temporal operators
and allows a succinct representation of the temporal ordering
of events.
Therefore, we use it as one of our vulnerability signature representation.
Here, we only provide a brief overview of \pltl but the detailed presentation can be found elsewhere \cite{pastpltl1985}.
The syntax of \pltl is defined inductively below where $\Phi, \Psi$ (possibly, with subscripts) are meta-variables
denoting well-formed \pltl formulas.
\begin{equation*}
  \Phi,\Psi  \quad::=\quad \top\quad|\quad \bot\quad|\quad p \quad|\quad  \circ^{1} \Phi_1 \quad|\quad \Phi_1 \circ^{2} \Psi_1
\end{equation*}
In the above presentation, $\top$ and $\bot$ refer to Boolean constants \true\; and \false, respectively.
The propositional variable $p$ is drawn from the set of a fixed alphabet
$\mathcal{A}$ (i.e., a set of propositions). \pltl supports unary operators
$\circ^{1}\in\{\neg, \yesterday, \once, \historically\}$,
 as well as binary operators  $\circ^{2}\in\{\wedge,\vee, \since\}$.
The Boolean logical operators include $\neg$ (not), $\vee$ (disjunction), and
$\wedge$ (conjunction) and the temporal operators include
$\yesterday$ (yesterday), $\once$ (once), $\historically$ (historically), and $\since$ (since).
We will now discuss the semantics of \pltl.

The Boolean logic operators in \pltl have their usual definition as in propositional logic.
We \emph{fix an alphabet $\mathcal{A}$} (i.e., a set of propositions)
for the PLTL formulas and consider it in the rest of the paper.
The semantics of \pltl is given with respect to a Kripke structure.
In a Kripke structure~\cite{Kri63}, a trace $\sigma$ is a finite sequence of states $(\sigma_0, \dots, \sigma_{n-1})$
that maps  propositions $p$ in $\mathcal{A}$ to  Boolean values
at each step $i\in[0,n-1]$\footnote{We write $i\in[0,n-1]$ to denote $0\leq i\leq n-1$.}
(i.e., $\sigma_i(p)\in\mathbb{B}$).
Although, the standard \pltl semantics are defined over (infinite) traces,
we are only required to reason about \emph{finite} traces.

\begin{mydef}[Semantics]\label{p_sem}
Given a \pltl formula $\Phi$ and a finite trace $\sigma = (\sigma_0, \dots, \sigma_{n-1})$ of length $n\in\mathbb{N}$, the satisfiability relation $(\sigma,i)\models$ $\Phi$ ($\Phi$
holds at position $i\in\mathbb{N}$) is inductively defined as follows:
  \begin{itemize}\setlength{\itemsep}{0em}
  \item $(\sigma,i)\models\top$ iff $\models\true$
 \item  $(\sigma,i)\models\bot$ iff  $\models\false$
  \item $(\sigma,i)\models p$ iff $\sigma_i(p) = \true$
  \item $(\sigma,i)\models \neg \Phi$ iff  $(\sigma,i)\not\models\Phi$
  \item $(\sigma,i)\models\Phi\wedge \Psi$ iff  $(\sigma,i)\models\Phi$  and $(\sigma, i)\models\Psi$
 \item  $(\sigma,i)\models\Phi\vee \Psi$ iff  $(\sigma,i)\models\Phi$  or $(\sigma, i)\models\Psi$
  \item $(\sigma,i)\models \yesterday \Phi$ iff  $i>0$ and $(\sigma,{i-1})\models\Phi$
 \item $(\sigma,i)\models \once \Phi$ iff  $\exists j\in[0, i].(\sigma, {j}) \models\Psi$
 \item $(\sigma,i)\models \historically \Phi$  iff  $\forall j\in[0, i].(\sigma, {j}) \models\Psi$.
  \item $(\sigma,i)\models \Phi \since \Psi$  iff $\exists j\in[0, i].(\sigma, {j})\models\Psi$ and $\forall k\in[j+1, i].(\sigma, k)\models\Phi$
  \end{itemize}
\end{mydef}

Intuitively, $\yesterday \Phi$ (read, Yesterday $\Phi$) holds in the current state if and only if
the current state is not the initial state and $\Phi$ held in the previous state. $\Phi \since \Psi$
holds  true currently if and only if
$\Psi$ held in any previous state (inclusive) and $\Phi$ held in all successive
states including the current one.
%
The rest of temporal operators \once (read, true once in the past) and \historically (read,
always true in the past)  can be defined through the following
equivalences: $\once\Phi\equiv (\top\since\Phi);
\historically\Phi\equiv \neg(\once (\neg\Phi))$.

\section{Overview of \systemtitle}
In this section, we discuss the scope, threat model, challenges, and requirements of
a \system like system. We conclude by presenting two concrete instantiations of \system, namely,
as a warning system and a full-fledged defense.

\subsection{Undesired Behavior and Scope}

In our presentation, we define an
\emph{undesired behavior/vulnerability}
broadly to include
inherent protocol flaws at the design-level,
an exploitable implementation vulnerability
of the baseband processor, an exploitable misconfiguration or
deployment choice of a network operator, and unsafe security
practices by a baseband manufacturer and network operator.
For instance, not using encryption for protecting traffic
is considered a vulnerability in our presentation. Even though null encryption
is permitted by the specification on the NAS layer \cite{3gppNAS},
we argue that this is an unsafe practice since subsequent NAS traffic (e.g., SMS over NAS~\cite{kim_ltefuzz_sp19, lteinspector})
would be exposed in plaintext. 

In this paper, we focus on the undesired behavior
of the 4G LTE control-plane protocols, i.e., protocols
running in the NAS and RRC layers \cite{lteinspector, TORPEDO, privacy_ndss16, kim_ltefuzz_sp19, 5Gformal_authentication_basin,
5g_reasoner, lte_redirection, how_not_to_break_crypto}.
Among these attacks, we focus on attacks that are detectable from the
device's perspective and can be viewed as undesired outcomes of protocols'
state-machines. \emph{One distinct advantage of a device-centric attack detection mechanism
is that certain attacks necessarily cannot be observed by the network operators, which is
observable only from the device vantage point}. Examples of such attacks include ones that require
an adversary setting up a fake base station that lures the victim device and then launch an attack \cite{lteinspector,kim_ltefuzz_sp19, TORPEDO}.
Attacks that target other network components
or employ adversary's passive sniffing capabilities are out of scope as they are not detectable
through in-device traffic monitoring \cite{TORPEDO, alter,guti_reallocation_demystified_ndss18}.
In addition, the current instantiations of \system do not support attacks that require reasoning about quantitative
aspects (e.g., the number of certain messages received in a time window) of the protocol (e.g., ToRPEDO attack \cite{TORPEDO}).
Please consult Table \ref{app:cellular_network_extensive_list} in the Appendix for an
exhaustive list of \system supported and unsupported attacks.

\subsection{Threat Model}
We consider an adversary with the following capabilities:
(1) He has access to malicious cellular devices with legitimate
credentials;
(2) He can setup a rouge base station, cloning parameters of a legitimate one,
provides a higher signal strength than legitimate base stations within the vicinity.
(3) He can setup a base station which acts as a relay between the device
and legitimate base station, enabling him to drop, replay, and inject messages
at will while respecting cryptographic assumptions;
(4) For targeted attacks, we assume the attacker has access to the victim's
soft identity such as phone number and social network profile.
We assume that the device in which \system  runs is not compromised.

\subsection{Example: A Privacy Attack on Radio Link Failure (RLF) Report}
\label{sec:running_example}

In cellular networks, there is essentially
no authentication mechanism between a device and the base station during the connection initiation with the core network.
The device trusts the base station emitting the highest signal strength and
establishes an unsafe connection with it using unprotected RRC layer messages.
The base station acts as the trusted intermediary to facilitate communication between
the device and core network. Once the device and core network mutually authenticate each other,
they setup a security context making all the following control-plane messages to be encrypted and integrity protected.
One  such control-plane message is the \rlfReport which contains neighboring base stations'
signal strengths (and, optionally the device's GPS coordinates). This is used to identify
potential failures and aids when identifying coverage problems.

A privacy attack against
this RLF report message \cite{privacy_ndss16} proceeds by luring a cellular device to connect to a rogue base station, which exploits
the lack of authentication of initial broadcast messages as well as the unprotected RRC connection setup in the bootstrapping phase.
Before setting up the security context (with protected \securityModeCommand and \securityModeComplete messages)
at the RRC layer, the rogue base station sends an unprotected \ueInformationRequest
message to the device. This triggers the device to respond with a \rlfReport message (if it posses one) in the clear.
Since the RLF report includes signal strength measurements of neighboring cells (and optionally GPS coordinates),
the attacker can use that information to triangulate  the victim's location.

\subsection{Challenges}
Realizing the vision of \system has the following challenges.
(C-1) An attack detection mechanism like \system has to be lightweight,
otherwise substantial overhead can impede adoption due to negatively impacting the user's Quality-of-service (QoS).
(C-2) The system must be able to operate in a standalone fashion without
requiring assistance from network operators.
(C-3) The system must be attack- and protocol-agnostic, and amenable to extension
to new attacks discovered after its deployment and future protocol versions (e.g., 5G).
(C-4) The detection accuracy of the system must be high (i.e., low false positives and negatives).
If the system incurs a large number of false positives, then in its instantiation
as part of the baseband processor, can create interoperability issue. In the same vein, false positives in
\system{}'s instantiation as a warning system can overwhelm the user, making her ignore the raised warnings.
A large number of false negatives, on the other hand, makes the system prone to vulnerabilities.
(C-5) The attack detection system should detect the attack as soon as it is feasible when
the malicious session is underway. As an example, let us consider the above attack on RLF report.
If a detection system identifies the attack only after the device has already sent the \rlfReport message in the clear
to the adversary then the attack has happened and this reduces the impact of a detection system like \system. An
effective detection mechanism will identify the attack as soon as the device receives the
unprotected  \ueInformationRequest before security context establishment in which case
it can thwart the attack.

\subsection{\systemtitle Architecture}
\begin{figure}[t]
	\centering
		\includegraphics[width=\columnwidth]{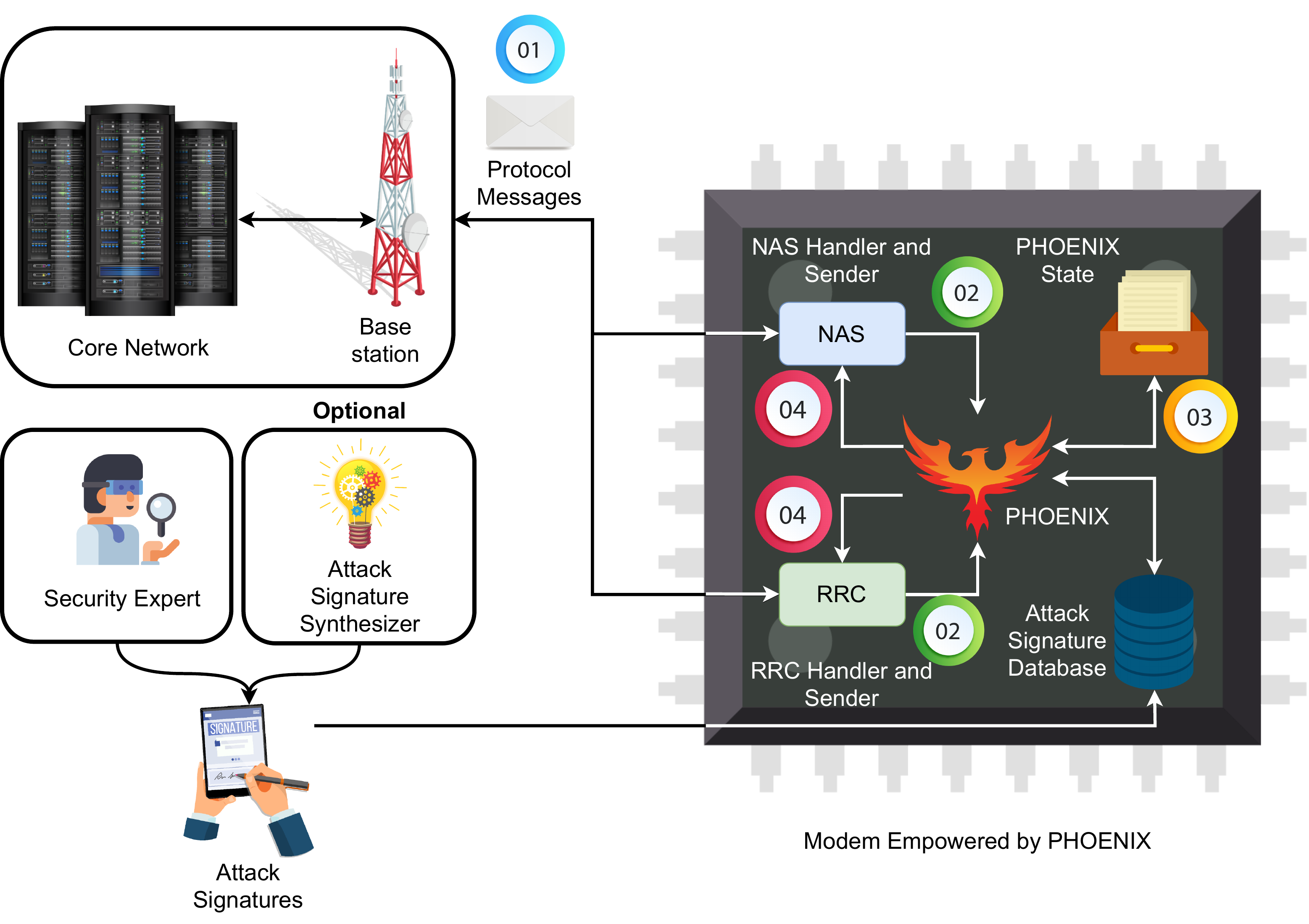}
		\caption{The envisioned architecture of \system inside a baseband processor.}
		\label{fig:baseband_implementation}
	\end{figure}

\begin{figure}[t]
\centering
	\includegraphics[width=\columnwidth]{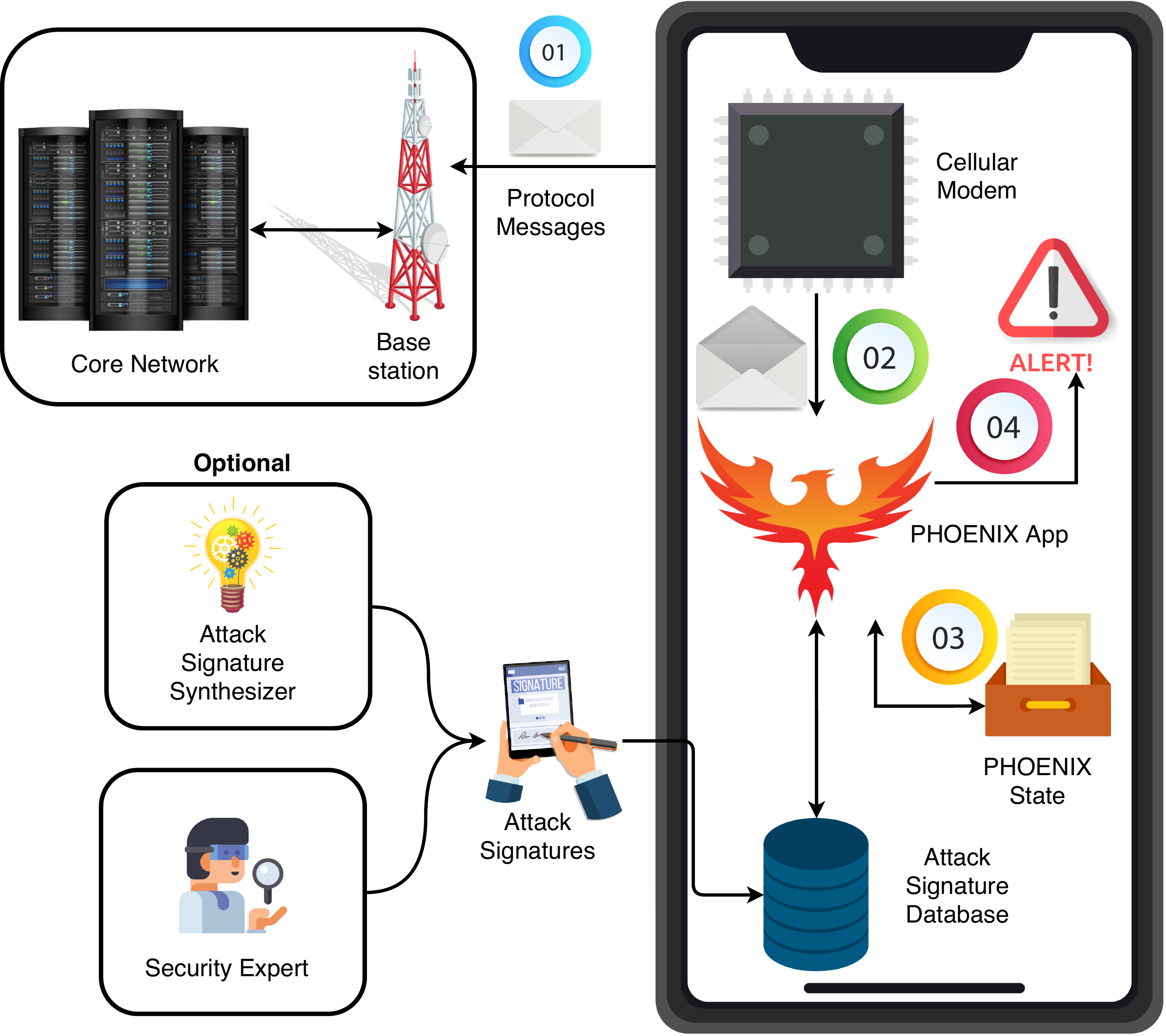}
	\caption{The envisioned architecture of \system as an Android app.}
	\label{fig:overview}
\end{figure}

We now discuss the architecture of \system in two settings:
(1) when it is deployed inside a baseband processor as a full-fledged
defense (see Figure \ref{fig:baseband_implementation});
(2) when it is deployed as an Android application and serves as a warning system (see Figure \ref{fig:overview}).



\paragraph{\systemtitle Components.}
In its purest form (Figure~\ref{fig:baseband_implementation}), \system has two main components, namely,
\textit{Attack Signature Database} and  \textit{Monitor}.

\textbf{Attack Signature Database.} \system expects a pre-populated attack signature database containing
the signatures of attacks it is tasked to detect. An example attack signature for the privacy attack on
RLF report above is: \emph{receiving the unprotected  \ueInformationRequest message
before security context establishment in a session.} Note that, a signature that requires the device to send a \rlfReport
message before security context establishment is ineffective as it detects the attack only after it has occurred. Signatures
can be generated by cellular network security experts, possibly in collaboration with an optional \system component that
can automatically generate candidate signatures from benign and attack traces.

\textbf{Monitor.} The \monitor  component analyzes the decoded messages and payloads (potentially,
received from the message extractor component discussed below in case of Android app deployment),
and matches them with its pre-populated undesired behavioral signature database.
In case a behavioral signature is identified, the action of \monitor component depends on the deployment scenario.
For its baseband processor deployment, the \monitor communicates the violation information to a corrective action
module who can either terminate the session or drop the particular message depending on the signature. In its
Android app deployment, it identifies which vulnerabilities have occurred and returns this information to the user along
with possible remedies, if any exists.

For its instantiation as an Android app, \system requires an additional component called \textit{message extractor}.
It gathers information about incoming/outgoing traffic (e.g., decoding a protocol message)
between the baseband processor and network. This collected information
(e.g., message type, payload) is then fed into the \textit{\monitor} component
for vulnerability detection. Note that, in the baseband deployment, \system does
not require this component as the baseband processor inherently decodes and
interprets the messages.

\paragraph{Workflow of \systemtitle.}
The workflow of \system deployed as an Android app is given below.
The baseband deployment does not require step (1) of the workflow.

(1) The \textit{message extractor} intercepts
an incoming/outgoing protocol message and decodes it.
(2) Pre-defined predicates over this message
(and, its payload) are then calculated and sent to the \monitor.
(3) The \monitor then classifies the ongoing trace as either benign or
vulnerable (with label). 
(4) If \system identifies a vulnerability, it either drops the message/terminates
the connection when implemented inside a baseband processor, or alerts the user
of the undesired behavior with possible remedies when deployed as an Android
app (see Figure \ref{fig:phoenix_app_screenshots} for an example)

\section{Vulnerability Signatures and Monitors}
\label{sec:approaches}
In this section, we discuss the possible vulnerability signature representations and their monitors that we consider.

\subsection{Insight on Vulnerability Signatures}
After analyzing existing control-plane attacks on 4G LTE \cite{lteinspector, TORPEDO, privacy_ndss16, kim_ltefuzz_sp19,
5Gformal_authentication_basin, 5g_reasoner, lte_redirection, how_not_to_break_crypto}, we observed that
a substantial amount of these attacks have very specific behavioral signatures
when considering protocol messages, their payloads, and predicates over them.
Precisely, considering the relative ordering of events often are sufficient
to synthesize a discernible and precise vulnerability signature. For instance,
in the running example described in Section \ref{sec:running_example},
not seeing both the \securityModeCommand and \securityModeComplete
messages prior to the \rlfReport being exposed,
can serve as a confident indicator for such vulnerability.

\subsection{Vulnerability Signature Representations}
To precisely capture the behavioral signatures of cellular network vulnerabilities,
we consider regular languages and \pltl as two possible representations.
These formalisms are chosen
due to their effectiveness in capturing relative temporal ordering of events
as well as being efficiently monitorable at real-time. In addition, there is one
more representational question we have to address: \emph{Does one keep per-vulnerability`
signatures or one giant signature capturing all of the considered vulnerabilities}? These
design choices induce the following
signature representations.

\begin{figure}[t]
  \centering
	\includegraphics[width=.8\columnwidth]{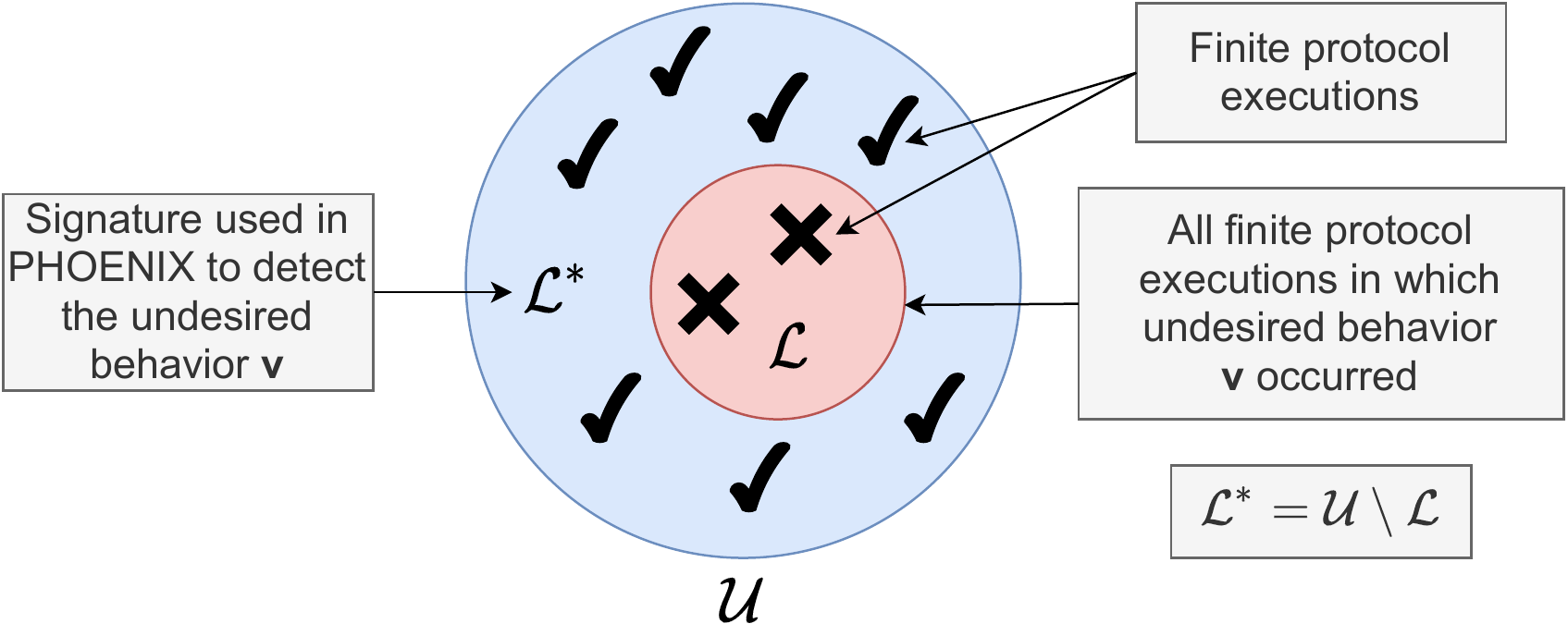}
	\caption{Universe of finite protocol executions separated into two regular languages,
  depending on if the undesired behavior
  \textbf{v} occurred or not (resp., $\lang$ and $\lang^*$).}
	\label{fig:universe_of_traces}
\end{figure}

\textbf{Signatures as Regular Languages.}
In this scheme, let us consider \calc{U} to be all finite protocol execution
traces. Let us denote all the finite protocol executions in which a given vulnerability $v$
occurs as a regular language \lang. Then the behavioral vulnerability signature we consider
is the language $\lang^* = \calc{U}-\lang$ which is the complement of \lang and accepts all finite protocol
execution traces where $v$  does not happen (See Figure \ref{fig:universe_of_traces}). This signifies that
$\lang^*$ will only reject traces in which $v$ happens. For representing $\lang^*$,
we consider the protocol message types, their payloads, and predicates over them
as the alphabet. For a given vulnerability whose behavioral
signature is denoted by $\lang^*$, we represent its signature as a deterministic
finite automata (DFA). For the case of having one giant signature for all vulnerabilities,
we use a Mealy Machine whose outputs in the transitions indicates whether a certain
execution is benign (labeled with output \texttt{benign}) or vulnerable in which case the output label identifies the vulnerability.

\textbf{Signatures as \pltl formulas.}
\pltl has been shown to be a natural candidate for succinctly representing the
temporal ordering of events of the past. We use message types, their payloads,
and predicates over them as propositions of the logic.
In this scheme, we keep one behavioral
signature as a \pltl formula for each vulnerability that rejects only those finite
traces in which the vulnerability in question occurs. We do not keep a giant \pltl
formula for all vulnerabilities as it would not allow us to identify the particular vulnerability
that occurs, impairing us to provide vulnerability-specific remedies and severity.

\subsection{Vulnerability Monitors}
We now discuss how we monitor vulnerability signatures based on their
representations.

\paragraph{Monitoring Regular Language Signatures.}
For monitoring a signature represented as a DFA, we need to store the DFA
along with the current state in the memory. When a new packet and its associated
information arrives to the \monitor, we try to take a transition in DFA. If the
transition lands us on a non-accepting state that means a vulnerability has been observed
in which case we raise an alarm and provide vulnerability-specific information (e.g.,
name of the vulnerability, severity, and remedies). In case of a benign scenario, we
just take the transition and update the current state. The monitoring with respect
to a Mealy Machine is very similar with the one difference is that the output
label of the transition indicates whether a vulnerability has been observed, and if so
which particular vulnerability was observed.

\paragraph{Monitoring \pltl Signatures.}
For monitoring \pltl formulas, we consider a standard dynamic programming (DP) based approach
from the literature of runtime verification \cite{d2005efficient, basin2010monitoring, basin2010policy,
 monpoly, bauer2011runtime, rosu2001synthesizing}.

In this approach,
to monitor a \pltl formula $\Phi$, the \monitor requires one bit of information
for each sub-formula of $\Phi$. This bit signifies whether the associated formula
holds true in the current state. If the truth value bit of $\Phi$ is true
in the current state, then there is no vulnerability.
For a given \pltl formula $\Phi$, let us assume
that $\llbracket\Phi\rrbracket^i$ represents the truth value bit of formula $\Phi$
at position $i$ of the trace. Adhering to the \pltl semantics,
the DP algorithm constructs
$\llbracket\Phi\rrbracket^i$ from $\llbracket\Phi\rrbracket^{(i-1)}$ and the current
state $\sigma_i$ in the following way. Note that, we just need to store $\llbracket\Phi\rrbracket^{(i-1)}$
to calculate $\llbracket\Phi\rrbracket^{(i)}$.
The current state $\sigma_i$ in our presentation is a total map which maps
each propositional variable in the alphabet \calc{A} to either true or false.
\begin{align*}
\llbracket p \rrbracket^i  &=  \sigma_i(p)\\
\llbracket\neg\Phi\rrbracket^i &= \neg\llbracket\Phi\rrbracket^{i}\\
\llbracket\Phi\wedge\Psi\rrbracket^i &= \llbracket\Phi\rrbracket^{i} \wedge \llbracket\Psi\rrbracket^{i}\\
\llbracket\yesterday\Phi\rrbracket^i &= i>0 \wedge  \llbracket\Phi\rrbracket^{(i-1)}\\
\llbracket\Phi\since\Psi\rrbracket^i &= \llbracket\Psi\rrbracket^{i} \vee (\llbracket\Phi\since\Psi\rrbracket^{(i-1)}\wedge \llbracket\Phi\rrbracket^{i})
\end{align*}
\label{sec:pltl_description}

\section{Automated Vulnerability signature Synthesis}\label{atk-syn}
We now discuss the design of the optional \system component called \signatureSynthesizer.

\subsection{Potential Application of the Signature Synthesizer}
For using the \system system, we want to emphasize it is not mandatory to have the
signature synthesizer component; a cellular network security expert will suffice for
generating signatures. Despite that, an automatic signature synthesizer can be useful to the expert
in the following three scenarios.

First, when a cellular network security expert knows the root cause of an attack but does not
know how to represent it one of the forms, then it can use the signature synthesizer to generate
a candidate signature. DFA and MM signatures can be particularly complex.
Please see Figure \ref{fig:aka_bypass_dfa} in the Appendix
for the DFA signature of the AKA bypass attack \cite{kim_ltefuzz_sp19}.
Second, when an expert neither knows the root cause of a newly discovered attack nor knows the signature representation, the
signature synthesizer, especially the PLTL synthesizer because of its ability to generate succinct signatures,
can be particularly helpful for not only
identifying the root cause but also to synthesize the signature in the appropriate representation.
Finally, the runtime and space overheads of monitors, especially the PLTL-based monitor, are proportional
to the length of the signature. As the PLTL signature synthesizer is guaranteed to generate the minimum length
signature, it induces an efficient monitor. \emph{We envision a more collaborative process between
the automatic signature synthesizer and a human expert; instead of completely bypassing the expert and only using
the synthesizer in a standalone fashion. In this envisioned process,
the human expert asks the synthesizer to
generate multiple candidate signatures and then chooses the one she finds more appropriate.}
Such a collaborative interaction reliefs the human expert to be also an expert of formal logic like PLTL.

\subsection{The Problem of Signature Synthesis}
The signature synthesis problem is
an instance of the \emph{language learning from the informant} problem \cite{informant_learning}.
In this problem, for a fixed alphabet $\mathcal{A}$,
an \emph{informed learning sample (i.e., training dataset)} \calc{D} is given which
comprises of two disjoint sets of strings
\calc{P} and \calc{N}, such that $\calc{P}\cap \calc{N}=\emptyset$.
The aim is to learn an \emph{observationally
consistent}  language $\mathcal{L}$ that accepts all strings in \calc{P} and rejects all strings in \calc{N}.
In our setting, without the loss of generality,
for a given vulnerability \textbf{v} the set \calc{N} are vulnerable execution traces in which \textbf{v} happens
and the set \calc{P} are (benign) traces in which \textbf{v} does not happen. Then the
learned observationally consistent
language $\mathcal{L}$ represents the vulnerability signature for \textbf{v}.


\subsection{Regular Language Signature Synthesis}
\label{sec:regular_languages}
The observationally consistent language $\mathcal{L}$ is considered to be
regular and
we used variations of the
RPNI (Regular Positive and Negative Inference) algorithm \cite{rpni}
to learn both DFA and Mealy machine based vulnerability signatures.
The complexity time of RPNI is the
following: $\bigO(l\cdot\vert\Sigma\vert\cdot k^4)$, where $l$ is the total
number of states in the negative traces, $\vert\Sigma\vert$ is the total size
of the alphabet, and $k$ is the number of unique prefixes \cite{rpni}.
Below we discuss how to prepare \calc{P} and \calc{N} that are
required inputs to the RPNI algorithm.


\paragraph{DFA Signature Synthesis.} For a given vulnerability $v$,
we are given two sets of traces $\Sigma_+$ (i.e., $v$ does not happen in these traces)
and $\Sigma_{-}$ (i.e., $v$ happens in these traces) such that
$\Sigma_+ \cap \Sigma_{-} = \emptyset$. For each  positive trace $\sigma_+\in\Sigma_+$,
we add $\sigma_+$ and all its prefixes to \calc{P}. We set $\calc{N}=\Sigma_{-}$.
We then invoke the RPNI \cite{rpni} algorithm for obtaining a DFA signature for $v$.

\paragraph{Mealy Machine Signature Synthesis.} We are given a
set of vulnerabilities $V.$ For each such vulnerability $v_i\in V$,
we are given two sets of traces $\Sigma^i_+$ (i.e., $v_i$ does not happen in these traces)
and $\Sigma^i_{-}$ (i.e., $v_i$ happens in these traces) such that
$\Sigma^i_+ \cap \Sigma^i_{-} = \emptyset$. For each positive trace $\sigma_+\in\Sigma_+$,
we add $\sigma_+$ to \calc{P} and assign the output label \texttt{benign}.
We add each negative trace $\sigma_{-}\in\Sigma_{-}$ to \calc{N} with output label
$\mathrm{\texttt{vulnerability}}_i$ and then invoke the RPNI algorithm for obtaining a combined
Mealy machine signature for all vulnerabilities in $V.$

\subsection{\pltl Signature Synthesis}
\label{sec:pltl_synthesis}
A \pltl formula represents the observationally consistent language $\fml{L}$ that constitutes a vulnerability signature.
For synthesizing \pltl signatures, we propose a syntax-guided synthesis algorithm that extends Neider and Gavran \cite{learning_ltl} to
learn PLTL formulas using only finite length traces.
The proposed algorithm reduces the signature synthesis problem to a Boolean satisfaction problem (SAT) and
then solve it using an off-the-shelf SAT solver. In this setting, any satisfiable assignment (or, \emph{a model}) of that SAT
problem instance is used to derive observationally consistent \pltl signature.
We aim to learn minimal consistent signatures as
they can  capture a concise
vulnerability behavior
even from a smaller training dataset and are also intellectually
manageable (readable).
This feature is inherent to this algorithm in contrast to other representations (i.e., DFA and
Mealy machine).
Precisely, a formula $\Phi$ is minimally consistent with $\fml{D}$ if and only if $\Phi$ is consistent
with $\fml{D}$ and for every other \pltl formula $\Psi$ such that $|\Psi|<|\Phi|$, $\Psi$  is
inconsistent. Here $|\cdot|$ is a function that takes
a \pltl formula as input and returns the number of its sub-formulas.
Also, this algorithm can provide different candidate signatures for a given sample \calc{D}
by enumerating different models of the SAT problem. Thus, it provides the user with more
flexibility to select the most desirable signature among the suggested candidates.

\begin{algorithm}
\caption{\pltl Syntax-Guided Synthesis Algorithm}\label{algo1}
\hspace*{\algorithmicindent} \textbf{Input: }{Training dataset $\mathcal{D=(P,N)}$
and alphabet $\mathcal{A}$}\\
\hspace*{\algorithmicindent} \textbf{Output: }{Minimally consistent signature $\Phi_\ell$ of size $\ell\in\mathbb{N}$}
\begin{algorithmic}[1]
\State $\ell \gets \text{1}$ 
\While{$\ell \leq \Delta$} //$\Delta$ is a constant threshold
\State $\varphi_\ell \gets \text{encode}(\mathcal{D}, \ell)$
\State $m \gets \textsf{SAT}(\varphi_{\ell})$
\If {$m \neq \emptyset$}
    \State $\Phi_{\ell} \gets \text{decode(m)}$ 
    \State \Return $\Phi_\ell$
\Else
    \State $\ell \gets \ell +1$ 
\EndIf
\EndWhile
\end{algorithmic}
\end{algorithm}
\paragraph{Algorithm.}
For a given training dataset \calc{D} and alphabet \calc{A} (i.e., a set of
propositional variables), our learning algorithm (Algorithm \ref{algo1})
iterates over the depth of the \pltl formula
abstract syntax tree (AST) in ascending order.
For a given depth of the formula AST $\ell$,
the algorithm has two main steps: \ding{182} Generate all possible \pltl formulas
whose AST depth is exactly $\ell$; \ding{183} Check whether one of the generated
formulas is consistent with $\mathcal{D}$. Although logically the algorithm has two steps,
one can use a SAT solver to perform both searches simultaneously.
The advantage of such an approach is that the constraints capturing the
restrictions in step \ding{183} can rule out formulas from search at step
\ding{182}. We now, at a high-level, describe
how both steps are encoded as a SAT formula.

The first set of constraints
are regarding the syntax of the \pltl formula. These constraints are conjunctions
of the following:
(1) constraints for generating all  ASTs of depth $\ell$;
(2) constraints for assigning labels (i.e., propositions and operators) to the
AST nodes. Example constraints in the label assignment
include operators cannot be assigned to leaf nodes,  and binary operators can only be
assigned to nodes having two children. These
constraints are required to be strong enough to ensure that
only syntactically well-formed \pltl formulas are considered~\cite{mc03}.
Based on \pltl semantics, the second set of constraints capture that the synthesized formula should satisfy
all traces in \calc{P} while rejecting all traces in \calc{N}.

The encode function in the algorithm, given the AST depth $\ell$  and the training dataset
$\mathcal{D}$, generates a propositional formula $\varphi_\ell$ that capture these constraints.
%
The algorithm then uses an off-the-shelf SAT solver to search for a model of $\varphi_\ell$.
If a model  $m$ is found, it is decoded to obtain an \pltl formula
$\Phi_\ell$ that represents the consistent vulnerability signature. If no model is found,
the algorithm increments the bound size (i.e., $\ell$) and the search procedure continues
until a satisfying assignment is found or the bound threshold  is exceeded (i.e., $\ell > \Delta$).

\section{Implementation of \system}
We instantiate \system in two settings: a full-fledged defense as part of the
baseband processor and also as an Android app serving as warning system.
To study the overhead of \system when
running inside a baseband processor, we implement \system by modifying srsUE distributed as part of
srsLTE open-source protocol stack \cite{gomez2016srslte}. To analyze the effectiveness of \system as a warning
system, we implement the message extractor and the monitor in an Android
application on different devices. The optional \signatureSynthesizer component of
\system is developed as a standalone program.

\subsection{\system Implementation With srsUE}
\label{sec:baseband_implementation}
To simulate \system{}'s integration into the baseband
processor, we extend srsUE \cite{gomez2016srslte} so that it can
detect an
undesired behavior. As a baseband processor (similarly, srsUE) parses a message,
\system does not need to parse messages
and instead need to focus on the monitor component. For this instantiation,
we used the PLTL-based monitor because it is the
most effective monitor instantiation according to our evaluation in Section \ref{sec:evaluation_of_phoenix}.

\textbf{PLTL monitor.} In order to achieve a highly efficient implementation,
both when considering memory and computational overhead, we leverage the work
by Rosu et al. \cite{rosu2001synthesizing} to synthesize dynamic programming
algorithm-based PLTL monitors in C++. The runtime and memory requirements of these
monitors are constant with respect to the signature size.

\textbf{Monitor integration.}
Depending on the information required to evaluate a signature,
the monitors are integrated in either the RRC or NAS namespace
files, which are responsible for the handling (and sending)
messages of each layer.
In each such message handling/sending function, prior to processing
or sending a message, the entry point of \system is invoked with the label of the
new event. In order to empower \system to drop messages or
close the connection altogether, \system returns a Boolean value
representing whether or not at least one signature was violated, in order
to let the function either proceed with the handling (or sending) process or drop
the connection to prevent a vulnerability.

\subsection{\system Implementation as an Android App}
When implemented as an Android app, we instantiated \system
with DFA-, MM-, and PLTL-based monitors. We now discuss the
major component implementations.

\textbf{Message Extractor.} The message extractor first reads events from the baseband processor.
For efficiently parsing protocol packets, we 
modified MobileInsight~\cite{mobile_insight} application's traffic
dissector to efficiently capture NAS and RRC layers' traffic. We then apply
any required propositions and forward the message to the \monitor.
Note that since we modified MobileInsight to implement the message extractor,
\system requires root privileges to function. These types of apps require root access since normal
applications do not have access to the virtual device where the modem information
is exposed \cite{mobile_insight}.

\textbf{Monitor Component.}
Since MobileInsight is written with Python and compiled into an Android App using Python for Android~\cite{python_for_android},
we implement our monitors in the same fashion.
We now discuss the implementation details of the monitors for each of the attack signature representations.

\textit{DFA.} For an attack signature,
our 
DFA-based \monitor
stores the set of transitions, list of accepting states,
current state, and the alphabet in memory. The transition relation in our
implementation is just a dictionary lookup. A transition to a non-accepting state
is considered an attack.

\textit{MM.} Mealy machine-based \monitor is similar to the one for DFA with one
exception. Since Mealy-machine does not have any accepting and non-accepting states,
the output symbol of the transition indicates which particular attack has occurred.

\textit{PLTL.} We implemented the dynamic programming  algorithm \cite{rosu2001synthesizing} for monitoring \pltl formulas in Python.
Our implementation stores a single bit for each sub-formulas truth value and uses bitwise operations to identify the truth values.

\subsection{Signature Synthesizer} The implementation details of the
optional \signatureSynthesizer component is as follows.

\textbf{DFA.} For learning DFA signatures, we use the RPNI passive automata learning algorithm implemented in LearnLib~\cite{learnlib}.
We provide the attack traces as well as non-attack traces and all their prefixes as input. We also include empty string ($\epsilon$) as part of the positive
sample because without it the initial state of the synthesized DFA is marked as non-accepting.

\textbf{Mealy Machine.} Similar to DFA, we invoke the RPNI algorithm of LearnLib \cite{learnlib} to serve as the \signatureSynthesizer for
Mealy Machine. Each message in the trace is also mapped with its corresponding output (i.e., $\mathsf{benign}$ or $\mathsf{vulnerability}_i$).

Note that, since Mealy Machine is a monitoring mechanism capable of detecting multiple attacks at the same time, the training set
contains all the traces for that corresponding layer.

\textbf{PLTL.} To instantiate our \pltl \signatureSynthesizer, we implement the algorithm  in Section~\ref{sec:pltl_synthesis}.
Our implementation uses PySMT, a Python-based solver-agnostic library
built on top of SMT-LIB~\cite{smtlib}. 
By leveraging our \pltl \signatureSynthesizer's capability of
generating different candidate signatures, we create 5
candidate signatures
for each attack with 80\% of the training data. We then evaluate the candidate
signatures on the remaining 20\% of training data to pick the best one.
In case of a tie, we choose the smallest signature.

\section{Evaluation Criteria and Setup}

In this section, we discuss the evaluation criteria, experimental setup, and
trace generation for our evaluation.

\subsection{Evaluation Criteria}
\label{sec:evaluation_criteria}

\paragraph{Research Questions.}
We first aim to address the following research questions
for \system{}'s \signatureSynthesizer:
\begin{itemize}[leftmargin=0.35in]
	\item [$\mathsf{QS_1}$.] How effective are the synthesized signatures?
	\item [$\mathsf{QS_2}$.] How scalable are the signature synthesizers?
	\item [$\mathsf{QS_3}$.] Does training set size impact the quality of signatures?
\end{itemize}

We next focus on evaluating the monitor component, when considering the warning
system implementation, by answering to the following research questions:

\begin{itemize}[leftmargin=0.49in]
	\item [$\mathsf{QWS_1}$.] How many messages/second can a monitor classify?
	\item [$\mathsf{QWS_2}$.] What is the energy consumption overhead for a monitor?
	\item [$\mathsf{QWS_3}$.] What type, and how many, warnings do the different
	monitors produce when \system is deployed on real cellular networks?
\end{itemize}

We then evaluate the monitor component, when considering the baseband implementation,
by answering the following research questions:

\begin{itemize}[leftmargin=0.46in]
	\item [$\mathsf{QBB_1}$.] What is the memory overhead induced by \system?
	\item [$\mathsf{QBB_2}$.] What is the computational overhead induced by \system?
\end{itemize}

\subsection{Experiment Setup}
In this subsection, we provide details on the experimental setup for both components.

\paragraph{Signature Synthesizer Evaluation Infrastructure.} We perform all the signature synthesizer evaluation on a 4.5GHz
Intel i7-7700K CPU running Ubuntu 16.04 on 16GB of RAM. We set a time out
of 3,600 seconds for these experiments.

\paragraph{PHOENIX Baseband Implementation.} We perform the baseband
implementation experiments by implementing \system into srsUE as described in
Section \ref{sec:baseband_implementation} on a 4.5 GHz Intel i7-7700K CPU
running Ubuntu 16.04 on 16GB of RAM
connected to a USRP board~\cite{usrp}.

Note that we do not measure the power consumption in this instantiation as any
meaningful measurement would require additional appropriate hardware. Additionally, the baseband implementation experiments do not leverage a stress test
as it is not clear how to achieve this with srsUE \cite{gomez2016srslte}.

\paragraph{Sample Sizes.} We consider different sizes of traces (50, 100, 250, 500, 1250, and 2500) in our evaluation. In each trace, $50\%$ are positive
and the rest are negative. To generate these traces, we used
the procedure mentioned in Section\ref{sec:trace_gathering}.

\paragraph{Training and Testing Separation.} To measure the effectiveness of the  signatures,
we create disjoint testing and training sets for each attack, containing
1000 benign and 1000 malicious traces using the procedure mentioned in Section~\ref{sec:trace_gathering}.

\paragraph{Monitor Evaluation Testbed.}
We perform all the monitor experiments on three different
COTS Android devices (see Table~\ref{tab:device_for_experiment}
for devices' details).
Also, following the prior work~\cite{alter,lteinspector,kim_ltefuzz_sp19} we set up a similar 4G LTE testbed (consisting of eNodeB and EPC) using
srsLTE \cite{gomez2016srslte} and USRP B210~\cite{usrp} connected to Intel
Core i7 machines running Ubuntu 16.04 with 16 GB of memory.

\paragraph{Effectiveness Evaluation.}
To evaluate effectiveness of the signatures, we implement \system to its entirety
and replay benign and malicious traces through srsLTE~\cite{gomez2016srslte}.

\paragraph{Efficiency Evaluation.}
To evaluate efficiency through a \emph{stress test}, we develop an application that serves
as an in-device network simulator by replaying the logs within the device. We use this  setup because software-defined radios
have inherent limitations on transmission bandwidth. Therefore, a high-volume of packets within a short time-interval cannot be injected to the device for stress testing, which is important for realizing our monitors' efficiency in real networks.

\paragraph{Set of Attacks}.
\label{sec:set_of_attacks}
We consider 15 attacks (Table~\ref{tab:all_attacks}) for our evaluation.
The reason for considering these 15 attacks are twofold:
(1) These attacks can serve as representatives of
most of the known vulnerabilities in 4G LTE control-plane layers;
and (2) They have at least one of the following characteristics:
(a) violation of temporal ordering of events;
(b)  triggered by rogue eNodeB or Mobility Management Entity (MME) at RRC or NAS layers.
%

\begin{table}[htbp]
  \centering
  \resizebox{.8\columnwidth}{!}{
	\begin{tabular}{|l|l|l|}
		\hline
		\multicolumn{1}{|c|}{\textbf{Phone Model}} & \multicolumn{1}{c|}{\textbf{CPU}} & \multicolumn{1}{c|}{\textbf{Operating System}} \\ \hline
		Pixel 3 & Qualcomm Snapdragon 845 \cite{snapdragon845} & Android 9 \\ \hline
		Nexus 6P & Qualcomm Snapdragon 810 \cite{snapdragon810} & Android 8.0.0 \\ \hline
		Nexus 6 & Qualcomm Snapdragon 805 \cite{snapdragon805} & Android 5.1.1 \\ \hline
	\end{tabular}
  }
	\caption{Specifications of devices used for evaluation.}
	\label{tab:device_for_experiment}
\end{table}

\subsection{Trace Generation for Evaluation}\label{sec:trace_gathering}
We now discuss how we generate traces for evaluating \system{}'s monitor and
optional signature synthesizer components. We use the following approach
to generate a large number of  traces containing undesired behavior to evaluate scalability of the
synthesizers. Also, a different set of traces generated with this approach is used to evaluate the
effectiveness of \system's monitor.

\subsubsection{Sessions, Traces, and Variants}
We now introduce the concepts of a \emph{session}, \emph{trace}, and \emph{variants of an attack session} used later.
A \textbf{session}, which can be logically viewed as a sequence of protocol messages,
starts off with the device sending a connection initiation request (e.g., \rrcConnectionRequest,
\attachRequest) and contains all messages (including the current connection initiation request message)
until the next connection initiation request is sent. Note that, we do not say that a session ends with a termination request to
facilitate sessions which end abruptly.  A \textbf{trace} is just a sequence of
sessions. We call a session \textbf{$\beta$-undesired-behavior-session} if the undesired behavior $\beta$
occurs in that session. For a canonical $\beta$-undesired-behavior session $s$ (obtained from the original source
of the undesired behavior discovery), we call another \textbf{$\beta$-undesired-behavior-session} $\hat{s}$ a variant of $s$, only if
$s\neq\hat{s}$.

\begin{figure}[t]
	\centering
	  \includegraphics[width=.9\columnwidth]{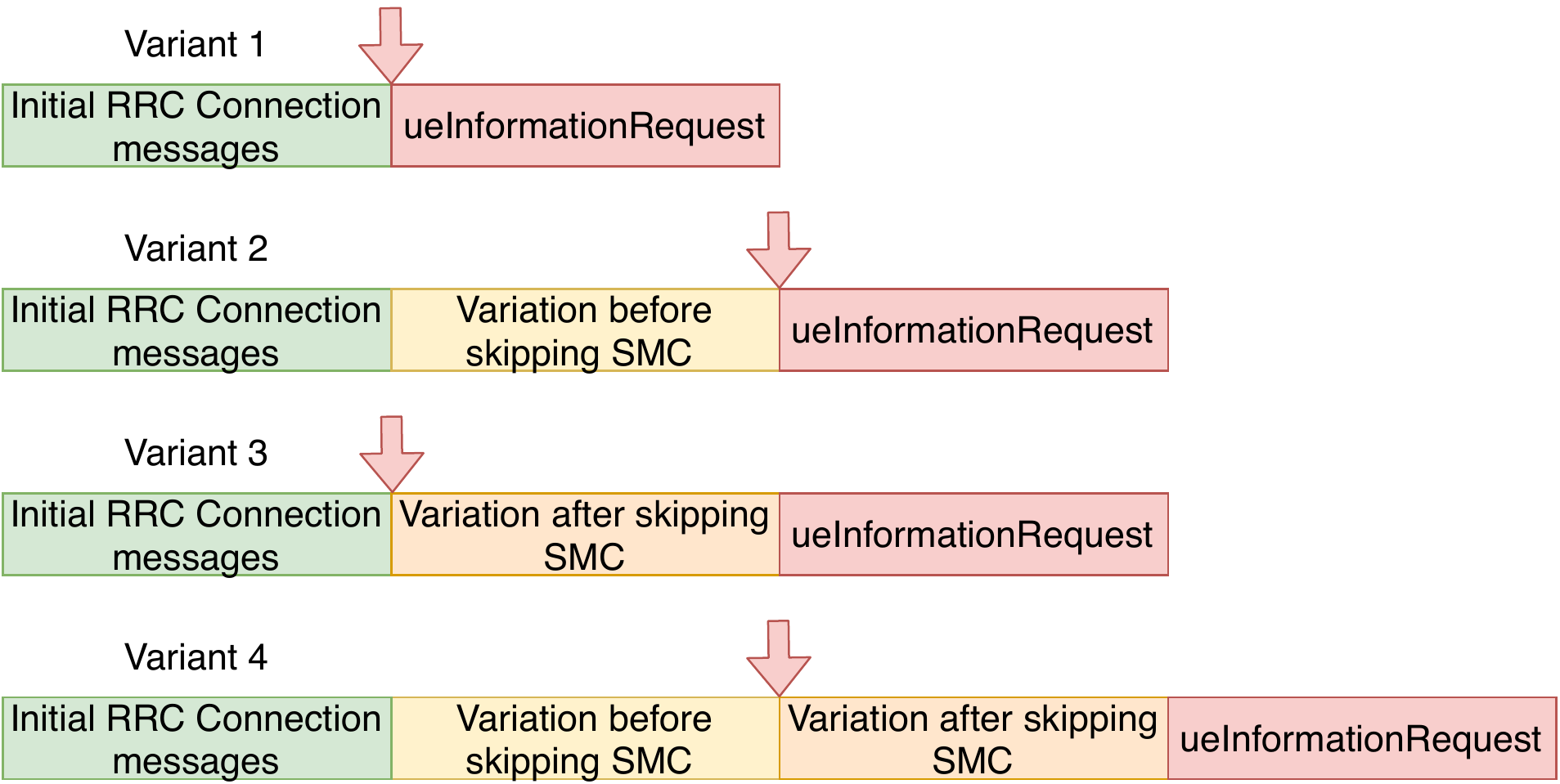}
	  \caption{$\beta$-undesired-behavior-session variants where $\beta$=privacy attack on the RLF report. The red arrow points to the location
	in a benign session where  both \securityModeCommand and \securityModeComplete
	would have appeared.}
	  \label{fig:measurement_report}
  \end{figure}

\begin{example}[$\beta$-undesired-behavior-session variants]
	%
	For this example, we consider $\beta$=the privacy attack on the RLF report \cite{privacy_ndss16}.
	In its canonical form, this attack happens in a session when a device responds with the RLF report message in plaintext
	due to an unprotected \ueInformationRequest message sent by the adversary before establishing a
	security context (i.e., before receiving \securityModeCommand and sending \securityModeComplete).
	$4$ example variants of this $\beta$-undesired-behavior-session is shown in Figure \ref{fig:measurement_report}.
 These different variations differ in what messages were sent before and after to the exclusion
of the \securityModeCommand and \securityModeComplete
messages.   Variant $1$, the canonical session,
does not introduce any messages before or after skipping the Security Mode
procedure and just sends the unprotected \ueInformationRequest message to induce the device to respond with an
unprotected RLF report message. Variant $2$ introduces a variation prior to the skipping of the Security
Mode procedure (e.g., sending an identity request message). Variant $3$ introduces a variation after
the skipping of the Security Mode procedure, possibly by inquiring about the UE’s capabilities through
the \ueCapabilityEnquiry message, before the plaintext \ueInformationRequest is sent by the adversary.
Variant $4$ combines both Variants $2$ and $3$.
\end{example}

\begin{figure}[t]
  \centering
	\includegraphics[width=.8\columnwidth]{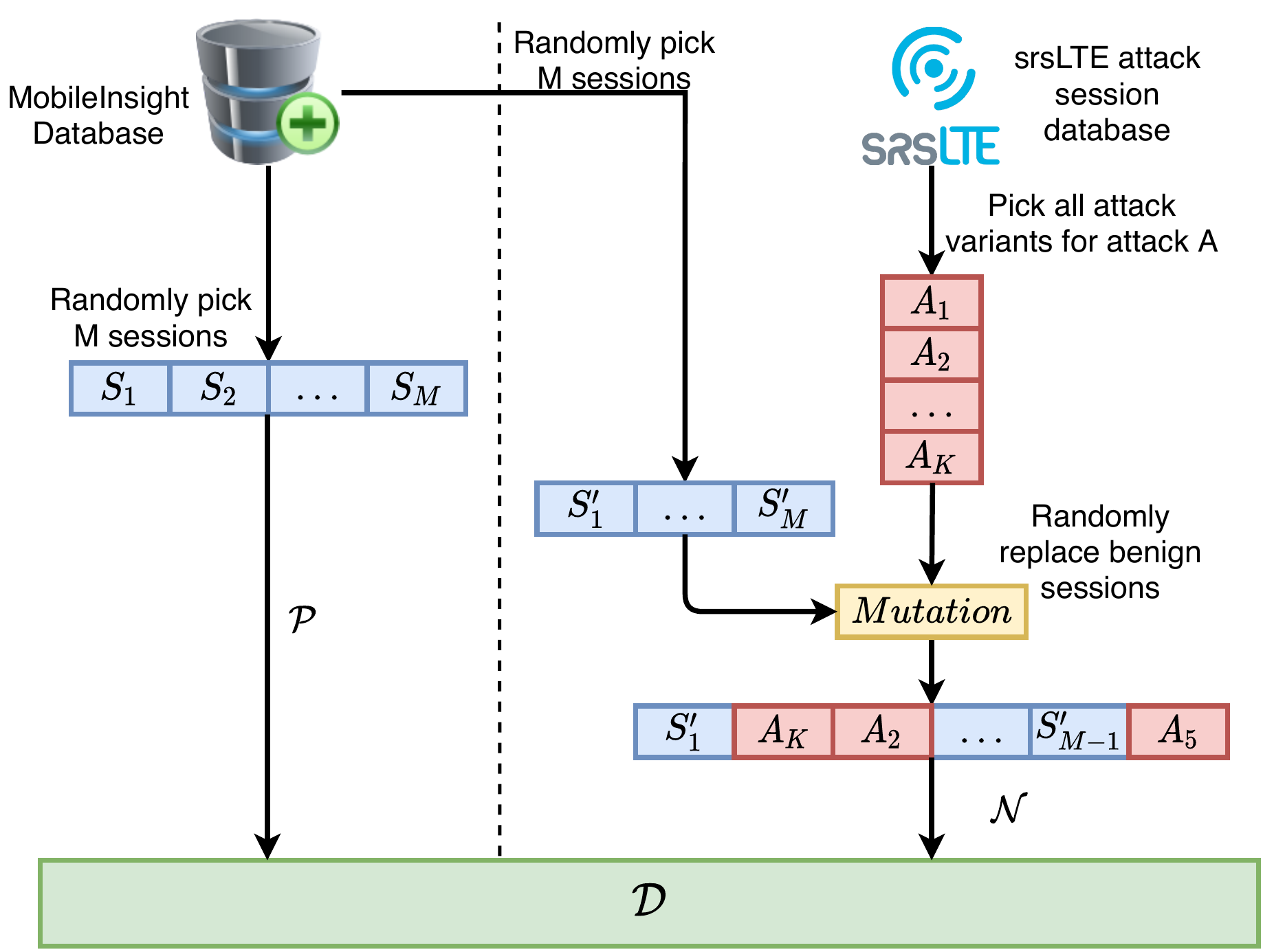}
	\caption{Trace Generation procedure.}
	\label{fig:trace_gathering}
\end{figure}

\subsubsection{Benign Trace Dataset}
To obtain benign traces, we use the MobileInsight \cite{mobile_insight}
crowd-sourced database. This database consists of log files captured by the
MobileInsight app and shared from users across the world; covering numerous
devices, networks, and countries. We decide to use this data rather than locally
captured benign traces to take into consideration other devices and networks,
which we do not have access to. We argue that this gives a better representation
as to how well the signatures would generalize in the real world trace, possibly
containing benign network failures.

From this dataset, we are able to obtain \textit{1,892} NAS layer traces which
contain over \textit{52K} messages, and as for RRC, we collect \textit{2,045}
RRC layer traces consisting of \textit{1.5M} messages. This large discrepancy
in the number of messages captured per layer can be attributed to the fact that
NAS traffic only serves as the communication between the UE and MME, while RRC
is responsible for the communication between the UE and the eNodeB and serves
as the backbone for NAS and other layers of the LTE protocol stack.

\textbf{Benign trace generation.} We use the collected MobileInsight
traces as seed traces and decompose them into individual sessions. In addition to
the message types in a session, we also capture relevant predicates from the data
(e.g., whether the identity request message warranted IMSI, IMEI, or GUTI).
After this step, suppose we have a total of $S$ number of sessions. If we want
 to generate $n$ benign traces of length $M$, then we will continue
the following process $n$ times. At each step, we will randomly pick $M$
benign sessions out of total $S$ sessions and concatenate them to create a new benign
trace. The process is shown on the left of the dotted vertical line in Figure \ref{fig:trace_gathering}.
After this process, we will obtain \emph{trace skeletons} comprising of individual message types and relevant
predicates. We then manually convert these trace skeletons to actual replayable benign traces
by choosing standard-compliant field values feasible in the testbed while respecting the different predicates. As an example,
if the benign trace skeleton in a session contained identity request with IMEI predicate, then we will create a
concrete packet reflecting that choice.

\subsubsection{Generating Malicious Traces}
A massive challenge with evaluating the effectiveness of
\system is the fact that no pre-existing
repository of vulnerable traces exists. To overcome this, we propose the generation of \emph{possibly} malicious
traces as shown in Figure \ref{fig:trace_gathering}. The trace generation has the following four steps.
(\ding{182}) The process starts with the manual implementation of all the attacks (and, their $\beta$-undesired-behavior-session variants)
as listed in Table~\ref{tab:all_attacks}. For doing so,
following the prior work \cite{kim_ltefuzz_sp19, privacy_ndss16, TORPEDO,
lteinspector, park2016white, how_not_to_break_crypto} we changed srsENB and srsEPC libraries in
srsLTE \cite{gomez2016srslte} to set up the rogue base station.
To collect the traces from the UE's perspective, we utilize SCAT \cite{SCAT}.
(\ding{183}) Once we have collected the concrete traces, we create skeletons of these traces akin to
to the benign trace generation process (i.e., capturing message types and relevant predicates).
After this process, for each attack, suppose we have $K$ skeletons  for $\beta$-undesired-behavior-session variants.
(\ding{184}) Suppose we want to generate $n$ possibly malicious traces of length $M$ for a given attack.
We will execute the following step $n$ times. At each step, we will first generate a
benign trace skeleton $bt$ of length $M$ using the procedure discussed above. Then,
we randomly choose $a_s$ attack variants out of $K$
(i.e., $1\leq a_s < min(M, K)$) and randomly replace $a_s$ of the benign sessions of $bt$ 
with the $a_s$ attack sessions to generate a possibly malicious  trace skeleton (see Figure \ref{fig:trace_gathering}).
(\ding{185}) For generating a concrete replayable malicious trace from a trace skeleton is a manual process and
attack-specific. Converting malicious trace skeletons to concrete traces require adding standard-compliant
field values while respecting the captured predicates.

\paragraph{Discussion.} Note that, all variants generated by the above process do not necessarily entail
an exploitable attack. This is not a limitation because the monitor has to be oblivious to whether a device is
susceptible to an attack or not, and instead should raise a warning irrespectively whenever it detects an attack attempt.
Taking the privacy attack on the RLF report as an example, the monitor should raise a warning whenever it receives
an unprotected \ueInformationRequest message before a security context is established without waiting for the
device to respond with an RLF report. For our evaluation, malicious traces that do not induce an attack are acceptable
as long as the trace contains an attack attempt. All variants can be found on
the following webpage \cite{phoenix_webpage}.

\begin{table}[]
	\centering
	\resizebox{\columnwidth}{!}{
	\begin{tabular}{|l|c|c|c|c|}
	\hline
	\multicolumn{1}{|c|}{\textbf{Attack}} & \textbf{Paper} & \textbf{Layer} & \textbf{\# of Variations} & \textbf{Implication} \\ \hline
	AKA Bypass & \cite{kim_ltefuzz_sp19} & \newmoon & 18 & Eavesdropping \\ \hline
	Measurement Report & \cite{privacy_ndss16} & \newmoon & 26 & Location Tracking \\ \hline
	RLF Report & \cite{privacy_ndss16} & \newmoon & 21 & Location Tracking \\ \hline
	IMSI Cracking & \cite{TORPEDO} & \newmoon & 2 & Information Leak \\ \hline
	Paging with IMSI & \cite{TORPEDO} & \newmoon & 2 & Information Leak \\ \hline
	Attach Reject & \cite{privacy_ndss16} & \fullmoon & 4 & Denial of Service \\ \hline
	Authentication Failure & \cite{lteinspector} & \fullmoon & 25 & Denial of Service \\ \hline
	EMM Information & \cite{park2016white} & \fullmoon & 32 & Spoofing \\ \hline
	IMEI Catching & \cite{3gppNAS} & \fullmoon & 2 & Information Leak \\ \hline
	IMSI Catching & \cite{3gppNAS} & \fullmoon & 2 & Information Leak \\ \hline
	Malformed Identity Request & \cite{how_not_to_break_crypto} & \fullmoon & 2 & Information Leak \\ \hline
	Null Encryption & \cite{3gppNAS} & \fullmoon & 49 & Eavesdropping \\ \hline
	Numb Attack & \cite{lteinspector} & \fullmoon & 2 & Denial of Service \\ \hline
	Service Reject & \cite{privacy_ndss16} & \fullmoon & 14 & Denial of Service \\ \hline
	TAU Reject & \cite{privacy_ndss16} & \fullmoon & 6 & Denial of Service \\ \hline
	\end{tabular}
	}
\caption{All attacks considered, total number of derived variants and their implication. (\newmoon = RRC, \fullmoon = NAS)}
\label{tab:all_attacks}
\end{table}

\section{Evaluation Results of \system}
\label{sec:evaluation_of_phoenix}

In this section, we discuss the evaluation results for both the signature synthesizer
and monitor components. In order to evaluate \system as both a warning system
and defense mechanism, we evaluate these two different implementations separately.
Due to space constraints, we report the results for 5 attacks here and
the rest can be found in the Appendix.
\subsection{Signature Synthesizer Evaluation}
We evaluate our \signatureSynthesizer{}s based on the research questions discussed in Section~\ref{sec:evaluation_criteria}.

\paragraph{Effectiveness of generated signatures ($\mathsf{QS_1}$).}
For evaluating the effectiveness of the synthesized signatures,
we replay the set of testing traces to a device running \system in our testbed (set up with srsLTE~\cite{gomez2016srslte} and
USRP~\cite{usrp}), and measure precision, recall, and F1 score for
identifying those vulnerability signatures at runtime.

\begin{table}[]
	\centering

	\begin{tabular}{|c|l|l|l|l|}
		\hline
		\textbf{Attack}                         & \textbf{Monitor} & \textbf{Precision} & \textbf{Recall} & \textbf{F1} \\ \hline
		\multirow{3}{*}{AKA Bypass}             & PLTL             & 1                  & 1               & 1           \\ \cline{2-5}
		& DFA              & 1                  & 0.95            & 0.97       \\ \cline{2-5}
		& MM               & 1                  & 1               & 1           \\ \hline
		\multirow{3}{*}{IMSI Cracking}          & PLTL             & 1                  & 1               & 1           \\ \cline{2-5}
		& DFA              & 1                  & 1               & 1           \\ \cline{2-5}
		& MM               & 0.67              & 1               & 0.80       \\ \hline
		\multirow{3}{*}{Measurement Report}     & PLTL             & 1                  & 1               & 1           \\ \cline{2-5}
		& DFA              & 0.95              & 0.83           & 0.89       \\ \cline{2-5}
		& MM               & 1                  & 1               & 1           \\ \hline
		\multirow{3}{*}{Numb Attack}            & PLTL             & 1                  & 1               & 1           \\ \cline{2-5}
		& DFA              & 1                  & 1               & 1           \\ \cline{2-5}
		& MM               & 1                  & 1               & 1           \\ \hline
		\multirow{3}{*}{RLF Report}             & PLTL             & 1                  & 1               & 1           \\ \cline{2-5}
		& DFA              & 0.83              & 0.64           & 0.72       \\ \cline{2-5}
		& MM               & 1                  & 1               & 1           \\ \hline
	\end{tabular}
	\caption{Effectiveness results for all monitors with maximum data each monitor can consume (MM stands for Mealy Machine). Note that all scores are in the range 0 to 1.}
	\label{tab:effectiveness_max_data}
\end{table}

Table~\ref{tab:effectiveness_max_data} presents the precision, recall and F1 score
achieved by our \signatureSynthesizer{}s for identifying different attacks at runtime.
The signatures used in this experiment were generated with $2,500$ traces for
DFA and Mealy Machine, and up $1,250$ for PLTL due to the synthesizer
timing out. The figure
demonstrates that all of the approaches were able to identify the
existing attacks with a high degree of success.
Among the different synthesizers, DFA, however,
produced a higher number of false positives (21.5\%)
and false negatives (17.1\%) on average whereas
Mealy Machine and PLTL turn out to be more reliable;
producing a significantly less number of false
positives ($\sim\!\!$ 0.03\%) and false negatives ($\sim\!\!$ 0.01\%).

The perfect F1 score for \pltl across different attacks can be attributed
to the fact that these control-plane attacks have
a highly discernible signature, which can be seen as the temporal property
which all variants of the attacks violate. For instance,
the signature synthesized for the RLF Report Attack \cite{privacy_ndss16} is the following:
$\ueInformationRequest \Rightarrow (\neg \rrcConnectionRequest \since \securityModeComplete)$. Since this signature precisely describes the behavior of the attack, regardless of the variant,
it enables \system to detect the attack with a perfect F1 score.

Another interesting result shown in Table~\ref{tab:effectiveness_max_data}
is that Mealy Machine based monitor outperforms the DFA based one in the
majority of the cases. This is because
DFA learns only on up to 2,500 traces for an individual attack whereas
Mealy Machine learns from all the attack traces (2,500 * 15) and therefore
has more information to learn from.

\begin{figure}[t]
 \centering
 \resizebox{.8\columnwidth}{!}{
\begin{tikzpicture}

\definecolor{color0}{rgb}{0.12156862745098,0.466666666666667,0.705882352941177}
\definecolor{color1}{rgb}{1,0.498039215686275,0.0549019607843137}
\definecolor{color2}{rgb}{0.172549019607843,0.627450980392157,0.172549019607843}
\definecolor{color3}{rgb}{0.83921568627451,0.152941176470588,0.156862745098039}
\definecolor{color4}{rgb}{0.580392156862745,0.403921568627451,0.741176470588235}
\definecolor{color5}{rgb}{0.549019607843137,0.337254901960784,0.294117647058824}

\begin{axis}[
legend columns = 2,
legend cell align={left},
legend style={fill opacity=0.8, draw opacity=1, text opacity=1, at={(0.45,-0.23)}, anchor=north, draw=white!80!black, font=\footnotesize},
log basis y={10},
tick align=outside,
tick pos=left,
x grid style={white!69.0196078431373!black},
xlabel={Training Dataset Size},
xmin=-10, xmax=2130,
xtick style={color=black},
y grid style={white!69.0196078431373!black},
ylabel={Training Time (seconds in log scale)},
ymin=-10.56314309339609, ymax=5204.89776935501,
ymode=log,
ytick style={color=black}
]
\addplot [semithick, red, forget plot]
table {%
-10 3600
1310 3600
};

\addplot [semithick, color1, mark=triangle, mark size=3, mark options={solid}]
table {%
50 57.05
100 144.87
250 359.68
500 558.54
1250 3600
2500 3600
};
\addlegendentry{Numb Attack / PLTL}

\addplot [semithick, color1, mark=triangle*, mark size=3, mark options={solid}]
table {%
50 0.07
100 0.15
250 0.15
500 0.11
1250 0.18
2500 0.13
};
\addlegendentry{Numb Attack / DFA}

\addplot [semithick, color2, mark=pentagon, mark size=3, mark options={solid}]
table {%
50 216.51
100 661.59
250 1428.20
500 3600
1250 3600
2500 3600
};
\addlegendentry{IMSI Cracking Attack (4G) / PLTL}

\addplot [semithick, color2, mark=pentagon*, mark size=3, mark options={solid}]
table {%
50 0.09
100 0.03
250 0.02
500 0.03
1250 0.03
2500 0.03
};
\addlegendentry{IMSI Cracking Attack (4G) / DFA}

\addplot [semithick, color3, mark=diamond, mark size=3, mark options={solid}]
table {%
50 2782.81
100 3600
250 3600
500 3600
1250 3600
2500 3600
};
\addlegendentry{AKA Bypass Attack / PLTL}

\addplot [semithick, color3, mark=diamond*, mark size=3, mark options={solid}]
table {%
50 0.10
100 0.13
250 0.05
500 0.10
1250 0.09
2500 0.05
};
\addlegendentry{AKA Bypass Attack / DFA}

\addplot [semithick, color4, mark=square, mark size=3, mark options={solid}]
table {%
50 0.10
100 0.07
250 0.08
500 0.10
1250 0.16
2500 0.21
};
\addlegendentry{NAS Layer Attacks / Mealy Machine}

\addplot [semithick, color4, mark=square*, mark size=3, mark options={solid}]
table {%
50 0.06
100 0.03
250 0.04
500 0.05
1250 0.06
2500 0.08
};
\addlegendentry{RRC Layer Attacks / Mealy Machine}

\end{axis}

\end{tikzpicture}}
 \caption{Time to learn DFA, PLTL and Mealy Machine.
 }
 \label{fig:learning_time}
\end{figure}
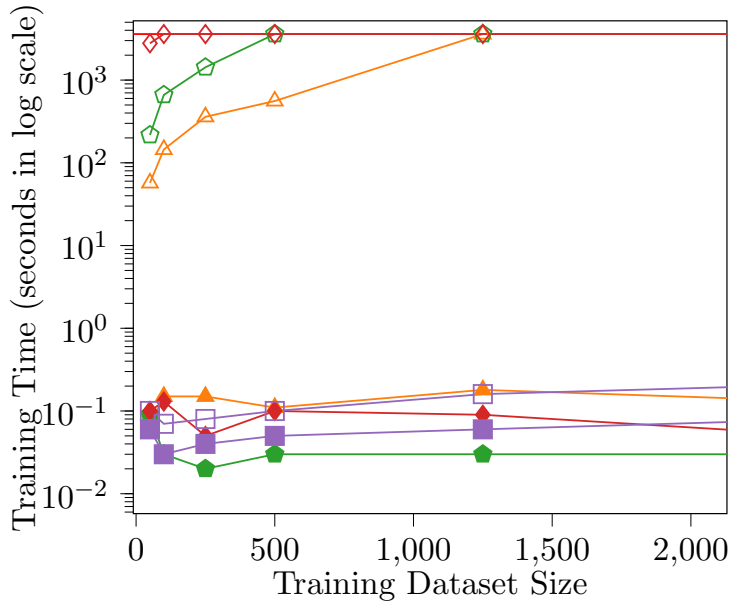

\paragraph{Scalability ($\mathsf{QS_2}$).}
We primarily consider signature learning time as an effective and
indirect indicator to the scalability of the corresponding signature synthesizer.
The lower the learning time, the higher the scalability. That signifies that scalability time is inversely
proportional to the signature learning time. Therefore, to evaluate the scalability of
the three proposed signature synthesizers (DFA, MM, and PLTL),
we vary the sample size of the training sets to 50, 100, 250, 500, 1250, and 2500, and measure
the learning time required by a synthesizer for each of the attacks.
Figure~\ref{fig:learning_time} presents the results of this evaluation in which the Y-axis is seconds in the logarithmic scale and
the X-axis is the training dataset size.

Figure~\ref{fig:learning_time} shows that our \pltl signature synthesizer takes considerably more time to
synthesize a signature as compared to DFA and MM synthesizers. This large discrepancy can be attributed to the fact
that the \pltl synthesizer is a \emph{search based algorithm}. The search space
grows very quickly as the depth of the abstract syntaxt tree (AST) increases. On the other hand,
RPNI \cite{rpni} proves to scale quite well because RPNI is a polynomial time algorithm while SAT is NP-Complete.
For instance, training the AKA Bypass~\cite{kim_ltefuzz_sp19} attack with \pltl synthesizer takes a significantly higher amount
of time than others. Though \pltl synthesizer for AKA Bypass attack quickly times out,
the same synthesizer does not time out for other attacks, such as the Numb Attack~\cite{lteinspector} until it reaches
$1250$ traces. This is due to the much deeper AST for AKA Bypass \pltl signature than that for the Numb Attack.

\paragraph{Impact of training set size on signature quality ($\mathsf{QS_3}$).}
Since real-life cellular attack traces are difficult to obtain,
we aim at evaluating whether or not more training data generate a
higher quality signature. We consider a high quality signature as
one that achieves a perfect F1 score. In other words, F1 score and signature quality are proportional to each other. To
evaluate this, we vary the size of the training datasets
and measure the synthesizers' effectiveness at detecting the attacks. 

\begin{figure}[t]
	\centering
	\resizebox{.8\columnwidth}{!}{
\begin{tikzpicture}

\definecolor{color0}{rgb}{0.12156862745098,0.466666666666667,0.705882352941177}
\definecolor{color1}{rgb}{1,0.498039215686275,0.0549019607843137}
\definecolor{color2}{rgb}{0.172549019607843,0.627450980392157,0.172549019607843}
\definecolor{color3}{rgb}{0.83921568627451,0.152941176470588,0.156862745098039}
\definecolor{color4}{rgb}{0.580392156862745,0.403921568627451,0.741176470588235}
\definecolor{color5}{rgb}{0.549019607843137,0.337254901960784,0.294117647058824}
\definecolor{color6}{rgb}{0.10588,0.28627,0.70980}

\begin{axis}[
legend columns = 3,
legend cell align={left},
legend style={fill opacity=0.8, draw opacity=1, text opacity=1, at={(0.44,-0.23)}, anchor=north, draw=white!80!black, font=\footnotesize},
tick align=outside,
tick pos=left,
title={},
x grid style={white!69.0196078431373!black},
xlabel={Training Dataset Size},
xmin=-72.5, xmax=2622.5,
xtick style={color=black},
y grid style={white!69.0196078431373!black},
ylabel={F1 score},
ymin=0.79, ymax=1.01,
ytick style={color=black},
tick label style={font=\footnotesize}
]
\addplot [semithick, color0, mark=square*, mark size=3, mark options={solid}]
table {%
50 1
100 1
250 1
500 1
1250 1
2500 1
};
\addlegendentry{Numb Attack / PLTL}
\addplot [semithick, color1, mark=pentagon*, mark size=3, mark options={solid}]
table {%
50 0.89
100 0.99
250 1
500 1
1250 0.99
2500 1
};
\addlegendentry{Numb Attack / DFA}
\addplot [semithick, color4, mark=triangle*, mark size=3, mark options={solid}]
table {%
50 1
100 1
250 1
500 0.98
1250 1
2500 1
};
\addlegendentry{Numb Attack / MM}
\addplot [semithick, color0, mark=square, mark size=3, mark options={solid}]
table {%
50 1
100 1
250 1
500 1
1250 1
2500 1
};
\addlegendentry{AKA Bypass / PLTL}
\addplot [semithick, color1, mark=pentagon, mark size=3, mark options={solid}]
table {%
50 0.89
100 0.8
250 0.81
500 0.99
1250 0.95
2500 0.97
};
\addlegendentry{AKA Bypass / DFA}
\addplot [semithick, color4, mark=triangle, mark size=3, mark options={solid}]
table {%
50 0.91
100 0.97
250 0.99
500 0.99
1250 0.99
2500 1
};
\addlegendentry{AKA Bypass / MM}
\end{axis}

\end{tikzpicture}}
	\caption{Training size and effectiveness comparison.
	}
	\label{fig:effectiveness_vs_sample_size}
\end{figure}
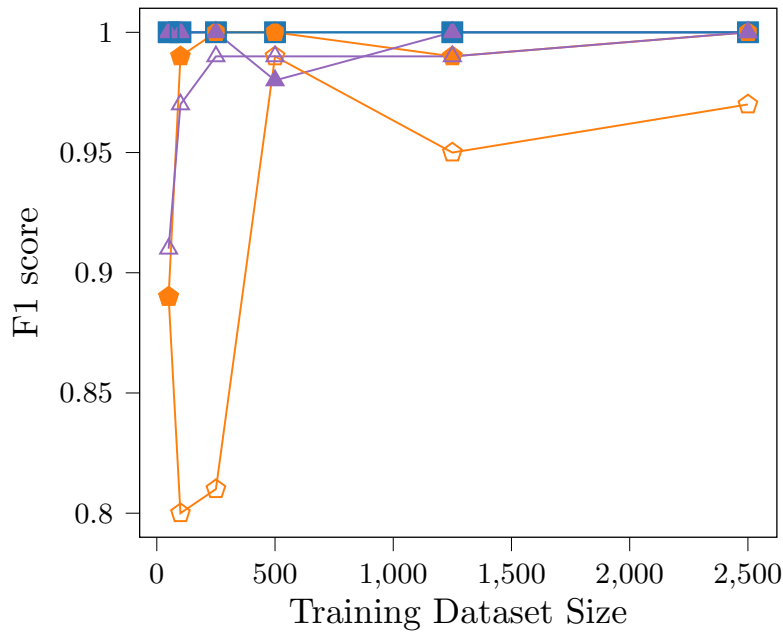

Figure \ref{fig:effectiveness_vs_sample_size}
shows that all three signature synthesizers achieve high F1 score when
training on 500 traces, with the exception of AKA Bypass for DFA, which
goes down as more training data is given. As the RPNI learning process
is highly dependent on the exact set of input traces, this discrepancy can be
attributed to the variability of the input traces. Note that, our \pltl signature synthesizer
achieves a perfect F1 score across all attacks, regardless of the training
dataset size, because of its usage of exhaustive search to learn a
precise but highly generalizable signature.

Since the \pltl synthesizer is able to produce a highly generalizable
signature regardless of the training dataset in the previous experiment,
we decide to analyze this further by discovering the minimum attack traces
required to generate a high quality signature. We consider a high quality signature is
one that achieves a perfect F1 score. To perform this experiment, we fix
the benign traces to $25$ and vary the number of attack traces from $1$ to $25$.
The results can be found in
Table \ref{tab:minimum_attack_traces}. These results show that the \pltl synthesizer
can rapidly produce a high quality signature. Both the RLF Report and Measurement
Report privacy attacks prove to require a larger number of attack traces.
This can be attributed to the fact that these signatures are more complex than
others, with the exception of the AKA Bypass attack, however, more variants exist.

 These results show that the \pltl
synthesizer can rapidly produce a high quality signature.
Another observation that is obvious is the fact that Measurement Report and
RLF Report require more attack traces than others. This can be attributed to a couple
of reasons. The first reason behind this result, is that these two attacks require a
larger search space since the alphabet is bigger than the others. The second
reason is that these attacks are more complex than others, with the exception
of the AKA Bypass attack which can be seen as a stepping stone for both. In addition,
these results can also attributed to the fact that our \pltl synthesizer blindly
searches for solutions instead of using the given traces to narrow down the search
space.

\begin{table}[]
  \centering

\begin{tabular}{|c|c|c|}
\hline
\textbf{Attack} & \textbf{Minimum Attack Trace} & \textbf{\# of Variations} \\ \hline
AKA Bypass & 3 & 2 \\ \hline
IMSI Cracking & 1 & 1 \\ \hline
Measurement Report & 11 & 5 \\ \hline
RLF Report & 8 & 2 \\ \hline
Numb Attack & 3 & 2 \\ \hline
\end{tabular}
\caption{Minimum attack traces (and variations), required to generate a high quality signature (Perfect F1 score) using PLTL synthesizer.}
\label{tab:minimum_attack_traces}
\end{table}

\paragraph{Signature Synthesizer Evaluation Conclusion.}
The \pltl synthesizer proved to not scale as well
as RPNI \cite{rpni} based approaches, however, it proved to quickly generated
highly generalizable signature. In fact, such a signature generation with a minimal number of traces is
critical since generating attack traces is a challenging task for cellular networks.
Therefore, we conclude that the \pltl synthesizer outperforms
the RPNI \cite{rpni} approaches.

\subsection{Monitor Evaluation (Warning System)}
\label{sec:monitor_evaluation_warning_system}
In this subsection, we answer the research questions driving the evaluation
of three different monitoring approaches (i.e., \pltl, DFA, and Mealy Machine)
when considering a warning system instantiation.

\paragraph{Efficiency ($\mathsf{QWS_1}$).}
One of the key factors in identifying the best monitor instantiation
is the number of messages each monitor can process per second.
For this, we perform a stress test by mimicking the modem through the
replaying of real traces captured
from MobileInsight's database \cite{mobile_insight} without any delay
between subsequent messages. We measure how long each monitor takes
to process and check for the presence of an attack by consulting
its entire signature database. Table \ref{tab:message_per_second_overall}
summarizes the processing speed (messages/second) of different
devices for different monitoring approaches running in two different layers
(RRC and NAS). In addition to this, we computed the CPU cycles required per call
to the monitor component for each device which can be found in Table \ref{tab:cpu_cycles_tab}.

\begin{table}[]
  \centering
	\begin{tabular}{|l|c|l|l|}
	\hline
	\multicolumn{1}{|c|}{\textbf{Layer}} & \textbf{Monitor} & \multicolumn{1}{c|}{\textbf{Device}} & \multicolumn{1}{c|}{\textbf{Avg. CPU Cycles}} \\ \hline
	\multirow{9}{*}{RRC} & \multirow{3}{*}{DFA} & Pixel 3 & 54,127.20 \\ \cline{3-4}
	 &  & Nexus 6P & 97,168.98 \\ \cline{3-4}
	 &  & Nexus 6 & 325,579.71 \\ \cline{2-4}
	 & \multirow{3}{*}{PLTL} & Pixel 3 & 384,282.80 \\ \cline{3-4}
	 &  & Nexus 6P & 560,255.48 \\ \cline{3-4}
	 &  & Nexus 6 & 4,065,040.65 \\ \cline{2-4}
	 & \multirow{3}{*}{MM} & Pixel 3 & 7,177.05 \\ \cline{3-4}
	 &  & Nexus 6P & 15,900.18 \\ \cline{3-4}
	 &  & Nexus 6 & 78,850.07 \\ \hline
	\multirow{9}{*}{NAS} & \multirow{3}{*}{DFA} & Pixel 3 & 81,957.62 \\ \cline{3-4}
	 &  & Nexus 6P & 136,431.23 \\ \cline{3-4}
	 &  & Nexus 6 & 599,973.33 \\ \cline{2-4}
	 & \multirow{3}{*}{PLTL} & Pixel 3 & 742,528.31 \\ \cline{3-4}
	 &  & Nexus 6P & 1,094,331.36 \\ \cline{3-4}
	 &  & Nexus 6 & 4,456,180.89 \\ \cline{2-4}
	 & \multirow{3}{*}{MM} & Pixel 3 & 7,586.85 \\ \cline{3-4}
	 &  & Nexus 6P & 14,812.58 \\ \cline{3-4}
	 &  & Nexus 6 & 78,853.99 \\ \hline
	\end{tabular}
\caption{CPU Cycles required per call to the monitor. The clock speed for the devices
are the following: Pixel 3 = 2.8 Ghz, Nexus 6P = 2.0 Ghz, and  Nexus 6 = 2.7 Ghz.}
\label{tab:cpu_cycles_tab}
\end{table}

As shown in Table \ref{tab:message_per_second_overall}, across all three
devices, Mealy Machine can process multiple orders of magnitude higher messages
per second than the other two monitoring approaches. This can be attributed to the
fact that Mealy Machine keeps only
a single internal state per layer, as compared to 10 internal states for
NAS and 5 for RRC. Moreover, Mealy Machine relies on a single dictionary
lookup to decide on the transition and whether to flag a trace as an attack.
Similar to Mealy Machine, DFA can also process messages at a much faster
rate than \pltl. This is because the DFA also relies on a simple
dictionary lookup similar to Mealy Machine for a single signature.
On the other hand, \pltl requires the evaluation of logical and
temporal operators to classify the incoming traces which is a
more expensive operation.

To put our results in perspective, we compare it with real traces.
We compute the mean, median, standard deviation, and
maximum number of messages of real NAS and RRC traces obtained from the
MobileInsight database \cite{mobile_insight}.
We observe that on average, there were 0.02 messages per second
for NAS traffic (median=0.011, standard deviation=0.069, maximum=0.8),
and 0.2 messages per second (median=0.122, standard deviation=0.273, maximum=2.76)
for RRC traffic.

In summary,
our slowest monitor (i.e., \pltl) can handle substantially more message
per second than the NAS and RRC traffic we observed in real traces.

\begin{table}[h!]
  \centering

  \begin{tabular}{|c|c|l|l|l|}
  \hline
  \textbf{Layer} & \textbf{Monitor} & \textbf{Device} & \multicolumn{1}{c|}{\textbf{Avg.}} & \multicolumn{1}{c|}{\textbf{SD}} \\ \hline
  \multirow{9}{*}{RRC} & \multirow{3}{*}{DFA} & Pixel 3 & 51730.6 & 158596.4 \\ \cline{3-5}
   &  & Nexus 6P & 20582.7 & 73663.6 \\ \cline{3-5}
   &  & Nexus 6 & 8292.9 & 8636.4 \\ \cline{2-5}
   & \multirow{3}{*}{PLTL} & Pixel 3 & 7286.3 & 55599.5 \\ \cline{3-5}
   &  & Nexus 6P & 3569.8 & 12976.3 \\ \cline{3-5}
   &  & Nexus 6 & 664.2 & 58.0 \\ \cline{2-5}
   & \multirow{3}{*}{MM} & Pixel 3 & 390132.6 & 790596.7 \\ \cline{3-5}
   &  & Nexus 6P & 125784.7 & 359847.7 \\ \cline{3-5}
   &  & Nexus 6 & 34242.2 & 14377.8 \\ \hline
  \multirow{9}{*}{NAS} & \multirow{3}{*}{DFA} & Pixel 3 & 34164.0 & 224904.7 \\ \cline{3-5}
   &  & Nexus 6P & 14659.4 & 110780.6 \\ \cline{3-5}
   &  & Nexus 6 & 4500.2 & 4170.7 \\ \cline{2-5}
   & \multirow{3}{*}{PLTL} & Pixel 3 & 3770.9 & 62512.7 \\ \cline{3-5}
   &  & Nexus 6P & 1827.6 & 22226.2 \\ \cline{3-5}
   &  & Nexus 6 & 605.9 & 1472.4 \\ \cline{2-5}
   & \multirow{3}{*}{MM} & Pixel 3 & 369059.5 & 723754.8 \\ \cline{3-5}
   &  & Nexus 6P & 135020.4 & 371327.7 \\ \cline{3-5}
   &  & Nexus 6 & 34240.5 & 20397.0 \\ \hline
  \end{tabular}
\caption{Measurement of how many messages per second can each monitor classify on different devices and layers.}
\label{tab:message_per_second_overall}
\end{table}

\begin{figure}[t]
  \centering
	\includegraphics[width=.8\columnwidth]{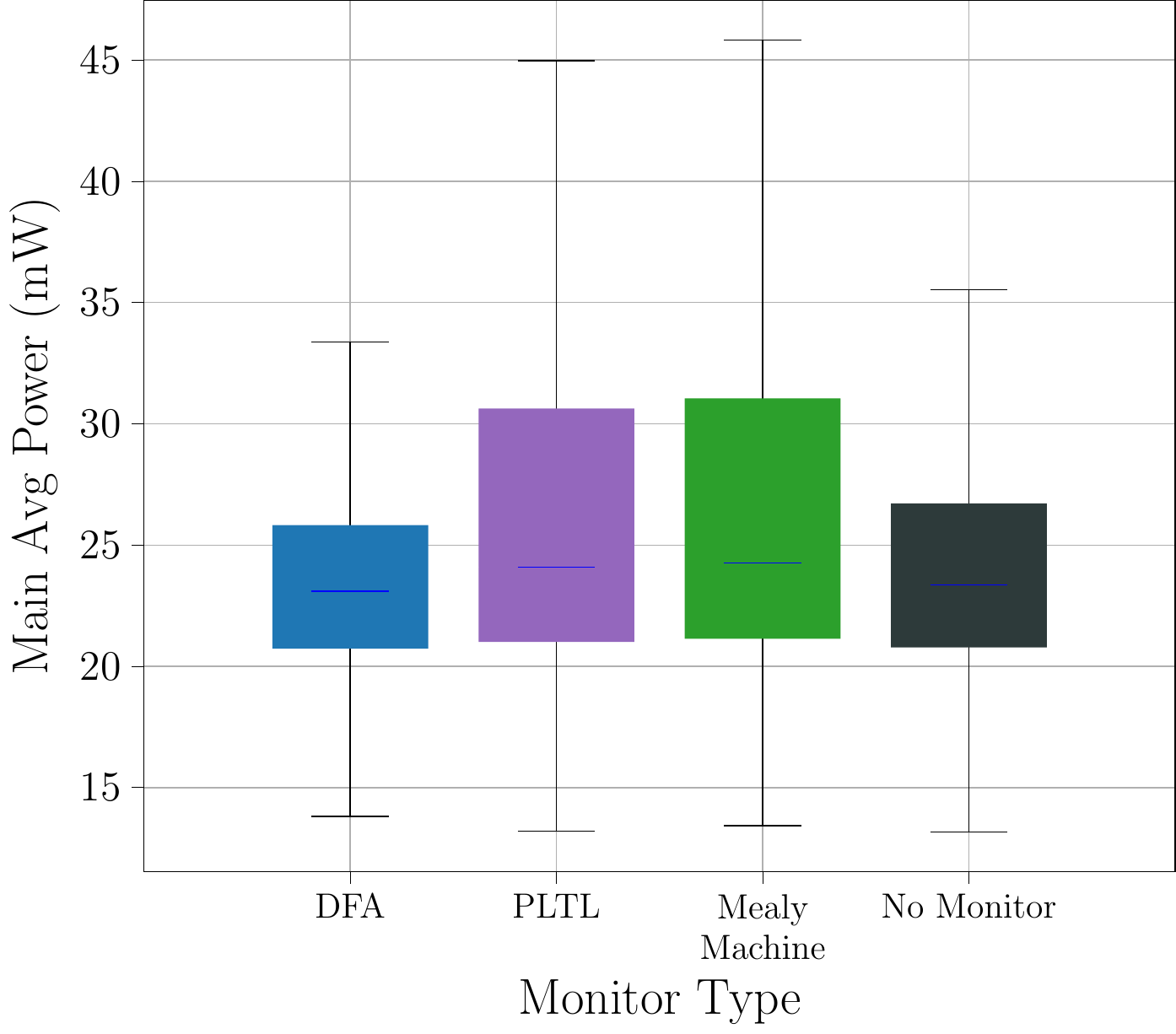}
	\caption{Power consumption on simulator in milliwatts (mW).}
	\label{fig:power_consumption_simulator}
\end{figure}

\paragraph{Energy Consumption ($\mathsf{QWS_2}$).}
To understand the energy consumption induced by each monitor component, we
measure the battery consumption induced by \system. We perform this
experiment by connecting
the Nexus 6 to a Monsoon Meter \cite{monsoon}. The Nexus 6, unlike the other
two devices, has a removable back which makes it easier to connect to the power
meter. In this experiment, the traffic is simulated to avoid the noise induced
by the cellular connection. In addition to the radio, we switch off the screen,
Bluetooth, and Wi-Fi. We then invoke each monitor with $10k$
messages to evaluate the average power consumption. Figure \ref{fig:power_consumption_simulator}
presents the average power consumption by three different monitors along with the case when no monitor is active.
The results match the trend with that of synthesizers' effectiveness,
except for the fact that Mealy Machine consumed slightly  more electricity than
\pltl and DFA, respectively. 
This discrepancy could be attributed to
the fact that even though we disabled many power hungry components of the Android system, we
have no control as to what other applications in the device are doing. Overall
though, all monitors add negligible overhead.

\paragraph{Real World Evaluation ($\mathsf{QWS_3}$).}
Vulnerability detection systems must balance false warnings with effectiveness.
If the user is bombarded with false warnings, the user would disable the
system in order to prevent continuously erroneous warnings. In light of this,
we aim to uncover how many warnings each different monitor produces
and the type of them. To carry out this experiment, we deploy \system
on two Pixel 3 devices running on four major U.S. cellular network carriers
on two different geographical areas.
In this experiment, we run \system for approximately 12 hours and use the
Pixel 3 as our daily devices, which includes driving approximately ~10 miles.
The results are shown in Table \ref{tab:real_world_warnings}. As expected by
previous results, DFA proves to be inadequate and produces a larger
amount of false warnings. We inspect each warning and uncover that the DFA
signature does not take into consideration the behavior seen by
these real networks. On the other hand, Mealy Machine produces no false warnings
and therefore would not bombard the user with these. Notably,
\pltl produces one warning on three different providers, specifically the
warning that is triggered when the EMM Information message is sent in plaintext.
After manual inspection, we discover that these in fact are not false warnings,
but misconfigurations by these three providers.

\begin{table}[]
  \centering
  \begin{tabular}{|c||*{4}{c|}}\hline
  \backslashbox{\textbf{Monitor}}{\textbf{Carrier}}
  &\textbf{US-1}&\textbf{US-2}
  &\textbf{US-3}&\textbf{US-4}\\\hline\hline
  DFA &6\xmark&7\xmark&4\xmark&4\xmark\\\hline
  PLTL &0&1\checkmark&1\checkmark&1\checkmark\\\hline
  MM &0&0&0&0\\\hline
  \end{tabular}
\caption{Number of warnings triggered by different monitor implementations in real networks
(\checkmark = Real Warnings, \xmark = False Warnings).}
\label{tab:real_world_warnings}
\end{table}

\paragraph{Evaluation Summary of Warning System Instantiation.}
Mealy Machine proved to be highly efficient, however,
all three monitors were able to parse a significantly high number of messages per second to not
induce any delay at runtime. We then measured power consumption and
discovered that all three monitors are highly efficient by imposing a negligible
overhead. We then carried
out a real world evaluation of \system by deploying it on cellular devices with
real SIM cards and uncover that \pltl and Mealy Machine produce no false warnings,
and in fact, \pltl uncovers real misconfigurations in three of the major U.S. cellular
network carriers. In summary,
\pltl proved to be the monitor component that best satisfies the core
requirements.

\subsection{Monitor Evaluation (Defense Mechanism)}
Understanding the requirements of \system when implemented in the baseband
is crucial in order to understand its deployability. Due to this,
this subsection answers the research questions driving the evaluation of the
baseband instantiation of \system. We perform these experiments on the baseband
implementation as discussed in Section \ref{sec:baseband_implementation}.
Due to the fact that \pltl is the monitor that performed the best as shown
previously (in Section \ref{sec:monitor_evaluation_warning_system}), we focus
on the \pltl monitor.

\paragraph{Memory overhead in baseband ($\mathsf{QBB_1}$).}
Low memory overhead is critical in order for a defense mechanism to be
feasible. To analyze this overhead, we measure the
memory using the \textbf{time} Linux command capable of extracting the maximum
resident set size. We then compare the implementation
of \system in srUE (dubbed $srsUE_{PHOENIX}$) and the vanilla version of srsUE
(dubbed $srsUE_{vanilla}$). To perform this experiment, we connect the srsUE
implementations 100 times to the eNodeB and EPC by running the corresponding
components of srsLTE \cite{gomez2016srslte} on a secondary machine.

\begin{figure}[t]
 \centering
 \resizebox{.8\columnwidth}{!}{
\pgfplotsset{every axis/.append style={
        scaled y ticks = false,
        scaled x ticks = false,
        y tick label style={/pgf/number format/.cd, fixed, fixed zerofill,
                            int detect,1000 sep={\;},precision=3},
        x tick label style={/pgf/number format/.cd, fixed, fixed zerofill,
                            int detect, 1000 sep={},precision=3}
    }
}

\begin{tikzpicture}

\definecolor{color0}{rgb}{0.12156862745098,0.466666666666667,0.705882352941177}
\definecolor{color1}{rgb}{1,0.498039215686275,0.0549019607843137}

\begin{axis}[
legend cell align={left},
legend style={fill opacity=0.8, draw opacity=1, text opacity=1, draw=white!80!black},
tick align=outside,
tick pos=left,
x grid style={white!69.0196078431373!black},
xmin=370046.41897224, xmax=372151.58102776,
xtick style={color=black},
y grid style={white!69.0196078431373!black},
ytick style={color=black},
ytick={0.0,0.05,0.1,0.15,0.2},
yticklabels={0.0,0.05,0.1,0.15,0.2},
xlabel={Maximum Resident Size},
ylabel={Density of Probability},
]
\addplot [semithick, color0]
table [x expr=\thisrow{X}, y expr=\thisrow{Y}*100] {%
X Y
370142.108156582 1.3894512036303e-05
370161.439304934 1.66344455597042e-05
370180.770453286 1.98416669724451e-05
370200.101601638 2.35804873887538e-05
370219.432749989 2.79210787760578e-05
370238.763898341 3.2939454739063e-05
370258.095046693 3.87173311481322e-05
370277.426195045 4.53418478706288e-05
370296.757343397 5.29051333314961e-05
370316.088491749 6.15036947613739e-05
370335.419640101 7.12376188637865e-05
370354.750788453 8.22095703106143e-05
370374.081936805 9.45235789943872e-05
370393.413085156 0.000108283611343211
370412.744233508 0.000123591926219817
370432.07538186 0.000140547221925702
370451.406530212 0.000159242587524956
370470.737678564 0.000179763278951167
370490.068826916 0.000202184347994531
370509.399975268 0.000226568160067656
370528.73112362 0.000252961844365656
370548.062271971 0.000281394727393808
370567.393420323 0.000311875807523102
370586.724568675 0.0003443913339443
370606.055717027 0.000378902557805655
370625.386865379 0.000415343726135848
370644.718013731 0.00045362039010676
370664.049162083 0.000493608098036862
370683.380310435 0.000535151540113367
370702.711458786 0.000578064206007965
370722.042607138 0.000622128608352207
370741.37375549 0.000667097114499698
370760.704903842 0.000712693416277966
370780.036052194 0.000758614652795441
370799.367200546 0.000804534185159072
370818.698348898 0.000850105004611316
370838.02949725 0.000894963737632258
370857.360645602 0.000938735193524802
370876.691793953 0.000981037382532676
370896.022942305 0.00102148691624495
370915.354090657 0.00105970468753159
370934.685239009 0.00109532171512103
370954.016387361 0.00112798502867617
370973.347535713 0.00115736346430035
370992.678684065 0.00118315323812866
371012.009832417 0.00120508316723941
371031.340980768 0.00122291941263066
371050.67212912 0.00123646962836307
371070.003277472 0.00124558641396476
371089.334425824 0.00125016998346237
371108.665574176 0.00125016998346237
371127.996722528 0.00124558641396476
371147.32787088 0.00123646962836307
371166.659019232 0.00122291941263066
371185.990167583 0.00120508316723941
371205.321315935 0.00118315323812866
371224.652464287 0.00115736346430035
371243.983612639 0.00112798502867617
371263.314760991 0.00109532171512103
371282.645909343 0.00105970468753159
371301.977057695 0.00102148691624495
371321.308206047 0.000981037382532676
371340.639354398 0.000938735193524802
371359.97050275 0.000894963737632258
371379.301651102 0.000850105004611316
371398.632799454 0.000804534185159072
371417.963947806 0.000758614652795441
371437.295096158 0.000712693416277966
371456.62624451 0.000667097114499698
371475.957392862 0.000622128608352207
371495.288541214 0.000578064206007965
371514.619689565 0.000535151540113367
371533.950837917 0.000493608098036862
371553.281986269 0.00045362039010676
371572.613134621 0.000415343726135848
371591.944282973 0.000378902557805655
371611.275431325 0.0003443913339443
371630.606579677 0.000311875807523102
371649.937728029 0.000281394727393808
371669.26887638 0.000252961844365656
371688.600024732 0.000226568160067656
371707.931173084 0.000202184347994531
371727.262321436 0.000179763278951167
371746.593469788 0.000159242587524956
371765.92461814 0.000140547221925702
371785.255766492 0.000123591926219817
371804.586914844 0.000108283611343211
371823.918063195 9.45235789943872e-05
371843.249211547 8.22095703106143e-05
371862.580359899 7.12376188637865e-05
371881.911508251 6.15036947613739e-05
371901.242656603 5.29051333314961e-05
371920.573804955 4.53418478706288e-05
371939.904953307 3.87173311481322e-05
371959.236101659 3.2939454739063e-05
371978.567250011 2.79210787760578e-05
371997.898398362 2.35804873887538e-05
372017.229546714 1.98416669724451e-05
372036.560695066 1.66344455597042e-05
372055.891843418 1.3894512036303e-05
};
\addlegendentry{PHOENIX}
\addplot [semithick, color1, dashed]
table [x expr=\thisrow{X}, y expr=\thisrow{Y}*100]{%
X Y
370397.776386324 2.45657153528959e-05
370408.710196702 2.94099608251851e-05
370419.644007079 3.50803786198825e-05
370430.577817456 4.16906717962593e-05
370441.511627834 4.93649054940989e-05
370452.445438211 5.82374729594822e-05
370463.379248588 6.84528491337684e-05
370474.313058965 8.01650986701708e-05
370485.246869343 9.35371060698518e-05
370496.18067972 0.000108739497631228
370507.114490097 0.000125949228217114
370518.048300475 0.000145347810578519
370528.982110852 0.00016711917119838
370539.915921229 0.000191447124353172
370550.849731606 0.000218512465317111
370561.783541984 0.000248489694236532
370572.717352361 0.000281543394037451
370583.651162738 0.000317824298548883
370594.584973116 0.000357465100513563
370605.518783493 0.000400576062959744
370616.45259387 0.000447240511044447
370627.386404247 0.000497510294488781
370638.320214625 0.00055140132255412
370649.254025002 0.000608889283594414
370660.187835379 0.000669905669030651
370671.121645757 0.000734334226578374
370682.055456134 0.00080200796922642
370692.989266511 0.000872706864448928
370703.923076888 0.000946156322059946
370714.856887266 0.00102202658887054
370725.790697643 0.00109993314380088
370736.72450802 0.00117943816844474
370747.658318398 0.00126005314561699
370758.592128775 0.00134124261250864
370769.525939152 0.00142242906642888
370780.459749529 0.00150299899044877
370791.393559907 0.00158230993448244
370802.327370284 0.001659698555488
370813.261180661 0.00173448948958213
370824.194991039 0.0018060049000358
370835.128801416 0.00187357451949374
370846.062611793 0.00193654598327702
370856.99642217 0.00199429523428914
370867.930232548 0.00204623676956399
370878.864042925 0.00209183349445971
370889.797853302 0.00213060595331502
370900.73166368 0.00216214071510457
370911.665474057 0.00218609770918842
370922.599284434 0.00220221632921908
370933.533094811 0.00221032015202945
370944.466905189 0.00221032015202945
370955.400715566 0.00220221632921908
370966.334525943 0.00218609770918842
370977.26833632 0.00216214071510457
370988.202146698 0.00213060595331502
370999.135957075 0.00209183349445971
371010.069767452 0.00204623676956399
371021.00357783 0.00199429523428914
371031.937388207 0.00193654598327702
371042.871198584 0.00187357451949374
371053.805008961 0.0018060049000358
371064.738819339 0.00173448948958213
371075.672629716 0.001659698555488
371086.606440093 0.00158230993448244
371097.540250471 0.00150299899044877
371108.474060848 0.00142242906642888
371119.407871225 0.00134124261250864
371130.341681602 0.00126005314561699
371141.27549198 0.00117943816844474
371152.209302357 0.00109993314380088
371163.143112734 0.00102202658887054
371174.076923112 0.000946156322059946
371185.010733489 0.000872706864448928
371195.944543866 0.00080200796922642
371206.878354243 0.000734334226578374
371217.812164621 0.000669905669030651
371228.745974998 0.000608889283594414
371239.679785375 0.00055140132255412
371250.613595753 0.000497510294488781
371261.54740613 0.000447240511044447
371272.481216507 0.000400576062959744
371283.415026884 0.000357465100513563
371294.348837262 0.000317824298548883
371305.282647639 0.000281543394037451
371316.216458016 0.000248489694236532
371327.150268394 0.000218512465317111
371338.084078771 0.000191447124353172
371349.017889148 0.00016711917119838
371359.951699525 0.000145347810578519
371370.885509903 0.000125949228217114
371381.81932028 0.000108739497631228
371392.753130657 9.35371060698518e-05
371403.686941035 8.01650986701708e-05
371414.620751412 6.84528491337684e-05
371425.554561789 5.82374729594822e-05
371436.488372166 4.93649054940989e-05
371447.422182544 4.16906717962593e-05
371458.355992921 3.50803786198825e-05
371469.289803298 2.94099608251851e-05
371480.223613676 2.45657153528959e-05
};
\addlegendentry{Vanilla}
\end{axis}

\end{tikzpicture}}
 \caption{Probability density function for the maximum resident size (kilobytes) for \system implementation in srsUE\cite{gomez2016srslte} and vanilla srsUE.}
 \label{fig:baseband_implementation_memory}
\end{figure}
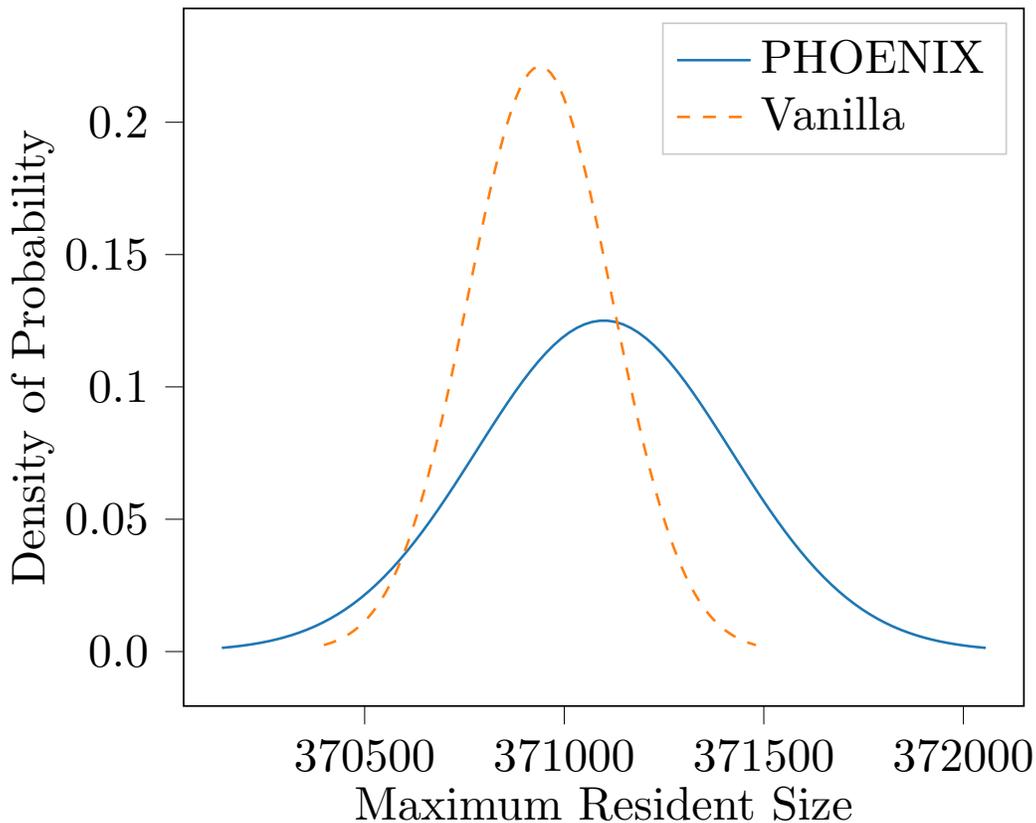

Figure
\ref{fig:baseband_implementation_memory} shows the distribution of both srsUE
implementations. The
distribution is similar in both implementations. The mean difference is only
159.25 KB.
To
put this result in perspective, $srsUE_{vanilla}$ on average consumes approximately
370MB, therefore, \system induces only a mere $0.04\%$ overhead.
Overall, we demonstrate that memory overhead of \system is not a major concern
in its baseband instantiation.


\paragraph{Computational overhead in baseband ($\mathsf{QBB_2}$).}
Another key point that must be analyzed is the computational overhead imposed by
\system in a baseband implementation. This is because 
any substantial
delay imposed by \system could affect the quality of service and result in a disruption
of service. In this experiment, we run the baseband implementation of \system running
all the monitors and measuring the time it takes for all monitors to run sequentially
by measuring the system time in microseconds with the \textbf{getrusage} c++ function.
We carried out this experiment connecting the modified version of srsUE 100 times
to an eNodeB and EPC running on a secondary machine. On average, calling all
15 monitors sequentially added an overhead of $5.43$ microseconds, with a
standard deviation of $10.8$. Overall, this experiment verifies that the overhead
induced by \system is negligible, and would unlikely to induce any
QoS or service disruption issues.

\paragraph{Evaluation Summary of Baseband Implementation.}
We evaluated the overhead induced by the baseband
implementation of \system in srsUE to serve as a proxy to understand the
real world requirements. \system showed to require minimal memory ($159.25$ kbytes)
and computational overhead ($5.4$ microseconds) which shows that \system could be deployed in a real baseband
implementation.

\section{Discussion}
We now discuss different salient aspects of \system.

\paragraph{Impacts of the attacks identified by \system.}
One of the questions that may come up is to ask how prevalent
are the different attacks and unsafe practices identified by \system.
Unfortunately, there are no public quantitative data on the attack
prevalence. We have seen anecdotal evidence
of some attacks and unsafe practices
in the literature. For instance, until two years ago, one of the major
US network operator did not use  encryption to protect their control-plane traffic.
Another major US operator used a persistent identifier in paging messages even though
this is discouraged. In addition, bidding down attacks are pretty commonplace
as warned by many media outlets and even Senators. We envision that \system can enable
more awareness on the security issues of cellular network and protect users from
such attacks. A possible instantiation of PHOENIX is to be deployed as cellular
network probes or honey pots that can log protocol sessions with undesired
behavior in order to perform statistical measurements, and in fact, its
application based instantiation has helped uncover unsafe practices on three
major U.S. cellular network providers.

\paragraph{Android and Qualcomm chipsets.}
Our current implementation of \system supports Qualcomm
baseband processors running on Android. We focus on Android
not only because it is the most popular mobile OS but also
it allows one to expose the cellular interface in the debug
mode with root access. We envision that OSes can expose the modem
information by requesting the permission from the user similar to how
other high privilege permissions can be granted to user
level applications. Additionally, in the future
we aim to extend this for other OSes
and baseband processors \cite{SCAT}.

\paragraph{\system deployed as an Android Application.}
\system{}'s application instantiation
is developed on top of MobileInsight which is written in Python.
This requires a python interpreter to be installed in Android.
Even then \system has shown its effectiveness without incurring a substantial
overhead. Going forward, however, writing the core functionality of \system in
C/C++ will not only positively
impact the performance but also the battery life of the device.


\paragraph{Supporting other protocol versions and device-specific attacks.}
\system is currently instantiated only for 4G LTE. Support of prior protocol
versions  (i.e., 2G/3G) and upcoming versions (i.e., 5G) can be effortlessly
added in the current instantiation by enhancing the current parsing
functionality of \system to include 2G, 3G, and 5G protocol packets. In addition,
signatures for device-specific attacks, due to implementation bugs in a device,
can be supported by the current version of
\system without any change.

\paragraph{User Study.} To be effective as a warning system,
it is paramount to ensure that warning messages \system generates
convey the right information regarding the warning. Performing
a user study with different warning designs is one of the effective
ways of obtaining some assurances on the efficacy of the warning messages.
To limit the scope of the paper to only the technical design of \system,
we leave the user study as future work.

\paragraph{False positives and the quality of signatures.}
The quality of synthesized signatures used in the empirical evaluation
of this paper heavily relies on the diversity of malicious traces used during training.
Having very similar malicious traces in training will likely induce the synthesizer
to come up with a non-generalized behavioral signature that will only be
effective in detecting undesired behavior similar to the ones observed in the malicious training traces and possibly,
a few of its variants. Such a non-generalized signature, however, is
unlikely to capture other diverse variants of the undesired behavior in question that are not present in the
malicious training traces.
This is because the synthesized signatures only guarantee \emph{observationally consistency},
that is, the synthesized signatures will not make any mistakes in correctly
classifying the sample traces given to it during training. Not having diverse
malicious traces during synthesis can thus induce a signature that incurs
false positives. In addition to this situation, false positives can also be incurred due to
having only coarse-grained information on the training traces (i.e., lacking fine-grained information).

Let us consider the downgrade attack through fabricated $\mathsf{attachReject}$ message  as an example \cite{privacy_ndss16}.
In this attack, the adversary establishes a malicious base
station which emits a higher signal strength and then lures the victim device into connecting to the malicious base station.
Then, the adversary during the mutual authentication of the attach procedure injects a fabricated $\mathsf{attachReject}$ message,
resulting in the device to downgrade to an insecure protocol version (e.g., moving from 4G LTE to 3G).
Just by observing the temporal ordering of events, without taking into consideration the migration of the device from 4G LTE to 3G, the best possible
signature a synthesizer can come up with is the existence of the $\mathsf{attachReject}$ message. Such a signature
may induce false positives because $\mathsf{attachReject}$ messages can be sent by the network in benign situations.

In our experiments, due to random selection of benign sessions from real-life traces and $\mathsf{attachReject}$
messages being rare, none of the benign traces during
training had any $\mathsf{attachReject}$ messages, whereas all the negative traces at least had one.
As a result, the synthesizer rightfully came up with a signature saying an existence of
the $\mathsf{attachReject}$ message is an attack, especially because this is the smallest signature that is
observationally consistent. Due to our session-level trace generation approach discussed before,
we did not have faithful information about the downgrade of the protocol version.
Even if we had that precise information as part of the training traces, the synthesizer will give us a precise
signature according to which the monitor will identify the attack only
when the connection downgrade \emph{has already happened} instead of before protocol version downgrade happens.
We argue that a proactive monitor, which notifies the user as soon as an $\mathsf{attachReject}$ message has occurred instead
of waiting for a downgrade, is more effective in protecting the user even  at the cost of a
few false positives.  Since $\mathsf{attachReject}$ messages are indeed rare in practice,
corroborated by the MobileInsight traffic, such a trade-off is a reasonable
choice.

\paragraph{Alternate approaches.}
One alternate method for in-device monitoring of cellular traffic for
attack detection is to construct a state machine representing the desired
cellular protocol behavior from the 3GPP standard  specification \cite{3gpp, 3gpp12}.
At runtime, one could then flag any state transitions that are not allowed by this
reference state machine as potential attacks.
Such an approach, however, suffers from the following limitations:
(1) Manually constructing the protocol state machine by reading natural languages specification
is an error-prone and time-consuming ordeal;
(2) The \emph{abstract} protocol state machine given by the standard specification is not always prescriptive, leaving
aspects to be decided by implementors;
(3) Not all deviant state transitions necessarily signify an attack
and thus can induce a high number of false warnings.

In another approach, one may consider manually writing one small program
for each attack using simple if-then-else constructs that checks for a certain
attack in the trace. The attacks we consider here, however, would require
storing memory and would require intensive manual effort to identify the portion
of the traces that need to be stored. In contrast, our approach is completely
automated in figuring out what portion of the history to store for careful attack
identification without requiring human intervention. Such automation comes in handy
when new attacks are discovered and their signatures need to be deployed promptly.

\section{Related Work}

\paragraph{Runtime Monitors.} Extensive work has been done in
developing efficient runtime monitors using different types of
logic~\cite{basin2010monitoring, basin2010policy, basin2008runtime,
basin2015monitoring, monpoly, bauer2011runtime, du2015trace,
rosu2001synthesizing,soewito2009self, wpse}. However, all but
~\cite{soewito2009self, wpse} attempt to create a deployable
system which tries to apply runtime monitoring to web protocols.
In contrast, \system aims to be a deployable system, similar
to ~\cite{soewito2009self, wpse}, however, we apply runtime
monitoring to 4G LTE cellular networks. In addition, we apply three different
runtime monitor approaches while~\cite{soewito2009self, wpse}
only rely on automata based approaches. \system not only serves
as the runtime monitor but also provide the learning component
to generate signatures, including \pltl formulas.

\paragraph{Anomaly Detection in Cellular Devices.} Some work has been done to detect anomalies in cellular networks within the cellular device, precisely to discover the presence of fake base stations proposed by Dabrowski et al.~\cite{imsi_catcher_catchers}. In addition,  multiple apps have attempted to enable the detection of fake base stations using an application, but unfortunately do not generalize well~\cite{white_stingray}. In contrast to these attempts at anomaly detection, \system looks for specific patterns of message flow to detect specific attacks and provide a possible remedy.

\paragraph{Modification of Protocol.} Another approach researchers have leveraged to provide a defense mechanism is the modification of the protocol, such as in \cite{concealing_imsi_in_5g, privacy_5g, PMSI_desynchronization, imsicatcher_ccs15, wisec_root}. Out of these works, only \cite{wisec_root} provides a wide array of coverage while the others mainly focus on the IMSI catching attack. In contrast to other work, \system is the first warning system for cellular networks that provides the device more intelligence about other components of the network by only relying on message flows.

\section{Conclusion}
In this paper, we develop \system, a general approach which can efficiently monitor
a device's cellular network traffic and identify the presence of attacks. We achieve
this by instantiating two different implementations
of PHOENIX: a runtime monitor within an Android application, allowing the
cellular device to reason about malicious message flow and alert the user;
A modified version of srsUE \cite{gomez2016srslte} powered by a runtime monitor
allowing it to detect vulnerabilities and prevent potential undesired behavior.

Overall we observe that our best approach with PLTL can correctly identify all the
15 n-day 4G LTE attacks and unsafe practices used in the evaluation section
with a high packet processing speed ($\sim$68000 packets/second), while inducing
a moderate energy ($\sim$4mW) and negligible memory overhead ($0.04\%$) on the device.

\section*{Acknowledgment}
We thank the anonymous reviewers for their valuable suggestions. This
work was funded by DARPA YFA contract no. D19AP00039. The views and conclusions
contained herein are those of the authors and should not be interpreted as
necessarily representing the official policies or endorsements of DARPA.

\bibliographystyle{IEEEtranS}
\bibliography{LTE_Monitor}

\appendix

\section{Cellular Network Vulnerabilities}
\label{appendix:attack-list}
In this section, we provide an extensive list of vulnerabilities found on cellular
networks. Table \ref{app:cellular_network_extensive_list} shows an extensive list
of cellular network attacks uncovered in recent years summarizing which attacks
we can detect with the current iteration, could detect with one of the following
extensions to \system: \textbf{1)} Extend the message parser to decode another
cellular network generation traffic; \textbf{2)} Extend the message parser to
decode other layer traffic; \textbf{3)} Add a specific predicate to enable
the attack detection, or cannot detect due to one of the following reasons: \textbf{a)}
temporal ordering does not serve as a proxy to uncover vulnerable behavior;
\textbf{b)} Require statistical information (e.g., number of paging messages
received within a given time frame).

\begin{table*}[]
  \centering
  \resizebox{\columnwidth}{!}{
  \begin{tabular}{|l|c|c|c|c|c|}
      \hline
      \multicolumn{1}{|c|}{\textbf{Attack Name}}                    & \textbf{Paper}   & \textbf{Layer} & \multicolumn{1}{c|}{\textbf{Repercussion}}        & \textbf{Detectable} & \textbf{Cause} \\ \hline
      Authentication Synchronization Failure                        & \cite{lteinspector}     & NAS            & Denial of Service                                 & \newmoon                    & \ding{72} \\ \hline
      Traceability Attack                                           & \cite{lteinspector}     & NAS            & Location Leak                                     &
    \newmoon                    & \ding{72} \\ \hline
      Numb Attack                                                   & \cite{lteinspector}     & NAS            & Denial of Service                                 &
    \newmoon                    & \ding{72} \\ \hline
      Paging Channel Hijacking                                      & \cite{lteinspector}     & RRC            & Denial of Service                                 & \fullmoon                    & \ding{72} \\ \hline
      Stealthy Kicking Off                                          & \cite{lteinspector}     & RRC            & Denial of Service                                 &
    \newmoon                    & \ding{72} \\ \hline
      Panic Attack                                                  & \cite{lteinspector}     & RRC            & Artificial Chaos                                  & \fullmoon                    & \ding{72} \\ \hline
      Energy Depletion Attack                                       & \cite{lteinspector}     & RRC            & Battery Depletion                                 & \fullmoon                    & \ding{72} \\ \hline
      Linkability Attack                                            & \cite{lteinspector}     & RRC            & Location Leak                                     &
    \newmoon                    & \ding{72} \\ \hline
      Detach / Downgrade Attack                                     & \cite{lteinspector}     & NAS            & Denial of Service / Downgrade                     &
    \newmoon                    & \ding{72} \\ \hline
      Authentication Relay Attack                                   & \cite{lteinspector}     & NAS            & Location Poisoning                                & \fullmoon                    & \ding{72} \\ \hline
      TORPEDO Attack                                                & \cite{TORPEDO}          & RRC            & Identifier Leak / Location Leak                   & \fullmoon                    & \ding{72} \\ \hline
      IMSI Cracking Attack Against 4G                               & \cite{TORPEDO}          & RRC            & Identifier Leak                                   &
    \newmoon                    & \ding{72} \\ \hline
      IMSI Cracking Attack Against 5G                               & \cite{TORPEDO}          & NAS            & Identifier Leak                                   &  \leftmoon                   & \ding{72} \\ \hline
      Social Network to TMSI Mapping                                & \cite{privacy_ndss16} & RRC            & Identifier Leak                                   & \fullmoon                    & \ding{72} \\ \hline
      Link Subscriber Location Movement                             & \cite{privacy_ndss16} & RRC            & Location Leak                                     & \fullmoon                    & \ding{72} \\ \hline
      Leak Coarse Location                                          & \cite{privacy_ndss16} & RRC            & Location Leak                                     & \fullmoon                    & \ding{72} \\ \hline
      Measurement Report Location Leak                              & \cite{privacy_ndss16} & RRC            & Location Leak                                     &
    \newmoon                    & \ding{73} \\ \hline
      RLF Report Location Leak                                      & \cite{privacy_ndss16} & RRC            & Location Leak                                     &
    \newmoon                    & \ding{73} \\ \hline
      TAU Reject to Disrupt Service                                 & \cite{privacy_ndss16} & NAS            & Denial of Service                                 &
    \newmoon                    & \ding{73} \\ \hline
      Service Reject to Disrupt Service                             & \cite{privacy_ndss16} & NAS            & Denial of Service                                 &
    \newmoon                    & \ding{72} \\ \hline
      Attach Reject to Disrupt Service                              & \cite{privacy_ndss16} & NAS            & Denial of Service                                 &
    \newmoon                    & \ding{72} \\ \hline
      Denying Selected Service with Malicious Attach Request        & \cite{privacy_ndss16} & NAS            & Denial of Service                                 & \fullmoon                    & \ding{72} \\ \hline
      Counter Reset (Replay)                                        & \cite{5g_reasoner}       & NAS (5G)       & Denial of Service                                 & \leftmoon                    & \ding{72} \\ \hline
      Counter Reset (Reset)                                         & \cite{5g_reasoner}       & NAS (5G)       & Denial of Service / Overbilling                   & \leftmoon                    & \ding{72} \\ \hline
      Uplink NAS Counter Desynchronization                          & \cite{5g_reasoner}       & NAS (5G)       & Denial of Service                                 & \leftmoon                    & \ding{72} \\ \hline
      Exposing NAS Sequence Number                                  & \cite{5g_reasoner}       & NAS (5G)       & Service Profiling                                 & \fullmoon                    & \ding{72} \\ \hline
      Neutralizing TMSI Refreshment                                 & \cite{5g_reasoner}       & NAS (5G)       & Location Leak                                     & \leftmoon                    & \ding{72} \\ \hline
      Cutting off the Device                                        & \cite{5g_reasoner}       & NAS (5G)       & Denial of Service                                 & \fullmoon                    & \ding{72} \\ \hline
      Denial of Service with RRC Setup Request                      & \cite{5g_reasoner}       & RRC (5G)       & Denial of Service                                 & \fullmoon                    & \ding{72} \\ \hline
      Installing Null Cipher and Null Integrity                     & \cite{5g_reasoner}       & RRC (5G)       & Identifier Leak                                   & \leftmoon                    & \ding{72} \\ \hline
      Lullaby Attack                                                & \cite{5g_reasoner}       & RRC (5G)       & Battery Depletion                                 & \fullmoon                    & \ding{72} \\ \hline
      Incarceration using RRC messages                              & \cite{5g_reasoner}       & RRC (5G)       & Denial of Service                                 & \leftmoon                    & \ding{72} \\ \hline
      Exposing Device's TMSI and Paging Occasion                    & \cite{5g_reasoner}       & NAS / RRC (5G) & \makecell{Identifier Leak/\\ Denial of Service/\\ Location Leak} & \fullmoon                    & \ding{72} \\ \hline
      Exposing Device's I-RNTI                                      & \cite{5g_reasoner}       & NAS / RRC (5G) & Identifier Leak                                   & \fullmoon                    & \ding{72} \\ \hline
      Blind DoS Attack                                              & \cite{5g_reasoner}       & NAS / RRC (5G) & Denial of Service                                 & \fullmoon                    & \ding{72} \\ \hline
      AKA Bypass                                                    & \cite{kim_ltefuzz_sp19}          & RRC            & Eavesdropping                                     &
    \newmoon                    & \ding{73} \\ \hline
      RRC Connection Reestablishment - Unencrypted                  & \cite{kim_ltefuzz_sp19}          & RRC            & Denial of Service                                 &
    \newmoon                    & \ding{73} \\ \hline
      RRC Connection Reconfiguration - Unencrypted                  & \cite{kim_ltefuzz_sp19}          & RRC            & Eavesdropping / Location Leak                     &
    \newmoon                    & \ding{73} \\ \hline
      RRC Connection Reestablishment Reject - Unencrypted            & \cite{kim_ltefuzz_sp19}          & RRC            & Denial of Service                                 &
    \newmoon                    & \ding{73} \\ \hline
      RRC Connection Reconfiguration - Invalid Integrity Protection & \cite{kim_ltefuzz_sp19}          & RRC            & Eavesdropping / Location Leak                     &
    \newmoon                    & \ding{73} \\ \hline
      UP Capability Enquiry - Invalid Integrity Protection          & \cite{kim_ltefuzz_sp19}          & RRC            & Service Profiling                                 &
    \newmoon                    & \ding{73} \\ \hline
      UP Capability Enquiry - Invalid Sequence Number               & \cite{kim_ltefuzz_sp19}          & RRC            & Service Profiling                                 &
    \newmoon                    & \ding{73} \\ \hline
      RRC Connection Reject - Distinguishability                    & \cite{kim_ltefuzz_sp19}          & RRC            & Denial of Service                                 &
    \newmoon                    & \ding{73} \\ \hline
      RRC Connection Setup - Distinguishability                     & \cite{kim_ltefuzz_sp19}          & RRC            & Denial of Service                                 &
    \newmoon                    & \ding{73} \\ \hline
      GUTI Reallocation Command - Invalid Sequence Number           & \cite{kim_ltefuzz_sp19}          & NAS            & Identifier Poisoning                              &
    \newmoon                    & \ding{73} \\ \hline
      Identity Request - Invalid Sequence Number                     & \cite{kim_ltefuzz_sp19}          & NAS            & Identifier Leak                                   &
    \newmoon                    & \ding{73} \\ \hline
      IMSI Catcher with TAU Reject                                  & \cite{lte_redirection}  & NAS            & Identifier Leak                                   &
    \newmoon                    & \ding{72} \\ \hline
      Attach Reject to Disrupt Service                              & \cite{lte_redirection}  & NAS            & Denial of Service                                 &
    \newmoon                    & \ding{72} \\ \hline
      Redirection Attack                                            & \cite{lte_redirection}  & RRC/NAS       & Denial of Service                                 & \fullmoon                    & \ding{72} \\ \hline
      EMM Information - Unencrypted                                 & \cite{park2016white}     & NAS            & Time Desynchronization                            &
    \newmoon                    & \ding{72} \\ \hline
      Identity Request - Improper Type ID                           & \cite{how_not_to_break_crypto} & NAS            & Identifier Leak                                   &
    \newmoon                    & \ding{73} \\ \hline
      Identity Mapping                                              & \cite{alter}            & RRC            & Identifier Leak                                   & \fullmoon                    & \ding{72} \\ \hline
      Website Fingerprinting                                       & \cite{alter}            & RRC            & Service Profiling                                 & \fullmoon                    & \ding{72} \\ \hline
      alter                                                         & \cite{alter}            & RRC            & DNS Redirection                                   & \fullmoon                    & \ding{72} \\ \hline
      Traffic Fingerprinting                                        & \cite{wisec191}     & RRC            & Service Profiling                                 & \fullmoon                    & \ding{72} \\ \hline
      Identification and Localization                               & \cite{wisec191}     & RRC            & Location Leak                                     & \fullmoon                    & \ding{72} \\ \hline
      ReVoLTE                                                       & \cite{rupprecht2020call}    & RRC            & Eavesdropping                                     & \leftmoon                    & \ding{73} \\ \hline
    \end{tabular}}
\caption{Attacks on Cellular Networks describing whether or not, each attack is
detectable by \system ($\newmoon$), theoretically possible with the aid
of an extension to \system such as decoding other layer traffic ($\leftmoon$),
or not detectable by \system ($\fullmoon$) on the detectable column. Additionally,
the cause is classified either as an implementation slipup (\ding{73}) or
an error in the standard enabling this vulnerability (\ding{72}).}
\label{app:cellular_network_extensive_list}
\end{table*}

\section{\system Warning to User}
\label{app:phoenix_warning_to_user}
In this section we provide screenshots of the \system app and the warnings
provided to the user.

\subsection{\system screenshots}
\begin{figure}[t]
\centering
\subfloat[No attack has occurred.]{{\includegraphics[width=.3\columnwidth]{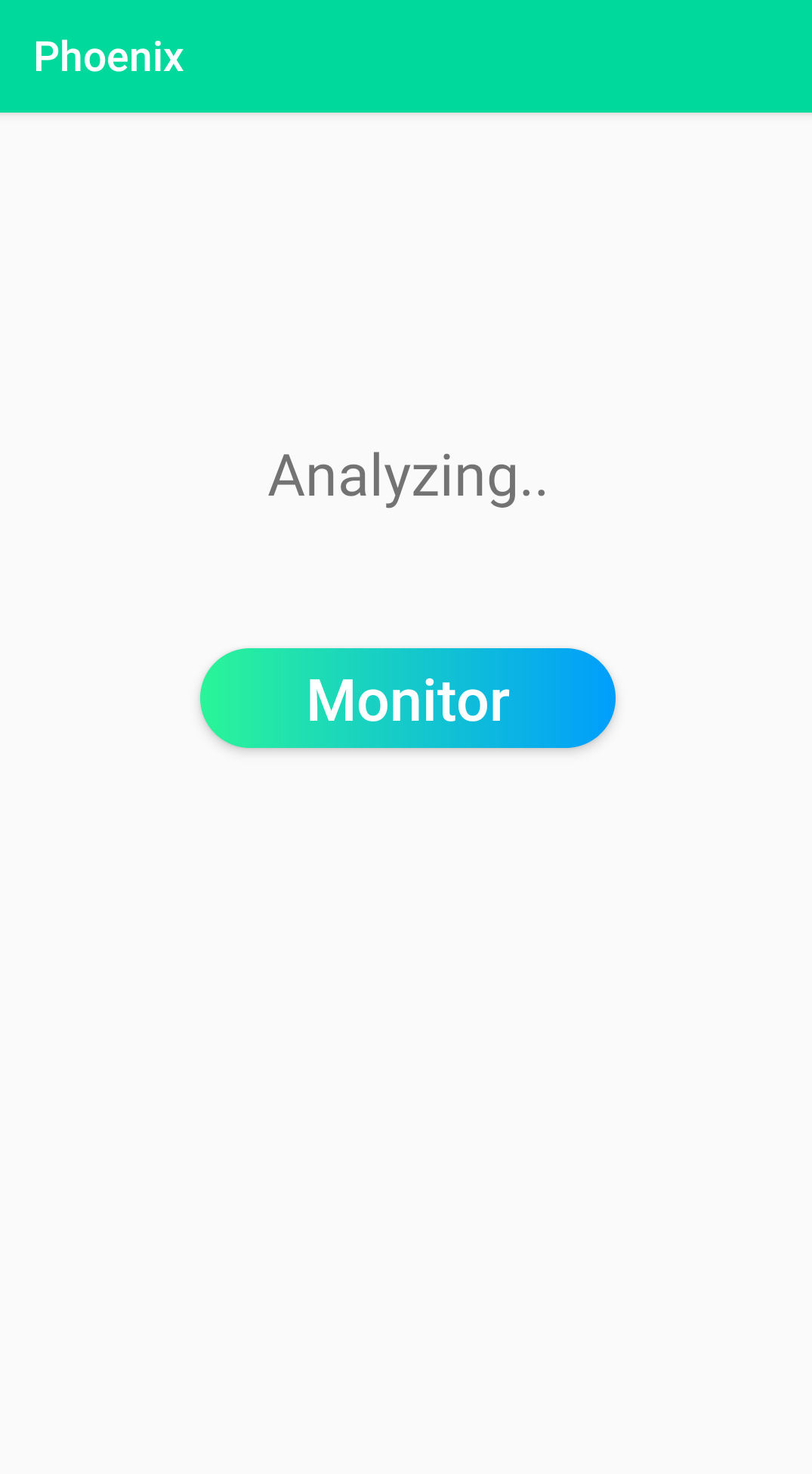}}}
\qquad
\subfloat[Numb Attack Detected by \system.]{{\includegraphics[width=.3\columnwidth]{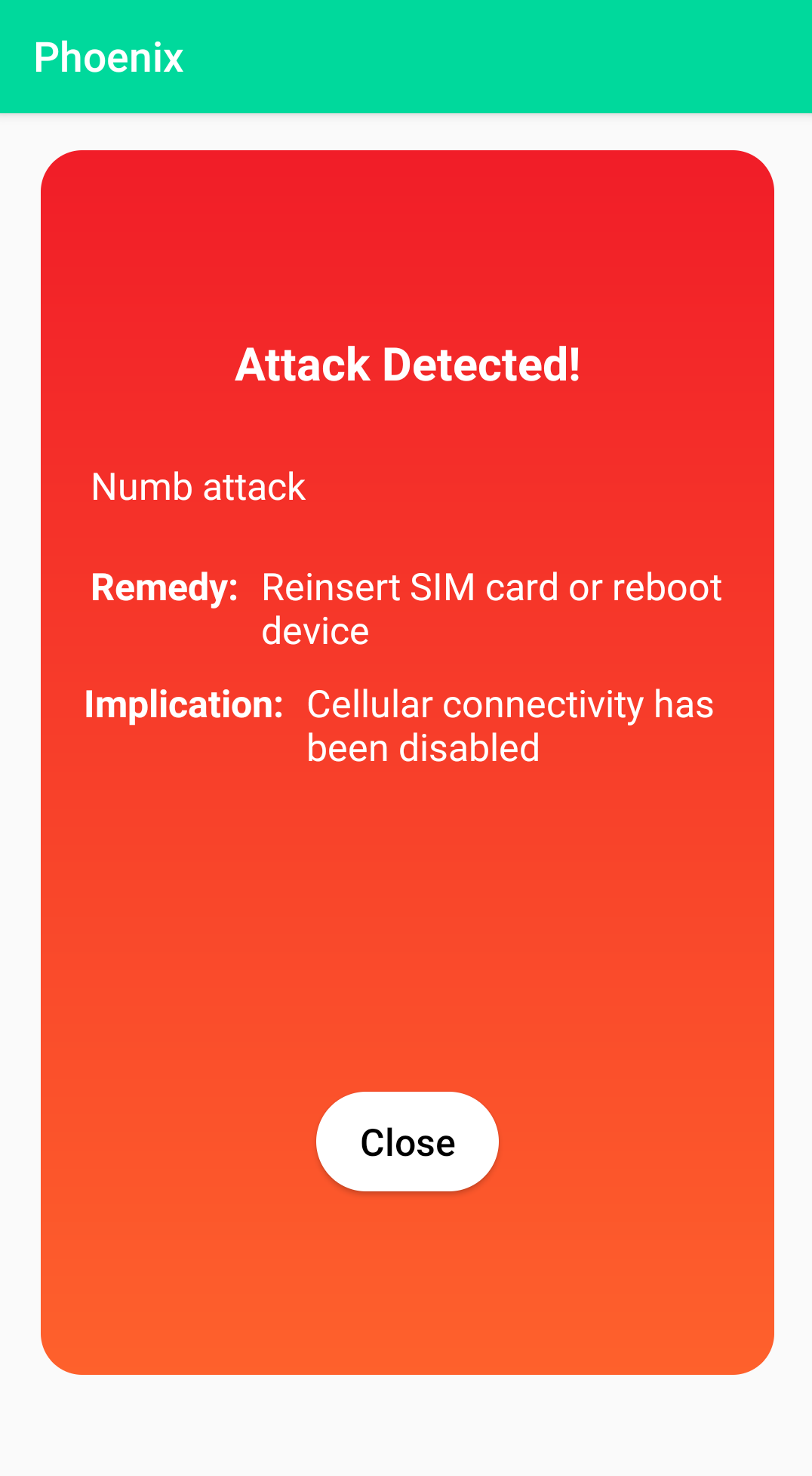}}}
	\caption{Screenshot of \system's application running.}
	\label{fig:phoenix_app_screenshots}
\end{figure}

In this subsection we present two screenshots of the \system application.
In Figure \ref{fig:phoenix_app_screenshots}(a), \system is running but no attack
has occurred. In Figure \ref{fig:phoenix_app_screenshots}(b), \system is running
when the Numb attack was performed and the attack was detected.

When \system detects the Numb Attack, it provides a possible remedy which is
to re-insert the SIM card or completely reboot the device. Additionally,
it provides a description of the implication of this attack.

\section{Evaluation}
\label{app:appendix_evaluation_section}
In this section we provide more in depth results for the experiments described
in Section \ref{sec:evaluation_of_phoenix}. Additionally, we provide a figure
for the DFA signature to detect the AKA Bypass Attack \cite{kim_ltefuzz_sp19}.

\clearpage
\begin{figure}[t]
\centering
	\resizebox{\textwidth}{!}{
	\includegraphics[width=.8\columnwidth]{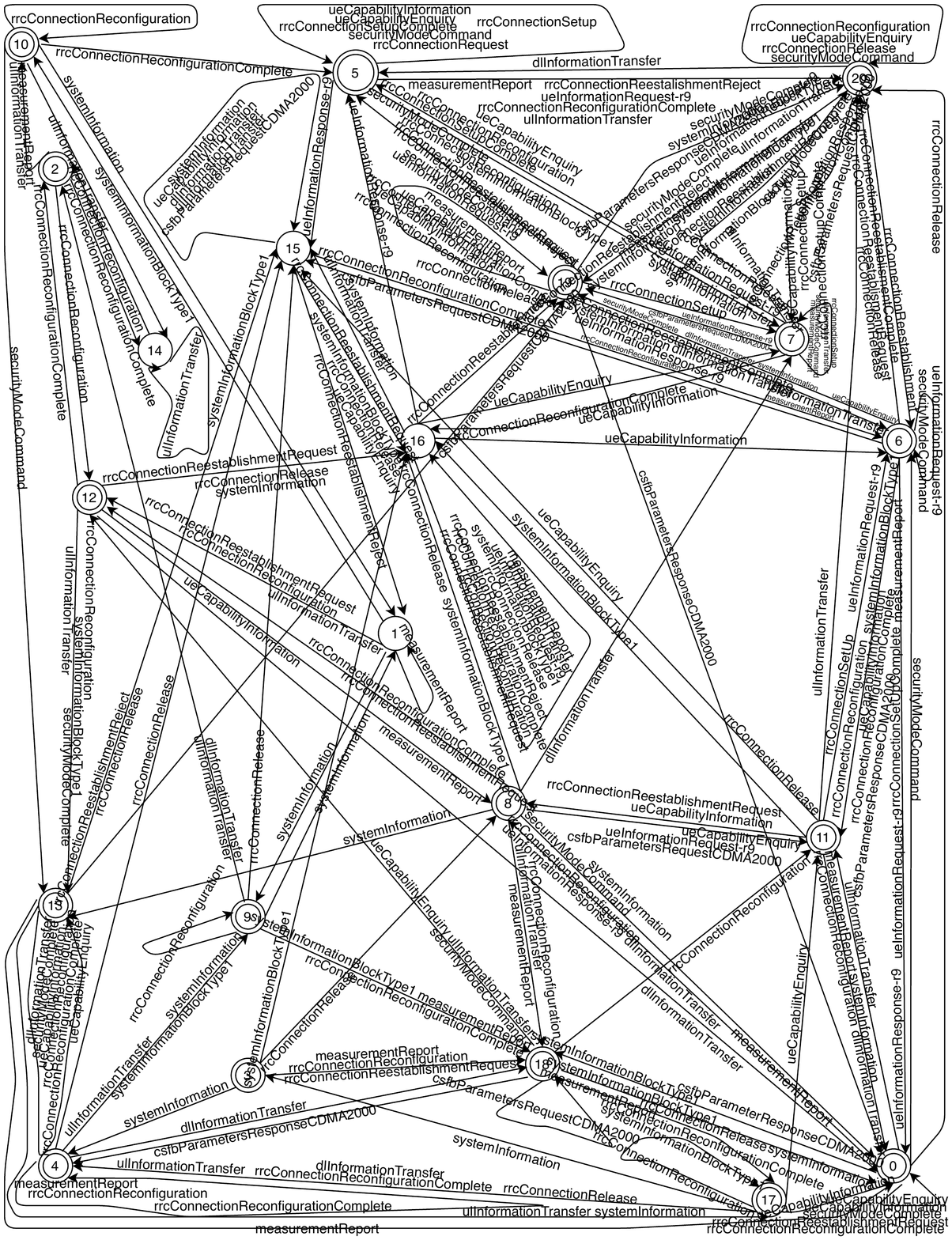}
	}
	\caption{DFA signature powering the AKA Bypass Attack \cite{kim_ltefuzz_sp19}}.
	\label{fig:aka_bypass_dfa}
\end{figure}
\clearpage

\subsection{Training Time and Size of Synthesized Signatures}
\label{app:training_time_and_size_all_dfa_pltl_section}

In this subsection we present the training time and size of all the synthesized
signatures.

Table \ref{app:training_time_and_size_all_mm} provides the training
time in seconds and the corresponding size for all the synthesized Mealy Machine
signatures.

Table \ref{app:training_time_and_size_all_dfa_pltl} provides the training time
in seconds and the corresponding size for all the synthesized \pltl and DFA
signatures. Note that the \pltl synthesizer produces $5$ different signatures
for each attack. An exception to this is the Measurement Report which we
had to invoke with $20$ traces due to it timing out. When the synthesizer
was invoked with $20$ traces it was not capable of synthesizing $5$ signatures,
however, it was able to synthesize one.

\begin{table}[]
  \centering
\begin{tabular}{cccccc}
\hline
\multicolumn{1}{|c|}{\textbf{Size}} & \multicolumn{1}{c|}{\textbf{States}} & \multicolumn{1}{c|}{\textbf{Transitions}} & \multicolumn{1}{c|}{\textbf{Input}} & \multicolumn{1}{c|}{\textbf{Output}} & \multicolumn{1}{c|}{\textbf{Training Time}} \\ \hline
\multicolumn{6}{|c|}{\textbf{NAS}}                                                                                                                                                                                                                \\ \hline
\multicolumn{1}{|c|}{50}            & \multicolumn{1}{c|}{2}               & \multicolumn{1}{c|}{60}                   & \multicolumn{1}{c|}{32}             & \multicolumn{1}{c|}{11}              & \multicolumn{1}{c|}{0.1}                    \\ \hline
\multicolumn{1}{|c|}{100}           & \multicolumn{1}{c|}{2}               & \multicolumn{1}{c|}{59}                   & \multicolumn{1}{c|}{32}             & \multicolumn{1}{c|}{11}              & \multicolumn{1}{c|}{0.07}                   \\ \hline
\multicolumn{1}{|c|}{250}           & \multicolumn{1}{c|}{2}               & \multicolumn{1}{c|}{55}                   & \multicolumn{1}{c|}{32}             & \multicolumn{1}{c|}{11}              & \multicolumn{1}{c|}{0.08}                   \\ \hline
\multicolumn{1}{|c|}{500}           & \multicolumn{1}{c|}{5}               & \multicolumn{1}{c|}{118}                  & \multicolumn{1}{c|}{32}             & \multicolumn{1}{c|}{11}              & \multicolumn{1}{c|}{0.1}                    \\ \hline
\multicolumn{1}{|c|}{1250}          & \multicolumn{1}{c|}{4}               & \multicolumn{1}{c|}{113}                  & \multicolumn{1}{c|}{32}             & \multicolumn{1}{c|}{11}              & \multicolumn{1}{c|}{0.16}                   \\ \hline
\multicolumn{1}{|c|}{2500}          & \multicolumn{1}{c|}{4}               & \multicolumn{1}{c|}{108}                  & \multicolumn{1}{c|}{32}             & \multicolumn{1}{c|}{11}              & \multicolumn{1}{c|}{0.21}                   \\ \hline
                                    &                                      &                                           &                                     &                                      &                                             \\ \hline
\multicolumn{6}{|c|}{\textbf{RRC}}                                                                                                                                                                                                                \\ \hline
\multicolumn{1}{|c|}{50}            & \multicolumn{1}{c|}{25}              & \multicolumn{1}{c|}{435}                  & \multicolumn{1}{c|}{33}             & \multicolumn{1}{c|}{6}               & \multicolumn{1}{c|}{0.06}                   \\ \hline
\multicolumn{1}{|c|}{100}           & \multicolumn{1}{c|}{30}              & \multicolumn{1}{c|}{466}                  & \multicolumn{1}{c|}{33}             & \multicolumn{1}{c|}{6}               & \multicolumn{1}{c|}{0.03}                   \\ \hline
\multicolumn{1}{|c|}{250}           & \multicolumn{1}{c|}{23}              & \multicolumn{1}{c|}{409}                  & \multicolumn{1}{c|}{33}             & \multicolumn{1}{c|}{6}               & \multicolumn{1}{c|}{0.04}                   \\ \hline
\multicolumn{1}{|c|}{500}           & \multicolumn{1}{c|}{28}              & \multicolumn{1}{c|}{525}                  & \multicolumn{1}{c|}{33}             & \multicolumn{1}{c|}{6}               & \multicolumn{1}{c|}{0.05}                   \\ \hline
\multicolumn{1}{|c|}{1250}          & \multicolumn{1}{c|}{46}              & \multicolumn{1}{c|}{955}                  & \multicolumn{1}{c|}{33}             & \multicolumn{1}{c|}{6}               & \multicolumn{1}{c|}{0.06}                   \\ \hline
\multicolumn{1}{|c|}{2500}          & \multicolumn{1}{c|}{2}               & \multicolumn{1}{c|}{65}                   & \multicolumn{1}{c|}{33}             & \multicolumn{1}{c|}{6}               & \multicolumn{1}{c|}{0.08}                   \\ \hline
\end{tabular}%
\caption{Mealy Machine Sizes and training time in seconds.}
\label{app:training_time_and_size_all_mm}
\end{table}

\begin{table*}[]
  \centering
  \resizebox{1\columnwidth}{!}{%
	\begin{minipage}{2\columnwidth}
		\centering
  \begin{tabular}{|l|l|l|l|l|l|l|l|l|}
  \hline
  \multicolumn{1}{|c|}{\textbf{Attack name}} & \multicolumn{1}{c|}{\textbf{Dataset Size}} & \multicolumn{1}{c|}{\textbf{\begin{tabular}[c]{@{}c@{}}DFA \\ Training Time\end{tabular}}} & \multicolumn{1}{c|}{\textbf{\begin{tabular}[c]{@{}c@{}}PLTL \\ Training Time\end{tabular}}} & \multicolumn{1}{c|}{\textbf{\begin{tabular}[c]{@{}c@{}}DFA \\ States\end{tabular}}} & \multicolumn{1}{c|}{\textbf{\begin{tabular}[c]{@{}c@{}}DFA \\ Transitions\end{tabular}}} & \multicolumn{1}{c|}{\textbf{\begin{tabular}[c]{@{}c@{}}DFA Alphabet\\ Size\end{tabular}}} & \multicolumn{1}{c|}{\textbf{\begin{tabular}[c]{@{}c@{}}PLTL \\ Propositions\end{tabular}}} & \multicolumn{1}{c|}{\textbf{\begin{tabular}[c]{@{}c@{}}PLTL \\ Operators\end{tabular}}} \\ \hline
  \multicolumn{9}{|c|}{\textbf{NAS}}                                                                                                                                                                                                                \\ \hline
  \multirow{6}{*}{Attach Reject} & 50 & 0.05 & 21.44 & 7 & 80 & 17 & 1 & 1 \\ \cline{2-9}
   & 100 & 0.67 & 49.7 & 2 & 33 & 17 & 1 & 1 \\ \cline{2-9}
   & 250 & 0.67 & 137.25 & 18 & 140 & 17 & 1 & 1 \\ \cline{2-9}
   & 500 & 0.67 & 389.94 & 22 & 286 & 19 & 1 & 1 \\ \cline{2-9}
   & 1250 & 1 & TIMEOUT & 4 & 60 & 18 & N/A & N/A \\ \cline{2-9}
   & 2500 & 0.5 & TIMEOUT & 2 & 35 & 18 & N/A & N/A \\ \hline
  \multirow{6}{*}{Authentication Failure} & 50 & 0.29 & 14.07 & 3 & 36 & 17 & 1 & 1 \\ \cline{2-9}
   & 100 & 0.6 & 43.77 & 5 & 64 & 17 & 1 & 1 \\ \cline{2-9}
   & 250 & 0.43 & 114.26 & 9 & 101 & 17 & 1 & 1 \\ \cline{2-9}
   & 500 & 0.375 & 379.76 & 17 & 184 & 18 & 1 & 1 \\ \cline{2-9}
   & 1250 & 0.5 & 1677.92 & 5 & 62 & 18 & 1 & 1 \\ \cline{2-9}
   & 2500 & 0.38 & TIMEOUT & 4 & 58 & 18 & N/A & N/A \\ \hline
  \multirow{6}{*}{EMM Information} & 50 & 0.17 & 26.62 & 10 & 104 & 18 & 1 & 1 \\ \cline{2-9}
   & 100 & 0.5 & 50.25 & 4 & 53 & 18 & 1 & 1 \\ \cline{2-9}
   & 250 & 0.2 & 463.9 & 3 & 47 & 18 & 1 & 1 \\ \cline{2-9}
   & 500 & 0.375 & 1372.09 & 11 & 120 & 18 & 1 & 1 \\ \cline{2-9}
   & 1250 & 0.375 & TIMEOUT & 2 & 35 & 19 & N/A & N/A \\ \cline{2-9}
   & 2500 & 0.6 & TIMEOUT & 5 & 64 & 18 & N/A & N/A \\ \hline
  \multirow{6}{*}{IMEI Catching} & 50 & 0.5 & 22.42 & 12 & 106 & 20 & 1 & 1 \\ \cline{2-9}
   & 100 & 0.3 & 46.78 & 5 & 81 & 19 & 1 & 1 \\ \cline{2-9}
   & 250 & 0.33 & 142.52 & 4 & 69 & 19 & 1 & 1 \\ \cline{2-9}
   & 500 & 0.3 & 370.02 & 4 & 65 & 20 & 1 & 1 \\ \cline{2-9}
   & 1250 & 0.8 & TIMEOUT & 6 & 99 & 19 & N/A & N/A \\ \cline{2-9}
   & 2500 & 0.3 & TIMEOUT & 3 & 46 & 19 & N/A & N/A \\ \hline
  \multirow{6}{*}{IMSI Catching} & 50 & 0.15 & 26.61 & 5 & 64 & 20 & 1 & 1 \\ \cline{2-9}
   & 100 & 0.27 & 58.55 & 5 & 80 & 19 & 1 & 1 \\ \cline{2-9}
   & 250 & 0.25 & 149.23 & 14 & 166 & 20 & 1 & 1 \\ \cline{2-9}
   & 500 & 0.5 & 329.04 & 7 & 107 & 20 & 1 & 1 \\ \cline{2-9}
   & 1250 & 0.6 & TIMEOUT & 3 & 49 & 20 & N/A & N/A \\ \cline{2-9}
   & 2500 & 0.38 & TIMEOUT & 2 & 36 & 19 & N/A & N/A \\ \hline
  \multirow{6}{*}{Malformed Identity Request} & 50 & 0.33 & 21.94 & 8 & 101 & 20 & 1 & 1 \\ \cline{2-9}
   & 100 & 0.23 & 51.89 & 18 & 191 & 20 & 1 & 1 \\ \cline{2-9}
   & 250 & 0.5 & 190.94 & 12 & 146 & 21 & 1 & 1 \\ \cline{2-9}
   & 500 & 0.23 & 370.37 & 9 & 116 & 21 & 1 & 1 \\ \cline{2-9}
   & 1250 & 0.33 & TIMEOUT & 4 & 69 & 21 & N/A & N/A \\ \cline{2-9}
   & 2500 & 0.43 & TIMEOUT & 4 & 66 & 20 & N/A & N/A \\ \hline
  \multirow{6}{*}{Null Encryption} & 50 & 0.38 & 18.29 & 4 & 53 & 20 & 1 & 1 \\ \cline{2-9}
   & 100 & 0.6 & 47.22 & 4 & 54 & 20 & 1 & 1 \\ \cline{2-9}
   & 250 & 0.6 & 161.44 & 12 & 158 & 21 & 1 & 1 \\ \cline{2-9}
   & 500 & 0.3 & 385.91 & 13 & 150 & 21 & 1 & 1 \\ \cline{2-9}
   & 1250 & 0.43 & TIMEOUT & 10 & 119 & 21 & N/A & N/A \\ \cline{2-9}
   & 2500 & 0.5 & TIMEOUT & 5 & 87 & 21 & N/A & N/A \\ \hline
  \multirow{6}{*}{Numb Attack} & 50 & 0.43 & 57.05 & 4 & 44 & 18 & 2 & 2 \\ \cline{2-9}
   & 100 & 0.2 & 144.87 & 4 & 39 & 18 & 2 & 2 \\ \cline{2-9}
   & 250 & 0.2 & 359.68 & 3 & 31 & 19 & 2 & 2 \\ \cline{2-9}
   & 500 & 0.27 & 558.54 & 3 & 35 & 19 & 2 & 2 \\ \cline{2-9}
   & 1250 & 0.17 & TIMEOUT & 6 & 75 & 20 & N/A & N/A \\ \cline{2-9}
   & 2500 & 0.23 & TIMEOUT & 3 & 33 & 20 & N/A & N/A \\ \hline
  \multirow{6}{*}{Service Reject} & 50 & 0.21 & 33.93 & 7 & 100 & 18 & 1 & 1 \\ \cline{2-9}
   & 100 & 0.33 & 82.2 & 2 & 34 & 18 & 1 & 1 \\ \cline{2-9}
   & 250 & 0.23 & 153.34 & 5 & 71 & 18 & 1 & 1 \\ \cline{2-9}
   & 500 & 0.33 & 1110.77 & 25 & 279 & 18 & 1 & 1 \\ \cline{2-9}
   & 1250 & 0.67 & TIMEOUT & 2 & 35 & 18 & N/A & N/A \\ \cline{2-9}
   & 2500 & 0.6 & TIMEOUT & 2 & 35 & 18 & N/A & N/A \\ \hline
  \multirow{6}{*}{TAU Reject} & 50 & 0.43 & 21.97 & 2 & 35 & 19 & 1 & 1 \\ \cline{2-9}
   & 100 & 0.43 & 55.51 & 9 & 118 & 19 & 1 & 1 \\ \cline{2-9}
   & 250 & 0.375 & 156.81 & 2 & 37 & 19 & 1 & 1 \\ \cline{2-9}
   & 500 & 0.43 & 348.59 & 2 & 36 & 19 & 1 & 1 \\ \cline{2-9}
   & 1250 & 0.33 & TIMEOUT & 15 & 187 & 19 & N/A & N/A \\ \cline{2-9}
   & 2500 & 0.3 & TIMEOUT & 6 & 66 & 19 & N/A & N/A \\ \hline
   \multicolumn{9}{|c|}{\textbf{RRC}}                                                                                                                                                                                                                \\ \hline
  \multirow{6}{*}{AKA Bypass} & 50 & 0.3 & 2782.81 & 10 & 98 & 19 & 3 & 3 \\ \cline{2-9}
   & 100 & 0.2 & TIMEOUT & 15 & 166 & 22 & N/A & N/A \\ \cline{2-9}
   & 250 & 0.8 & TIMEOUT & 38 & 519 & 22 & N/A & N/A \\ \cline{2-9}
   & 500 & 0.3 & TIMEOUT & 28 & 323 & 22 & N/A & N/A \\ \cline{2-9}
   & 1250 & 0.33 & TIMEOUT & 76 & 886 & 22 & N/A & N/A \\ \cline{2-9}
   & 2500 & 0.6 & TIMEOUT & 118 & 1447 & 22 & N/A & N/A \\ \hline
  \multirow{6}{*}{IMSI Cracking (4G)} & 50 & 0.56 & 216.51 & 5 & 92 & 23 & 2 & 2 \\ \cline{2-9}
   & 100 & 0.33 & 661.59 & 17 & 245 & 28 & 2 & 2 \\ \cline{2-9}
   & 250 & 1 & 1428.2 & 4 & 82 & 24 & 2 & 2 \\ \cline{2-9}
   & 500 & 0.67 & TIMEOUT & 3 & 80 & 32 & N/A & N/A \\ \cline{2-9}
   & 1250 & 0.33 & TIMEOUT & 7 & 156 & 33 & N/A & N/A \\ \cline{2-9}
   & 2500 & 0.67 & TIMEOUT & 4 & 100 & 33 & N/A & N/A \\ \hline
  \multirow{7}{*}{Measurement Report} & 20 & 0.71 & TIMEOUT* & 14 & 202 & 23 & 3 & 3 \\ \cline{2-9}
   & 50 & 0.38 & TIMEOUT & 13 & 182 & 21 & N/A & N/A \\ \cline{2-9}
   & 100 & 0.3 & TIMEOUT & 6 & 89 & 23 & N/A & N/A \\ \cline{2-9}
   & 250 & 0.33 & TIMEOUT & 43 & 537 & 27 & N/A & N/A \\ \cline{2-9}
   & 500 & 0.25 & TIMEOUT & 53 & 712 & 27 & N/A & N/A \\ \cline{2-9}
   & 1250 & 0.33 & TIMEOUT & 122 & 1646 & 27 & N/A & N/A \\ \cline{2-9}
   & 2500 & 0.22 & TIMEOUT & 161 & 2184 & 27 & N/A & N/A \\ \hline
  \multirow{6}{*}{Paging with IMSI} & 50 & 0.25 & 57.31 & 8 & 135 & 23 & 1 & 1 \\ \cline{2-9}
   & 100 & 0.25 & 51.66 & 3 & 65 & 24 & 1 & 1 \\ \cline{2-9}
   & 250 & 0.15 & 146.74 & 29 & 487 & 24 & 1 & 1 \\ \cline{2-9}
   & 500 & 0.12 & 517.27 & 2 & 46 & 24 & 1 & 1 \\ \cline{2-9}
   & 1250 & 0.25 & TIMEOUT & 3 & 73 & 27 & N/A & N/A \\ \cline{2-9}
   & 2500 & 0.18 & TIMEOUT & 111 & 1835 & 27 & N/A & N/A \\ \hline
  \multirow{6}{*}{RLF Report} & 50 & 0.25 & 1538.16 & 15 & 188 & 22 & 4 & 3 \\ \cline{2-9}
   & 100 & 0.18 & TIMEOUT & 19 & 229 & 22 & N/A & N/A \\ \cline{2-9}
   & 250 & 0.38 & TIMEOUT & 31 & 429 & 22 & N/A & N/A \\ \cline{2-9}
   & 500 & 0.25 & TIMEOUT & 50 & 744 & 22 & N/A & N/A \\ \cline{2-9}
   & 1250 & 0.14 & TIMEOUT & 97 & 1416 & 22 & N/A & N/A \\ \cline{2-9}
   & 2500 & 0.08 & TIMEOUT & 117 & 1633 & 22 & N/A & N/A \\ \hline
  \end{tabular}
\end{minipage}}
\caption{Training time in seconds and size of the synthesized DFA and PLTL signatures. (* = PLTL synthesizer generated at least one signature but less than five before timing out.)}
\label{app:training_time_and_size_all_dfa_pltl}
\end{table*}
\newpage

\subsection{Evaluation Scores For All Synthesized Monitors}
Table \ref{app:evaluation_all_generated_monitors} provides the precision,
recall, and F1 score for all the generated signatures. Note that since Mealy Machine
builds a signature for all attacks, the Measurement Report attack was not evaluated
when trained on $20$ traces since this was only a requirement for \pltl for that
specific attack.

\clearpage
\begin{table*}[]
	\centering
  \resizebox{.75\columnwidth}{!}{
  \captionsetup{justification=centering}
		\begin{tabular}{ccccccccccc}
			\hline
			\multicolumn{1}{|c|}{\textbf{Attack Experiment}}                             & \multicolumn{1}{c|}{\textbf{Size}} & \multicolumn{1}{c|}{\textbf{DFA Precision}} & \multicolumn{1}{c|}{\textbf{DFA Recall}} & \multicolumn{1}{c|}{\textbf{DFA F1}} & \multicolumn{1}{c|}{\textbf{PLTL Precision}} & \multicolumn{1}{c|}{\textbf{PLTL Recall}} & \multicolumn{1}{c|}{\textbf{PLTL F1}} & \multicolumn{1}{c|}{\textbf{MM Precision}} & \multicolumn{1}{c|}{\textbf{MM Recall}} & \multicolumn{1}{c|}{\textbf{MM F1}} \\ \hline
			\multicolumn{11}{|c|}{\textbf{NAS}}                                                                                                                                                                                                                                                                                                                                                                                                                                                                             \\ \hline
			\multicolumn{1}{|c|}{\multirow{6}{*}{Attach Reject}}              & \multicolumn{1}{c|}{50}            & \multicolumn{1}{c|}{0.35}                    & \multicolumn{1}{c|}{0.799}                & \multicolumn{1}{c|}{0.487}            & \multicolumn{1}{c|}{1}                      & \multicolumn{1}{c|}{1}                   & \multicolumn{1}{c|}{1}               & \multicolumn{1}{c|}{1}                      & \multicolumn{1}{c|}{0.979}               & \multicolumn{1}{c|}{0.989}           \\ \cline{2-11}
			\multicolumn{1}{|c|}{}                                            & \multicolumn{1}{c|}{100}           & \multicolumn{1}{c|}{1}                       & \multicolumn{1}{c|}{1}                    & \multicolumn{1}{c|}{1}                & \multicolumn{1}{c|}{1}                        & \multicolumn{1}{c|}{1}                     & \multicolumn{1}{c|}{1}                 & \multicolumn{1}{c|}{1}                      & \multicolumn{1}{c|}{1}                   & \multicolumn{1}{c|}{1}               \\ \cline{2-11}
			\multicolumn{1}{|c|}{}                                            & \multicolumn{1}{c|}{250}           & \multicolumn{1}{c|}{0.874}                   & \multicolumn{1}{c|}{0.931}                & \multicolumn{1}{c|}{0.902}            & \multicolumn{1}{c|}{1}                        & \multicolumn{1}{c|}{1}                     & \multicolumn{1}{c|}{1}                 & \multicolumn{1}{c|}{1}                      & \multicolumn{1}{c|}{0.988}               & \multicolumn{1}{c|}{0.994}           \\ \cline{2-11}
			\multicolumn{1}{|c|}{}                                            & \multicolumn{1}{c|}{500}           & \multicolumn{1}{c|}{0.855}                   & \multicolumn{1}{c|}{0.808}                & \multicolumn{1}{c|}{0.831}            & \multicolumn{1}{c|}{N/A}                      & \multicolumn{1}{c|}{N/A}                   & \multicolumn{1}{c|}{N/A}               & \multicolumn{1}{c|}{1}                      & \multicolumn{1}{c|}{1}                   & \multicolumn{1}{c|}{1}               \\ \cline{2-11}
			\multicolumn{1}{|c|}{}                                            & \multicolumn{1}{c|}{1250}          & \multicolumn{1}{c|}{0.697}                   & \multicolumn{1}{c|}{0.674}                & \multicolumn{1}{c|}{0.685}            & \multicolumn{1}{c|}{N/A}                        & \multicolumn{1}{c|}{N/A}                     & \multicolumn{1}{c|}{N/A}                 & \multicolumn{1}{c|}{1}                      & \multicolumn{1}{c|}{1}                   & \multicolumn{1}{c|}{1}               \\ \cline{2-11}
			\multicolumn{1}{|c|}{}                                            & \multicolumn{1}{c|}{2500}          & \multicolumn{1}{c|}{1}                       & \multicolumn{1}{c|}{1}                    & \multicolumn{1}{c|}{1}                & \multicolumn{1}{c|}{N/A}                        & \multicolumn{1}{c|}{N/A}                     & \multicolumn{1}{c|}{N/A}                 & \multicolumn{1}{c|}{1}                      & \multicolumn{1}{c|}{0.767}               & \multicolumn{1}{c|}{0.868}           \\ \hline
			\multicolumn{1}{|c|}{\multirow{6}{*}{Authentication Failure}}     & \multicolumn{1}{c|}{50}            & \multicolumn{1}{c|}{0.983}                   & \multicolumn{1}{c|}{0.77}                 & \multicolumn{1}{c|}{0.864}            & \multicolumn{1}{c|}{1}                        & \multicolumn{1}{c|}{1}                     & \multicolumn{1}{c|}{1}                 & \multicolumn{1}{c|}{1}                      & \multicolumn{1}{c|}{1}                   & \multicolumn{1}{c|}{1}               \\ \cline{2-11}
			\multicolumn{1}{|c|}{}                                            & \multicolumn{1}{c|}{100}           & \multicolumn{1}{c|}{0.943}                   & \multicolumn{1}{c|}{0.891}                & \multicolumn{1}{c|}{0.916}            & \multicolumn{1}{c|}{1}                        & \multicolumn{1}{c|}{1}                     & \multicolumn{1}{c|}{1}                 & \multicolumn{1}{c|}{1}                      & \multicolumn{1}{c|}{0.996}               & \multicolumn{1}{c|}{0.998}           \\ \cline{2-11}
			\multicolumn{1}{|c|}{}                                            & \multicolumn{1}{c|}{250}           & \multicolumn{1}{c|}{0.751}                   & \multicolumn{1}{c|}{0.962}                & \multicolumn{1}{c|}{0.844}            & \multicolumn{1}{c|}{1}                        & \multicolumn{1}{c|}{1}                     & \multicolumn{1}{c|}{1}                 & \multicolumn{1}{c|}{1}                      & \multicolumn{1}{c|}{1}                   & \multicolumn{1}{c|}{1}               \\ \cline{2-11}
			\multicolumn{1}{|c|}{}                                            & \multicolumn{1}{c|}{500}           & \multicolumn{1}{c|}{0.72}                    & \multicolumn{1}{c|}{0.824}                & \multicolumn{1}{c|}{0.768}            & \multicolumn{1}{c|}{N/A}                      & \multicolumn{1}{c|}{N/A}                   & \multicolumn{1}{c|}{N/A}               & \multicolumn{1}{c|}{1}                      & \multicolumn{1}{c|}{1}                   & \multicolumn{1}{c|}{1}               \\ \cline{2-11}
			\multicolumn{1}{|c|}{}                                            & \multicolumn{1}{c|}{1250}          & \multicolumn{1}{c|}{0.671}                   & \multicolumn{1}{c|}{0.997}                & \multicolumn{1}{c|}{0.802}            & \multicolumn{1}{c|}{N/A}                        & \multicolumn{1}{c|}{N/A}                     & \multicolumn{1}{c|}{N/A}                 & \multicolumn{1}{c|}{1}                      & \multicolumn{1}{c|}{1}                   & \multicolumn{1}{c|}{1}               \\ \cline{2-11}
			\multicolumn{1}{|c|}{}                                            & \multicolumn{1}{c|}{2500}          & \multicolumn{1}{c|}{0.914}                   & \multicolumn{1}{c|}{1}                    & \multicolumn{1}{c|}{0.955}            & \multicolumn{1}{c|}{N/A}                        & \multicolumn{1}{c|}{N/A}                     & \multicolumn{1}{c|}{N/A}                 & \multicolumn{1}{c|}{1}                      & \multicolumn{1}{c|}{1}                   & \multicolumn{1}{c|}{1}               \\ \hline
			\multicolumn{1}{|c|}{\multirow{6}{*}{EMM Information}}            & \multicolumn{1}{c|}{50}            & \multicolumn{1}{c|}{0.242}                   & \multicolumn{1}{c|}{0.949}                & \multicolumn{1}{c|}{0.386}            & \multicolumn{1}{c|}{1}                      & \multicolumn{1}{c|}{1}                   & \multicolumn{1}{c|}{1}               & \multicolumn{1}{c|}{1}                      & \multicolumn{1}{c|}{1}                   & \multicolumn{1}{c|}{1}               \\ \cline{2-11}
			\multicolumn{1}{|c|}{}                                            & \multicolumn{1}{c|}{100}           & \multicolumn{1}{c|}{0.624}                   & \multicolumn{1}{c|}{0.85}                 & \multicolumn{1}{c|}{0.72}             & \multicolumn{1}{c|}{1}                        & \multicolumn{1}{c|}{1}                     & \multicolumn{1}{c|}{1}                 & \multicolumn{1}{c|}{1}                      & \multicolumn{1}{c|}{1}                   & \multicolumn{1}{c|}{1}               \\ \cline{2-11}
			\multicolumn{1}{|c|}{}                                            & \multicolumn{1}{c|}{250}           & \multicolumn{1}{c|}{0.278}                   & \multicolumn{1}{c|}{1}                    & \multicolumn{1}{c|}{0.435}            & \multicolumn{1}{c|}{1}                        & \multicolumn{1}{c|}{1}                     & \multicolumn{1}{c|}{1}                 & \multicolumn{1}{c|}{1}                      & \multicolumn{1}{c|}{1}                   & \multicolumn{1}{c|}{1}               \\ \cline{2-11}
			\multicolumn{1}{|c|}{}                                            & \multicolumn{1}{c|}{500}           & \multicolumn{1}{c|}{0.353}                   & \multicolumn{1}{c|}{0.989}                & \multicolumn{1}{c|}{0.52}             & \multicolumn{1}{c|}{N/A}                      & \multicolumn{1}{c|}{N/A}                   & \multicolumn{1}{c|}{N/A}               & \multicolumn{1}{c|}{1}                      & \multicolumn{1}{c|}{1}                   & \multicolumn{1}{c|}{1}               \\ \cline{2-11}
			\multicolumn{1}{|c|}{}                                            & \multicolumn{1}{c|}{1250}          & \multicolumn{1}{c|}{1}                       & \multicolumn{1}{c|}{1}                    & \multicolumn{1}{c|}{1}                & \multicolumn{1}{c|}{N/A}                        & \multicolumn{1}{c|}{N/A}                     & \multicolumn{1}{c|}{N/A}                 & \multicolumn{1}{c|}{1}                      & \multicolumn{1}{c|}{1}                   & \multicolumn{1}{c|}{1}               \\ \cline{2-11}
			\multicolumn{1}{|c|}{}                                            & \multicolumn{1}{c|}{2500}          & \multicolumn{1}{c|}{0.81}                    & \multicolumn{1}{c|}{0.998}                & \multicolumn{1}{c|}{0.894}            & \multicolumn{1}{c|}{N/A}                        & \multicolumn{1}{c|}{N/A}                     & \multicolumn{1}{c|}{N/A}                 & \multicolumn{1}{c|}{1}                      & \multicolumn{1}{c|}{1}                   & \multicolumn{1}{c|}{1}               \\ \hline
			\multicolumn{1}{|c|}{\multirow{6}{*}{IMEI Catching}}              & \multicolumn{1}{c|}{50}            & \multicolumn{1}{c|}{0.821}                   & \multicolumn{1}{c|}{0.688}                & \multicolumn{1}{c|}{0.749}            & \multicolumn{1}{c|}{1}                      & \multicolumn{1}{c|}{1}                   & \multicolumn{1}{c|}{1}               & \multicolumn{1}{c|}{1}                      & \multicolumn{1}{c|}{1}                   & \multicolumn{1}{c|}{1}               \\ \cline{2-11}
			\multicolumn{1}{|c|}{}                                            & \multicolumn{1}{c|}{100}           & \multicolumn{1}{c|}{0.965}                   & \multicolumn{1}{c|}{0.659}                & \multicolumn{1}{c|}{0.783}            & \multicolumn{1}{c|}{1}                        & \multicolumn{1}{c|}{1}                     & \multicolumn{1}{c|}{1}                 & \multicolumn{1}{c|}{1}                      & \multicolumn{1}{c|}{1}                   & \multicolumn{1}{c|}{1}               \\ \cline{2-11}
			\multicolumn{1}{|c|}{}                                            & \multicolumn{1}{c|}{250}           & \multicolumn{1}{c|}{0.999}                   & \multicolumn{1}{c|}{0.972}                & \multicolumn{1}{c|}{0.985}            & \multicolumn{1}{c|}{1}                        & \multicolumn{1}{c|}{1}                     & \multicolumn{1}{c|}{1}                 & \multicolumn{1}{c|}{1}                      & \multicolumn{1}{c|}{1}                   & \multicolumn{1}{c|}{1}               \\ \cline{2-11}
			\multicolumn{1}{|c|}{}                                            & \multicolumn{1}{c|}{500}           & \multicolumn{1}{c|}{0.999}                   & \multicolumn{1}{c|}{0.972}                & \multicolumn{1}{c|}{0.985}            & \multicolumn{1}{c|}{N/A}                      & \multicolumn{1}{c|}{N/A}                   & \multicolumn{1}{c|}{N/A}               & \multicolumn{1}{c|}{1}                      & \multicolumn{1}{c|}{1}                   & \multicolumn{1}{c|}{1}               \\ \cline{2-11}
			\multicolumn{1}{|c|}{}                                            & \multicolumn{1}{c|}{1250}          & \multicolumn{1}{c|}{0.632}                   & \multicolumn{1}{c|}{0.635}                & \multicolumn{1}{c|}{0.633}            & \multicolumn{1}{c|}{N/A}                        & \multicolumn{1}{c|}{N/A}                     & \multicolumn{1}{c|}{N/A}                 & \multicolumn{1}{c|}{1}                      & \multicolumn{1}{c|}{1}                   & \multicolumn{1}{c|}{1}               \\ \cline{2-11}
			\multicolumn{1}{|c|}{}                                            & \multicolumn{1}{c|}{2500}          & \multicolumn{1}{c|}{0.5}                     & \multicolumn{1}{c|}{0.7}                  & \multicolumn{1}{c|}{0.583}            & \multicolumn{1}{c|}{N/A}                        & \multicolumn{1}{c|}{N/A}                     & \multicolumn{1}{c|}{N/A}                 & \multicolumn{1}{c|}{1}                      & \multicolumn{1}{c|}{1}                   & \multicolumn{1}{c|}{1}               \\ \hline
			\multicolumn{1}{|c|}{\multirow{6}{*}{IMSI Catching}}              & \multicolumn{1}{c|}{50}            & \multicolumn{1}{c|}{0.538}                   & \multicolumn{1}{c|}{0.876}                & \multicolumn{1}{c|}{0.667}            & \multicolumn{1}{c|}{1}                      & \multicolumn{1}{c|}{1}                   & \multicolumn{1}{c|}{1}               & \multicolumn{1}{c|}{1}                      & \multicolumn{1}{c|}{1}                   & \multicolumn{1}{c|}{1}               \\ \cline{2-11}
			\multicolumn{1}{|c|}{}                                            & \multicolumn{1}{c|}{100}           & \multicolumn{1}{c|}{0.653}                   & \multicolumn{1}{c|}{0.985}                & \multicolumn{1}{c|}{0.785}            & \multicolumn{1}{c|}{1}                        & \multicolumn{1}{c|}{1}                     & \multicolumn{1}{c|}{1}                 & \multicolumn{1}{c|}{1}                      & \multicolumn{1}{c|}{1}                   & \multicolumn{1}{c|}{1}               \\ \cline{2-11}
			\multicolumn{1}{|c|}{}                                            & \multicolumn{1}{c|}{250}           & \multicolumn{1}{c|}{0.942}                   & \multicolumn{1}{c|}{0.943}                & \multicolumn{1}{c|}{0.942}            & \multicolumn{1}{c|}{1}                        & \multicolumn{1}{c|}{1}                     & \multicolumn{1}{c|}{1}                 & \multicolumn{1}{c|}{1}                      & \multicolumn{1}{c|}{1}                   & \multicolumn{1}{c|}{1}               \\ \cline{2-11}
			\multicolumn{1}{|c|}{}                                            & \multicolumn{1}{c|}{500}           & \multicolumn{1}{c|}{0.981}                   & \multicolumn{1}{c|}{0.966}                & \multicolumn{1}{c|}{0.973}            & \multicolumn{1}{c|}{N/A}                      & \multicolumn{1}{c|}{N/A}                   & \multicolumn{1}{c|}{N/A}               & \multicolumn{1}{c|}{1}                      & \multicolumn{1}{c|}{0.999}               & \multicolumn{1}{c|}{0.999}           \\ \cline{2-11}
			\multicolumn{1}{|c|}{}                                            & \multicolumn{1}{c|}{1250}          & \multicolumn{1}{c|}{0.977}                   & \multicolumn{1}{c|}{1}                    & \multicolumn{1}{c|}{0.988}            & \multicolumn{1}{c|}{N/A}                        & \multicolumn{1}{c|}{N/A}                     & \multicolumn{1}{c|}{N/A}                 & \multicolumn{1}{c|}{1}                      & \multicolumn{1}{c|}{1}                   & \multicolumn{1}{c|}{1}               \\ \cline{2-11}
			\multicolumn{1}{|c|}{}                                            & \multicolumn{1}{c|}{2500}          & \multicolumn{1}{c|}{1}                       & \multicolumn{1}{c|}{1}                    & \multicolumn{1}{c|}{1}                & \multicolumn{1}{c|}{N/A}                        & \multicolumn{1}{c|}{N/A}                     & \multicolumn{1}{c|}{N/A}                 & \multicolumn{1}{c|}{1}                      & \multicolumn{1}{c|}{0.997}               & \multicolumn{1}{c|}{0.998}           \\ \hline
			\multicolumn{1}{|c|}{\multirow{6}{*}{Malformed Identity Request}} & \multicolumn{1}{c|}{50}            & \multicolumn{1}{c|}{0.739}                   & \multicolumn{1}{c|}{0.502}                & \multicolumn{1}{c|}{0.598}            & \multicolumn{1}{c|}{1}                      & \multicolumn{1}{c|}{1}                   & \multicolumn{1}{c|}{1}               & \multicolumn{1}{c|}{1}                      & \multicolumn{1}{c|}{1}                   & \multicolumn{1}{c|}{1}               \\ \cline{2-11}
			\multicolumn{1}{|c|}{}                                            & \multicolumn{1}{c|}{100}           & \multicolumn{1}{c|}{0.805}                   & \multicolumn{1}{c|}{0.504}                & \multicolumn{1}{c|}{0.62}             & \multicolumn{1}{c|}{1}                        & \multicolumn{1}{c|}{1}                     & \multicolumn{1}{c|}{1}                 & \multicolumn{1}{c|}{1}                      & \multicolumn{1}{c|}{1}                   & \multicolumn{1}{c|}{1}               \\ \cline{2-11}
			\multicolumn{1}{|c|}{}                                            & \multicolumn{1}{c|}{250}           & \multicolumn{1}{c|}{0.746}                   & \multicolumn{1}{c|}{0.702}                & \multicolumn{1}{c|}{0.723}            & \multicolumn{1}{c|}{1}                        & \multicolumn{1}{c|}{1}                     & \multicolumn{1}{c|}{1}                 & \multicolumn{1}{c|}{1}                      & \multicolumn{1}{c|}{1}                   & \multicolumn{1}{c|}{1}               \\ \cline{2-11}
			\multicolumn{1}{|c|}{}                                            & \multicolumn{1}{c|}{500}           & \multicolumn{1}{c|}{0.97}                    & \multicolumn{1}{c|}{0.662}                & \multicolumn{1}{c|}{0.787}            & \multicolumn{1}{c|}{N/A}                      & \multicolumn{1}{c|}{N/A}                   & \multicolumn{1}{c|}{N/A}               & \multicolumn{1}{c|}{1}                      & \multicolumn{1}{c|}{1}                   & \multicolumn{1}{c|}{1}               \\ \cline{2-11}
			\multicolumn{1}{|c|}{}                                            & \multicolumn{1}{c|}{1250}          & \multicolumn{1}{c|}{0.978}                   & \multicolumn{1}{c|}{0.5}                  & \multicolumn{1}{c|}{0.662}            & \multicolumn{1}{c|}{N/A}                        & \multicolumn{1}{c|}{N/A}                     & \multicolumn{1}{c|}{N/A}                 & \multicolumn{1}{c|}{1}                      & \multicolumn{1}{c|}{1}                   & \multicolumn{1}{c|}{1}               \\ \cline{2-11}
			\multicolumn{1}{|c|}{}                                            & \multicolumn{1}{c|}{2500}          & \multicolumn{1}{c|}{0.417}                   & \multicolumn{1}{c|}{0.466}                & \multicolumn{1}{c|}{0.44}             & \multicolumn{1}{c|}{N/A}                        & \multicolumn{1}{c|}{N/A}                     & \multicolumn{1}{c|}{N/A}                 & \multicolumn{1}{c|}{1}                      & \multicolumn{1}{c|}{1}                   & \multicolumn{1}{c|}{1}               \\ \hline
			\multicolumn{1}{|c|}{\multirow{6}{*}{Null Encryption}}            & \multicolumn{1}{c|}{50}            & \multicolumn{1}{c|}{0.524}                   & \multicolumn{1}{c|}{0.868}                & \multicolumn{1}{c|}{0.653}            & \multicolumn{1}{c|}{1}                      & \multicolumn{1}{c|}{1}                   & \multicolumn{1}{c|}{1}               & \multicolumn{1}{c|}{1}                      & \multicolumn{1}{c|}{1}                   & \multicolumn{1}{c|}{1}               \\ \cline{2-11}
			\multicolumn{1}{|c|}{}                                            & \multicolumn{1}{c|}{100}           & \multicolumn{1}{c|}{0.437}                   & \multicolumn{1}{c|}{0.944}                & \multicolumn{1}{c|}{0.597}            & \multicolumn{1}{c|}{1}                        & \multicolumn{1}{c|}{1}                     & \multicolumn{1}{c|}{1}                 & \multicolumn{1}{c|}{1}                      & \multicolumn{1}{c|}{0.967}               & \multicolumn{1}{c|}{0.983}           \\ \cline{2-11}
			\multicolumn{1}{|c|}{}                                            & \multicolumn{1}{c|}{250}           & \multicolumn{1}{c|}{0.822}                   & \multicolumn{1}{c|}{0.965}                & \multicolumn{1}{c|}{0.888}            & \multicolumn{1}{c|}{1}                        & \multicolumn{1}{c|}{1}                     & \multicolumn{1}{c|}{1}                 & \multicolumn{1}{c|}{1}                      & \multicolumn{1}{c|}{1}                   & \multicolumn{1}{c|}{1}               \\ \cline{2-11}
			\multicolumn{1}{|c|}{}                                            & \multicolumn{1}{c|}{500}           & \multicolumn{1}{c|}{0.528}                   & \multicolumn{1}{c|}{0.967}                & \multicolumn{1}{c|}{0.683}            & \multicolumn{1}{c|}{N/A}                      & \multicolumn{1}{c|}{N/A}                   & \multicolumn{1}{c|}{N/A}               & \multicolumn{1}{c|}{1}                      & \multicolumn{1}{c|}{1}                   & \multicolumn{1}{c|}{1}               \\ \cline{2-11}
			\multicolumn{1}{|c|}{}                                            & \multicolumn{1}{c|}{1250}          & \multicolumn{1}{c|}{0.467}                   & \multicolumn{1}{c|}{0.89}                 & \multicolumn{1}{c|}{0.613}            & \multicolumn{1}{c|}{N/A}                        & \multicolumn{1}{c|}{N/A}                     & \multicolumn{1}{c|}{N/A}                 & \multicolumn{1}{c|}{1}                      & \multicolumn{1}{c|}{1}                   & \multicolumn{1}{c|}{1}               \\ \cline{2-11}
			\multicolumn{1}{|c|}{}                                            & \multicolumn{1}{c|}{2500}          & \multicolumn{1}{c|}{0.709}                   & \multicolumn{1}{c|}{0.989}                & \multicolumn{1}{c|}{0.826}            & \multicolumn{1}{c|}{N/A}                        & \multicolumn{1}{c|}{N/A}                     & \multicolumn{1}{c|}{N/A}                 & \multicolumn{1}{c|}{1}                      & \multicolumn{1}{c|}{1}                   & \multicolumn{1}{c|}{1}               \\ \hline
			\multicolumn{1}{|c|}{\multirow{6}{*}{Numb Attack}}                & \multicolumn{1}{c|}{50}            & \multicolumn{1}{c|}{0.817}                   & \multicolumn{1}{c|}{1}                    & \multicolumn{1}{c|}{0.899}            & \multicolumn{1}{c|}{1}                        & \multicolumn{1}{c|}{1}                     & \multicolumn{1}{c|}{1}                 & \multicolumn{1}{c|}{0.997}                  & \multicolumn{1}{c|}{1}                   & \multicolumn{1}{c|}{0.999}           \\ \cline{2-11}
			\multicolumn{1}{|c|}{}                                            & \multicolumn{1}{c|}{100}           & \multicolumn{1}{c|}{0.98}                    & \multicolumn{1}{c|}{1}                    & \multicolumn{1}{c|}{0.99}                 & \multicolumn{1}{c|}{1}                        & \multicolumn{1}{c|}{1}                     & \multicolumn{1}{c|}{1}                 & \multicolumn{1}{c|}{0.968}                  & \multicolumn{1}{c|}{0.981}               & \multicolumn{1}{c|}{0.975}           \\ \cline{2-11}
			\multicolumn{1}{|c|}{}                                            & \multicolumn{1}{c|}{250}           & \multicolumn{1}{c|}{1}                       & \multicolumn{1}{c|}{1}                    & \multicolumn{1}{c|}{1}                 & \multicolumn{1}{c|}{1}                        & \multicolumn{1}{c|}{1}                     & \multicolumn{1}{c|}{1}                 & \multicolumn{1}{c|}{1}                      & \multicolumn{1}{c|}{1}                   & \multicolumn{1}{c|}{1}               \\ \cline{2-11}
			\multicolumn{1}{|c|}{}                                            & \multicolumn{1}{c|}{500}           & \multicolumn{1}{c|}{1}                       & \multicolumn{1}{c|}{1}                    & \multicolumn{1}{c|}{1}                 & \multicolumn{1}{c|}{1}                        & \multicolumn{1}{c|}{1}                     & \multicolumn{1}{c|}{1}                 & \multicolumn{1}{c|}{0.98}                   & \multicolumn{1}{c|}{0.987}               & \multicolumn{1}{c|}{0.984}           \\ \cline{2-11}
			\multicolumn{1}{|c|}{}                                            & \multicolumn{1}{c|}{1250}          & \multicolumn{1}{c|}{0.989}                   & \multicolumn{1}{c|}{1}                    & \multicolumn{1}{c|}{0.994}                 & \multicolumn{1}{c|}{N/A}                      & \multicolumn{1}{c|}{N/A}                   & \multicolumn{1}{c|}{N/A}               & \multicolumn{1}{c|}{1}                      & \multicolumn{1}{c|}{1}                   & \multicolumn{1}{c|}{1}               \\ \cline{2-11}
			\multicolumn{1}{|c|}{}                                            & \multicolumn{1}{c|}{2500}          & \multicolumn{1}{c|}{1}                       & \multicolumn{1}{c|}{1}                    & \multicolumn{1}{c|}{1}                 & \multicolumn{1}{c|}{N/A}                      & \multicolumn{1}{c|}{N/A}                   & \multicolumn{1}{c|}{N/A}               & \multicolumn{1}{c|}{1}                      & \multicolumn{1}{c|}{1}                   & \multicolumn{1}{c|}{1}               \\ \hline
			\multicolumn{1}{|c|}{\multirow{6}{*}{Service Reject}}             & \multicolumn{1}{c|}{50}            & \multicolumn{1}{c|}{0.704}                   & \multicolumn{1}{c|}{0.721}                & \multicolumn{1}{c|}{0.712}            & \multicolumn{1}{c|}{N/A}                      & \multicolumn{1}{c|}{N/A}                   & \multicolumn{1}{c|}{N/A}               & \multicolumn{1}{c|}{1}                      & \multicolumn{1}{c|}{0.944}               & \multicolumn{1}{c|}{0.971}           \\ \cline{2-11}
			\multicolumn{1}{|c|}{}                                            & \multicolumn{1}{c|}{100}           & \multicolumn{1}{c|}{1}                       & \multicolumn{1}{c|}{1}                    & \multicolumn{1}{c|}{1}                & \multicolumn{1}{c|}{1}                        & \multicolumn{1}{c|}{1}                     & \multicolumn{1}{c|}{1}                 & \multicolumn{1}{c|}{1}                      & \multicolumn{1}{c|}{1}                   & \multicolumn{1}{c|}{1}               \\ \cline{2-11}
			\multicolumn{1}{|c|}{}                                            & \multicolumn{1}{c|}{250}           & \multicolumn{1}{c|}{0.976}                   & \multicolumn{1}{c|}{0.84}                 & \multicolumn{1}{c|}{0.903}            & \multicolumn{1}{c|}{1}                        & \multicolumn{1}{c|}{1}                     & \multicolumn{1}{c|}{1}                 & \multicolumn{1}{c|}{1}                      & \multicolumn{1}{c|}{1}                   & \multicolumn{1}{c|}{1}               \\ \cline{2-11}
			\multicolumn{1}{|c|}{}                                            & \multicolumn{1}{c|}{500}           & \multicolumn{1}{c|}{0.765}                   & \multicolumn{1}{c|}{0.857}                & \multicolumn{1}{c|}{0.808}            & \multicolumn{1}{c|}{N/A}                      & \multicolumn{1}{c|}{N/A}                   & \multicolumn{1}{c|}{N/A}               & \multicolumn{1}{c|}{1}                      & \multicolumn{1}{c|}{0.975}               & \multicolumn{1}{c|}{0.987}           \\ \cline{2-11}
			\multicolumn{1}{|c|}{}                                            & \multicolumn{1}{c|}{1250}          & \multicolumn{1}{c|}{1}                       & \multicolumn{1}{c|}{1}                    & \multicolumn{1}{c|}{1}                & \multicolumn{1}{c|}{N/A}                        & \multicolumn{1}{c|}{N/A}                     & \multicolumn{1}{c|}{N/A}                 & \multicolumn{1}{c|}{1}                      & \multicolumn{1}{c|}{1}                   & \multicolumn{1}{c|}{1}               \\ \cline{2-11}
			\multicolumn{1}{|c|}{}                                            & \multicolumn{1}{c|}{2500}          & \multicolumn{1}{c|}{1}                       & \multicolumn{1}{c|}{1}                    & \multicolumn{1}{c|}{1}                & \multicolumn{1}{c|}{N/A}                        & \multicolumn{1}{c|}{N/A}                     & \multicolumn{1}{c|}{N/A}                 & \multicolumn{1}{c|}{1}                      & \multicolumn{1}{c|}{0.902}               & \multicolumn{1}{c|}{0.948}           \\ \hline
			\multicolumn{1}{|c|}{\multirow{6}{*}{TAU Reject}}                 & \multicolumn{1}{c|}{50}            & \multicolumn{1}{c|}{1}                       & \multicolumn{1}{c|}{0.877}                & \multicolumn{1}{c|}{0.934}            & \multicolumn{1}{c|}{1}                        & \multicolumn{1}{c|}{1}                     & \multicolumn{1}{c|}{1}                 & \multicolumn{1}{c|}{1}                      & \multicolumn{1}{c|}{1}                   & \multicolumn{1}{c|}{1}               \\ \cline{2-11}
			\multicolumn{1}{|c|}{}                                            & \multicolumn{1}{c|}{100}           & \multicolumn{1}{c|}{0.627}                   & \multicolumn{1}{c|}{0.951}                & \multicolumn{1}{c|}{0.756}            & \multicolumn{1}{c|}{1}                        & \multicolumn{1}{c|}{1}                     & \multicolumn{1}{c|}{1}                 & \multicolumn{1}{c|}{1}                      & \multicolumn{1}{c|}{1}                   & \multicolumn{1}{c|}{1}               \\ \cline{2-11}
			\multicolumn{1}{|c|}{}                                            & \multicolumn{1}{c|}{250}           & \multicolumn{1}{c|}{1}                       & \multicolumn{1}{c|}{0.902}                & \multicolumn{1}{c|}{0.948}            & \multicolumn{1}{c|}{1}                        & \multicolumn{1}{c|}{1}                     & \multicolumn{1}{c|}{1}                 & \multicolumn{1}{c|}{1}                      & \multicolumn{1}{c|}{1}                   & \multicolumn{1}{c|}{1}               \\ \cline{2-11}
			\multicolumn{1}{|c|}{}                                            & \multicolumn{1}{c|}{500}           & \multicolumn{1}{c|}{1}                       & \multicolumn{1}{c|}{1}                    & \multicolumn{1}{c|}{1}                & \multicolumn{1}{c|}{N/A}                      & \multicolumn{1}{c|}{N/A}                   & \multicolumn{1}{c|}{N/A}               & \multicolumn{1}{c|}{1}                      & \multicolumn{1}{c|}{1}                   & \multicolumn{1}{c|}{1}               \\ \cline{2-11}
			\multicolumn{1}{|c|}{}                                            & \multicolumn{1}{c|}{1250}          & \multicolumn{1}{c|}{0.98}                    & \multicolumn{1}{c|}{0.67}                 & \multicolumn{1}{c|}{0.796}            & \multicolumn{1}{c|}{N/A}                        & \multicolumn{1}{c|}{N/A}                     & \multicolumn{1}{c|}{N/A}                 & \multicolumn{1}{c|}{1}                      & \multicolumn{1}{c|}{1}                   & \multicolumn{1}{c|}{1}               \\ \cline{2-11}
			\multicolumn{1}{|c|}{}                                            & \multicolumn{1}{c|}{2500}          & \multicolumn{1}{c|}{1}                       & \multicolumn{1}{c|}{0.902}                & \multicolumn{1}{c|}{0.948}            & \multicolumn{1}{c|}{N/A}                        & \multicolumn{1}{c|}{N/A}                     & \multicolumn{1}{c|}{N/A}                 & \multicolumn{1}{c|}{1}                      & \multicolumn{1}{c|}{1}                   & \multicolumn{1}{c|}{1}               \\ \cline{1-11}
			&                                    &                                              &                                           &                                       &                                               &                                            &                                        &                                             &                                          &                                      \\ \hline
			\multicolumn{11}{|c|}{\textbf{RRC}}                                                                                                                                                                                                                                                                                                                                                                                                                                                                             \\ \hline
			\multicolumn{1}{|c|}{\multirow{6}{*}{AKA Bypass}}                 & \multicolumn{1}{c|}{50}            & \multicolumn{1}{c|}{0.984}                   & \multicolumn{1}{c|}{0.809}                & \multicolumn{1}{c|}{0.888}            & \multicolumn{1}{c|}{1}                      & \multicolumn{1}{c|}{1}                   & \multicolumn{1}{c|}{1}               & \multicolumn{1}{c|}{0.899}                  & \multicolumn{1}{c|}{0.93}                & \multicolumn{1}{c|}{0.914}           \\ \cline{2-11}
			\multicolumn{1}{|c|}{}                                            & \multicolumn{1}{c|}{100}           & \multicolumn{1}{c|}{0.781}                   & \multicolumn{1}{c|}{0.824}                & \multicolumn{1}{c|}{0.802}            & \multicolumn{1}{c|}{N/A}                      & \multicolumn{1}{c|}{N/A}                   & \multicolumn{1}{c|}{N/A}               & \multicolumn{1}{c|}{0.965}                  & \multicolumn{1}{c|}{0.975}               & \multicolumn{1}{c|}{0.97}            \\ \cline{2-11}
			\multicolumn{1}{|c|}{}                                            & \multicolumn{1}{c|}{250}           & \multicolumn{1}{c|}{0.817}                   & \multicolumn{1}{c|}{0.812}                & \multicolumn{1}{c|}{0.814}            & \multicolumn{1}{c|}{N/A}                      & \multicolumn{1}{c|}{N/A}                   & \multicolumn{1}{c|}{N/A}               & \multicolumn{1}{c|}{0.989}                  & \multicolumn{1}{c|}{0.996}               & \multicolumn{1}{c|}{0.993}           \\ \cline{2-11}
			\multicolumn{1}{|c|}{}                                            & \multicolumn{1}{c|}{500}           & \multicolumn{1}{c|}{1}                       & \multicolumn{1}{c|}{0.977}                & \multicolumn{1}{c|}{0.988}            & \multicolumn{1}{c|}{N/A}                      & \multicolumn{1}{c|}{N/A}                   & \multicolumn{1}{c|}{N/A}               & \multicolumn{1}{c|}{0.995}                  & \multicolumn{1}{c|}{0.997}               & \multicolumn{1}{c|}{0.996}           \\ \cline{2-11}
			\multicolumn{1}{|c|}{}                                            & \multicolumn{1}{c|}{1250}          & \multicolumn{1}{c|}{1}                       & \multicolumn{1}{c|}{0.908}                & \multicolumn{1}{c|}{0.952}            & \multicolumn{1}{c|}{N/A}                      & \multicolumn{1}{c|}{N/A}                   & \multicolumn{1}{c|}{N/A}               & \multicolumn{1}{c|}{0.993}                  & \multicolumn{1}{c|}{0.988}               & \multicolumn{1}{c|}{0.99}            \\ \cline{2-11}
			\multicolumn{1}{|c|}{}                                            & \multicolumn{1}{c|}{2500}          & \multicolumn{1}{c|}{1}                       & \multicolumn{1}{c|}{0.95}                 & \multicolumn{1}{c|}{0.974}            & \multicolumn{1}{c|}{N/A}                      & \multicolumn{1}{c|}{N/A}                   & \multicolumn{1}{c|}{N/A}               & \multicolumn{1}{c|}{1}                      & \multicolumn{1}{c|}{1}                   & \multicolumn{1}{c|}{1}               \\ \hline
			\multicolumn{1}{|c|}{\multirow{6}{*}{IMSI Cracking}}              & \multicolumn{1}{c|}{50}            & \multicolumn{1}{c|}{1}                       & \multicolumn{1}{c|}{1}                    & \multicolumn{1}{c|}{1}                & \multicolumn{1}{c|}{1}                        & \multicolumn{1}{c|}{1}                     & \multicolumn{1}{c|}{1}                 & \multicolumn{1}{c|}{0.92}                   & \multicolumn{1}{c|}{0.994}               & \multicolumn{1}{c|}{0.956}           \\ \cline{2-11}
			\multicolumn{1}{|c|}{}                                            & \multicolumn{1}{c|}{100}           & \multicolumn{1}{c|}{1}                       & \multicolumn{1}{c|}{1}                    & \multicolumn{1}{c|}{1}                & \multicolumn{1}{c|}{1}                        & \multicolumn{1}{c|}{1}                     & \multicolumn{1}{c|}{1}                 & \multicolumn{1}{c|}{0.736}                  & \multicolumn{1}{c|}{1}                   & \multicolumn{1}{c|}{0.848}           \\ \cline{2-11}
			\multicolumn{1}{|c|}{}                                            & \multicolumn{1}{c|}{250}           & \multicolumn{1}{c|}{1}                       & \multicolumn{1}{c|}{0.5}                  & \multicolumn{1}{c|}{0.667}            & \multicolumn{1}{c|}{1}                        & \multicolumn{1}{c|}{1}                     & \multicolumn{1}{c|}{1}                 & \multicolumn{1}{c|}{0.682}                  & \multicolumn{1}{c|}{1}                   & \multicolumn{1}{c|}{0.811}           \\ \cline{2-11}
			\multicolumn{1}{|c|}{}                                            & \multicolumn{1}{c|}{500}           & \multicolumn{1}{c|}{1}                       & \multicolumn{1}{c|}{1}                    & \multicolumn{1}{c|}{1}                & \multicolumn{1}{c|}{N/A}                      & \multicolumn{1}{c|}{N/A}                   & \multicolumn{1}{c|}{N/A}               & \multicolumn{1}{c|}{0.66}                   & \multicolumn{1}{c|}{0.998}               & \multicolumn{1}{c|}{0.795}           \\ \cline{2-11}
			\multicolumn{1}{|c|}{}                                            & \multicolumn{1}{c|}{1250}          & \multicolumn{1}{c|}{1}                       & \multicolumn{1}{c|}{1}                    & \multicolumn{1}{c|}{1}                & \multicolumn{1}{c|}{N/A}                      & \multicolumn{1}{c|}{N/A}                   & \multicolumn{1}{c|}{N/A}               & \multicolumn{1}{c|}{0.708}                  & \multicolumn{1}{c|}{1}                   & \multicolumn{1}{c|}{0.829}           \\ \cline{2-11}
			\multicolumn{1}{|c|}{}                                            & \multicolumn{1}{c|}{2500}          & \multicolumn{1}{c|}{1}                       & \multicolumn{1}{c|}{1}                    & \multicolumn{1}{c|}{1}                & \multicolumn{1}{c|}{N/A}                      & \multicolumn{1}{c|}{N/A}                   & \multicolumn{1}{c|}{N/A}               & \multicolumn{1}{c|}{0.671}                  & \multicolumn{1}{c|}{1}                   & \multicolumn{1}{c|}{0.803}           \\ \hline
			\multicolumn{1}{|c|}{\multirow{6}{*}{Measurement Report}}         & \multicolumn{1}{c|}{20}            & \multicolumn{1}{c|}{0.434}                   & \multicolumn{1}{c|}{0.456}                & \multicolumn{1}{c|}{0.445}            & \multicolumn{1}{c|}{1}                      & \multicolumn{1}{c|}{1}                   & \multicolumn{1}{c|}{1}               & \multicolumn{1}{c|}{N/A}                  & \multicolumn{1}{c|}{N/A}               & \multicolumn{1}{c|}{N/A}           \\ \cline{2-11}

      \multicolumn{1}{|c|}{}                                            & \multicolumn{1}{c|}{50}           & \multicolumn{1}{c|}{0.687}                   & \multicolumn{1}{c|}{0.565}                    & \multicolumn{1}{c|}{0.62}            & \multicolumn{1}{c|}{N/A}                      & \multicolumn{1}{c|}{N/A}                   & \multicolumn{1}{c|}{N/A}               & \multicolumn{1}{c|}{0.878}                  & \multicolumn{1}{c|}{0.864}               & \multicolumn{1}{c|}{0.871}           \\ \cline{2-11}

      \multicolumn{1}{|c|}{}                                            & \multicolumn{1}{c|}{100}           & \multicolumn{1}{c|}{0.998}                   & \multicolumn{1}{c|}{1}                    & \multicolumn{1}{c|}{0.792}            & \multicolumn{1}{c|}{N/A}                      & \multicolumn{1}{c|}{N/A}                   & \multicolumn{1}{c|}{N/A}               & \multicolumn{1}{c|}{0.948}                  & \multicolumn{1}{c|}{0.937}               & \multicolumn{1}{c|}{0.943}           \\ \cline{2-11}

      \multicolumn{1}{|c|}{}                                            & \multicolumn{1}{c|}{250}           & \multicolumn{1}{c|}{0.87}                    & \multicolumn{1}{c|}{0.689}                & \multicolumn{1}{c|}{0.769}            & \multicolumn{1}{c|}{N/A}                      & \multicolumn{1}{c|}{N/A}                   & \multicolumn{1}{c|}{N/A}               & \multicolumn{1}{c|}{0.984}                  & \multicolumn{1}{c|}{0.964}               & \multicolumn{1}{c|}{0.974}           \\ \cline{2-11}
			\multicolumn{1}{|c|}{}                                            & \multicolumn{1}{c|}{500}           & \multicolumn{1}{c|}{0.84}                    & \multicolumn{1}{c|}{0.759}                & \multicolumn{1}{c|}{0.887}            & \multicolumn{1}{c|}{N/A}                      & \multicolumn{1}{c|}{N/A}                   & \multicolumn{1}{c|}{N/A}               & \multicolumn{1}{c|}{0.989}                  & \multicolumn{1}{c|}{0.985}               & \multicolumn{1}{c|}{0.987}           \\ \cline{2-11}
			\multicolumn{1}{|c|}{}                                            & \multicolumn{1}{c|}{1250}          & \multicolumn{1}{c|}{0.854}                   & \multicolumn{1}{c|}{0.739}                & \multicolumn{1}{c|}{0.445}            & \multicolumn{1}{c|}{N/A}                      & \multicolumn{1}{c|}{N/A}                   & \multicolumn{1}{c|}{N/A}               & \multicolumn{1}{c|}{0.993}                  & \multicolumn{1}{c|}{0.976}               & \multicolumn{1}{c|}{0.984}           \\ \cline{2-11}
			\multicolumn{1}{|c|}{}                                            & \multicolumn{1}{c|}{2500}          & \multicolumn{1}{c|}{0.948}                   & \multicolumn{1}{c|}{0.834}                & \multicolumn{1}{c|}{0.62}             & \multicolumn{1}{c|}{N/A}                      & \multicolumn{1}{c|}{N/A}                   & \multicolumn{1}{c|}{N/A}               & \multicolumn{1}{c|}{1}                      & \multicolumn{1}{c|}{1}                   & \multicolumn{1}{c|}{1}               \\ \hline
			\multicolumn{1}{|c|}{\multirow{6}{*}{RLF Report}}                 & \multicolumn{1}{c|}{50}            & \multicolumn{1}{c|}{0.826}                   & \multicolumn{1}{c|}{0.632}                & \multicolumn{1}{c|}{0.716}            & \multicolumn{1}{c|}{1}                        & \multicolumn{1}{c|}{1}                     & \multicolumn{1}{c|}{1}                 & \multicolumn{1}{c|}{0.932}                  & \multicolumn{1}{c|}{0.816}               & \multicolumn{1}{c|}{0.87}            \\ \cline{2-11}
			\multicolumn{1}{|c|}{}                                            & \multicolumn{1}{c|}{100}           & \multicolumn{1}{c|}{0.268}                   & \multicolumn{1}{c|}{0.519}                & \multicolumn{1}{c|}{0.353}            & \multicolumn{1}{c|}{N/A}                      & \multicolumn{1}{c|}{N/A}                   & \multicolumn{1}{c|}{N/A}               & \multicolumn{1}{c|}{0.94}                   & \multicolumn{1}{c|}{0.896}               & \multicolumn{1}{c|}{0.918}           \\ \cline{2-11}
			\multicolumn{1}{|c|}{}                                            & \multicolumn{1}{c|}{250}           & \multicolumn{1}{c|}{0.515}                   & \multicolumn{1}{c|}{0.518}                & \multicolumn{1}{c|}{0.516}            & \multicolumn{1}{c|}{N/A}                      & \multicolumn{1}{c|}{N/A}                   & \multicolumn{1}{c|}{N/A}               & \multicolumn{1}{c|}{0.989}                  & \multicolumn{1}{c|}{0.957}               & \multicolumn{1}{c|}{0.973}           \\ \cline{2-11}
			\multicolumn{1}{|c|}{}                                            & \multicolumn{1}{c|}{500}           & \multicolumn{1}{c|}{0.55}                    & \multicolumn{1}{c|}{0.545}                & \multicolumn{1}{c|}{0.547}            & \multicolumn{1}{c|}{N/A}                      & \multicolumn{1}{c|}{N/A}                   & \multicolumn{1}{c|}{N/A}               & \multicolumn{1}{c|}{0.996}                  & \multicolumn{1}{c|}{0.956}               & \multicolumn{1}{c|}{0.976}           \\ \cline{2-11}
			\multicolumn{1}{|c|}{}                                            & \multicolumn{1}{c|}{1250}          & \multicolumn{1}{c|}{0.511}                   & \multicolumn{1}{c|}{0.515}                & \multicolumn{1}{c|}{0.513}            & \multicolumn{1}{c|}{N/A}                      & \multicolumn{1}{c|}{N/A}                   & \multicolumn{1}{c|}{N/A}               & \multicolumn{1}{c|}{0.995}                  & \multicolumn{1}{c|}{0.966}               & \multicolumn{1}{c|}{0.98}            \\ \cline{2-11}
			\multicolumn{1}{|c|}{}                                            & \multicolumn{1}{c|}{2500}          & \multicolumn{1}{c|}{0.829}                   & \multicolumn{1}{c|}{0.639}                & \multicolumn{1}{c|}{0.722}            & \multicolumn{1}{c|}{N/A}                      & \multicolumn{1}{c|}{N/A}                   & \multicolumn{1}{c|}{N/A}               & \multicolumn{1}{c|}{1}                      & \multicolumn{1}{c|}{1}                   & \multicolumn{1}{c|}{1}               \\ \hline
			\multicolumn{1}{|c|}{\multirow{6}{*}{Paging with IMSI}}           & \multicolumn{1}{c|}{50}            & \multicolumn{1}{c|}{50}                      & \multicolumn{1}{c|}{0.634}                & \multicolumn{1}{c|}{0.918}            & \multicolumn{1}{c|}{1}                        & \multicolumn{1}{c|}{1}                     & \multicolumn{1}{c|}{1}                 & \multicolumn{1}{c|}{1}                      & \multicolumn{1}{c|}{0.998}               & \multicolumn{1}{c|}{0.999}           \\ \cline{2-11}
			\multicolumn{1}{|c|}{}                                            & \multicolumn{1}{c|}{100}           & \multicolumn{1}{c|}{100}                     & \multicolumn{1}{c|}{0.653}                & \multicolumn{1}{c|}{1}                & \multicolumn{1}{c|}{1}                        & \multicolumn{1}{c|}{1}                     & \multicolumn{1}{c|}{1}                 & \multicolumn{1}{c|}{1}                      & \multicolumn{1}{c|}{1}                   & \multicolumn{1}{c|}{1}               \\ \cline{2-11}
			\multicolumn{1}{|c|}{}                                            & \multicolumn{1}{c|}{250}           & \multicolumn{1}{c|}{250}                     & \multicolumn{1}{c|}{0.591}                & \multicolumn{1}{c|}{0.963}            & \multicolumn{1}{c|}{1}                        & \multicolumn{1}{c|}{1}                     & \multicolumn{1}{c|}{1}                 & \multicolumn{1}{c|}{1}                      & \multicolumn{1}{c|}{1}                   & \multicolumn{1}{c|}{1}               \\ \cline{2-11}
			\multicolumn{1}{|c|}{}                                            & \multicolumn{1}{c|}{500}           & \multicolumn{1}{c|}{500}                     & \multicolumn{1}{c|}{0.653}                & \multicolumn{1}{c|}{1}                & \multicolumn{1}{c|}{1}                        & \multicolumn{1}{c|}{1}                     & \multicolumn{1}{c|}{1}                 & \multicolumn{1}{c|}{1}                      & \multicolumn{1}{c|}{0.998}               & \multicolumn{1}{c|}{0.999}           \\ \cline{2-11}
			\multicolumn{1}{|c|}{}                                            & \multicolumn{1}{c|}{1250}          & \multicolumn{1}{c|}{1250}                    & \multicolumn{1}{c|}{0.653}                & \multicolumn{1}{c|}{1}                & \multicolumn{1}{c|}{N/A}                      & \multicolumn{1}{c|}{N/A}                   & \multicolumn{1}{c|}{N/A}               & \multicolumn{1}{c|}{1}                      & \multicolumn{1}{c|}{1}                   & \multicolumn{1}{c|}{1}               \\ \cline{2-11}
			\multicolumn{1}{|c|}{}                                            & \multicolumn{1}{c|}{2500}          & \multicolumn{1}{c|}{2500}                    & \multicolumn{1}{c|}{0.632}                & \multicolumn{1}{c|}{0.571}            & \multicolumn{1}{c|}{N/A}                      & \multicolumn{1}{c|}{N/A}                   & \multicolumn{1}{c|}{N/A}               & \multicolumn{1}{c|}{1}                      & \multicolumn{1}{c|}{1}                   & \multicolumn{1}{c|}{1}               \\ \hline
		\end{tabular}}
  \caption{Effectiveness evaluation for all the synthesized signatures across all attacks. Where each row indicates the effectiveness on that specific attack, when trained on their respective training dataset with the size specified in the second (Size) column. Do note that Mealy Machine is also trained with other attacks at the same time.}
  \label{app:evaluation_all_generated_monitors}

\end{table*}
\clearpage

\subsection{Multithreading Evaluation}\label{app:threading}

\paragraph{Threading.}
Mealy Machine outperforms both DFA and
\pltl when it comes to monitoring all attack signatures at once.
The reason behind this is that Mealy Machine use a unified signature
for all attacks and thus requires only one internal state,
while DFA and \pltl require $N$ states for $N$ attack signatures.
To accelerate the DFA- and PLTL-based monitoring,
an intuitive approach is to employ multithreading.
We, therefore, investigate whether threading improves
the efficiency for these monitors. For this, we instantiate
three different versions of DFA and \pltl monitors.
The first instantiation is called ``\textit{No Threading}'' which
sequentially invokes each monitor per layer and then analyzes the results.
The second implementation is called ``\textit{N/2 Threading}`` and
it relies on $N/2$ threads, where each thread runs two monitors
and then communicates the results to the main thread.
The third implementation is called ``\textit{N Threading}``,
where $N$ threads are spawned, one per attack.
Similar to the \textit{N/2 Threading}, these spawned threads
converse back the results from each individual monitor to the main thread.

Table \ref{tab:threading_pixel_3} shows the average number of messages
processed per second by different monitoring approaches when different
levels of multithreading are employed. We observe that threading does \texttt{not}
improve the performance and in fact, there is a clear relation between
performance degradation and number of threads. For instance, NAS with no threading
was able to parse over $10K$ messages per second while $N$ threads dropped this
to $0.56$ messages per second. This result can be attributed to the fact that Python
prohibits two threads from executing simultaneously \cite{python_threading}.

\begin{table}[]
  \centering
  \begin{tabular}{|l|l|l|l|}
  \hline
  \multicolumn{1}{|c|}{\textbf{Monitor}} & \multicolumn{1}{c|}{\textbf{Threading}} & \multicolumn{1}{c|}{\textbf{Avg.}} & \multicolumn{1}{c|}{\textbf{SD.}} \\ \hline
  \multicolumn{4}{|c|}{\textbf{RRC}} \\ \hline
  \multirow{3}{*}{DFA} & \textit{N Threading} & 1.8 & 3.3 \\ \cline{2-4}
   & \textit{N/2 Threading} & 12.5 & 9.7 \\ \cline{2-4}
   & \textit{No Threading} & 51730.6 & 158596.4 \\ \hline
  \multirow{3}{*}{PLTL} & \textit{N Threading} & 2.25 & 4.6 \\ \cline{2-4}
   & \textit{N/2 Threading} & 13.7 & 10.19 \\ \cline{2-4}
   & \textit{No Threading} & 7286.3 & 55599 \\ \hline
  \multicolumn{4}{|c|}{\textbf{NAS}} \\ \hline
  \multirow{3}{*}{DFA} & \textit{N Threading} & 1.2 & 3.6 \\ \cline{2-4}
   & \textit{N/2 Threading} & 4.7 & 5.8 \\ \cline{2-4}
   & \textit{No Threading} & 34164 & 224904.7 \\ \hline
  \multirow{3}{*}{PLTL} & \textit{N Threading} & 1.23 & 3.73 \\ \cline{2-4}
   & \textit{N/2 Threading} & 4.7 & 6.2 \\ \cline{2-4}
   & \textit{No Threading} & 3771 & 62512.74 \\ \hline
  \end{tabular}
\caption{Average (and standard deviation) messages per second parsed across different layers with different levels of threading on Pixel 3. For NAS N = 10 and RRC, N = 5}
\label{tab:threading_pixel_3}
\end{table}

\subsection{Lower Bound Memory Requirement Formulas and Results}
\label{app:lower_bound_memory}
For a monitor to be viable, the static and dynamic memory
overhead are required to be  small. However, precisely measuring
the memory requirement is unreliable and hence we resort to
providing lower bound memory requirements for each of the
monitoring state. For measuring lower bound, we resort the
minimum number of bits one would require to represent the internal
monitoring state. This also serves as an indicator of possible update
size in bits when a new attack is discovered.
For this evaluation, we consider the signatures generated by
each monitor with the most data they can handle.

\paragraph{DFA.} For DFA, the internal data structure consists of
states, transitions, and an alphabet. If a monitor consists of
$N$ states, $M$ transitions, and an alphabet of size $A$,
then the following functions represent the minimum number of bits
required to represent a DFA.
For representing each transition, the number of bits required are
$log_2N + log_2N + log_2A$ since the monitor must keep in memory
the starting start, ending state, and the letter in that alphabet
which will triggers the transition. For each state, the number of
bits required are 2; one bit indicating whether the state is a start state
and another indicating whether it is an accepting state.
Also, during the monitoring process the monitor must keep in memory
the current state, requiring $log_2N$ bits. Therefore, DFA
requires $M(log_2N + log_2N + log_2A) + N(2) + log_2N$ bits of memory.
In addition, we need 12 bytes to account for the number of states
($N$), number of transitions ($M$), and the size of the alphabet $A$.
\paragraph{Mealy Machine.} Mealy Machine's internal structure consists
of states and transitions, similar to that of DFA except that
transitions are associated with an output letter and the states
are no longer are rejecting or accepting. If a monitor consists of
$N$ states, $M$ transitions, the size of the input alphabet is $I$,
and the output alphabet size is $O$ then the following functions
represent the lower bound of Mealy Machine's memory consumption.
For each transition, the number of bits required are
$log_2N + log_2N + log_2I + log_2O$. Similar to DFA, for
each transition we must keep in memory the starting state,
ending state, and the letter in the input alphabet that will
trigger this transition. In addition, it must also keep in
memory the output letter it generates when the transition is taken.
For each state the number of bits required is 1 since it must keep
track whether the state is a starting state. Like DFA, Mealy Machine
must also keep track of the current state while running,
requiring $log_2N$ bits. In total, Mealy Machine requires
a total of $M(log_2N + log_2N + log_2I + log_2O) + N + log_2N$ bits
of memory. To represent this structure in a signature file,
we add 16 bytes to accounts for the number of states ($N$),
the number of transitions ($M$), the size of input alphabet ($I$),
and the size of output alphabet ($O$).

\paragraph{\pltl.} To represent the size of the internal structure of a \pltl based monitor,
we rely on counting the number of propositions and operators in the formula.
If a formula has $P$ propositions, $T$ operators, $A$ is the size of the alphabet
and $O$ is the total number of distinct operators supported. Note that $O$ is
fixed at 9 as defined in Section \ref{sec:pltl_syntax_and_semantics}. The internal
structure then requires $P(log_2A) + T(log_2O)$ bits.

We also need 8 bytes to account
for the number of propositions ($P$), and the number of operators $T$.
In addition, we need $2$ bits for capturing the truth value of each subformula
of a signature (1 bit for the previous truth value and 1 bit for the current
truth value).

\begin{table}[h!]
  \centering
\begin{tabular}{ccccc}
\hline
\multicolumn{5}{|c|}{\textbf{Mealy Machine}}                                                                                                                                                                       \\ \hline
\multicolumn{1}{|c|}{\textbf{Layer}} & \multicolumn{1}{c|}{\textbf{States}}            & \multicolumn{1}{c|}{\textbf{Transition}}  & \multicolumn{1}{c|}{\textbf{Input}}    & \multicolumn{1}{c|}{\textbf{Output}} \\ \hline
\multicolumn{1}{|c|}{NAS}            & \multicolumn{1}{c|}{4}                          & \multicolumn{1}{c|}{108}                  & \multicolumn{1}{c|}{32}                & \multicolumn{1}{c|}{11}              \\ \hline
\multicolumn{1}{|c|}{RRC}            & \multicolumn{1}{c|}{2}                          & \multicolumn{1}{c|}{65}                   & \multicolumn{1}{c|}{33}                & \multicolumn{1}{c|}{6}               \\ \hline
                                     &                                                 &                                           &                                        &                                      \\ \cline{1-4}


 \multicolumn{4}{|c|}{\textbf{PLTL}}                                                                                                                                          &                                      \\ \cline{1-4}
 \multicolumn{1}{|c|}{\textbf{Layer}} & \multicolumn{1}{c|}{\textbf{PLTL Propositions}} & \multicolumn{1}{c|}{\textbf{PLTL Operators}} & \multicolumn{1}{c|}{\textbf{Alphabet}} &                                      \\ \cline{1-4}
 \multicolumn{1}{|c|}{NAS}            & \multicolumn{1}{c|}{11}                         & \multicolumn{1}{c|}{11}                  & \multicolumn{1}{c|}{32}               &                                      \\ \cline{1-4}
 \multicolumn{1}{|c|}{RRC}            & \multicolumn{1}{c|}{13}                        & \multicolumn{1}{c|}{12}                 & \multicolumn{1}{c|}{33}               &                                      \\ \cline{1-4}

\multicolumn{4}{|c|}{\textbf{DFA}}                                                                                                                                          &                                      \\ \cline{1-4}
\multicolumn{1}{|c|}{\textbf{Layer}} & \multicolumn{1}{c|}{\textbf{States}}            & \multicolumn{1}{c|}{\textbf{Transitions}} & \multicolumn{1}{c|}{\textbf{Alphabet}} &                                      \\ \cline{1-4}
\multicolumn{1}{|c|}{NAS}            & \multicolumn{1}{c|}{36}                         & \multicolumn{1}{c|}{526}                  & \multicolumn{1}{c|}{32}               &                                      \\ \cline{1-4}
\multicolumn{1}{|c|}{RRC}            & \multicolumn{1}{c|}{511}                        & \multicolumn{1}{c|}{7199}                 & \multicolumn{1}{c|}{33}               &                                      \\ \cline{1-4}
\end{tabular}
\caption{Memory Consumption required for all monitors per layer.}
\label{tab:memory_consumption_small}
\end{table}
\clearpage
\paragraph{Lower-bound Memory Requirements.}
We present the number of states, the number of transitions, the alphabet size,
and the formula size required for lower bound memory calculation, when considering
\texttt{all} 15 attacks,
in Table \ref{tab:memory_consumption_small} and the corresponding minimum bits
required in Table \ref{tab:memory_consumption_bits}.
Note that, we acknowledge that this lower bound does not account of
additional bytes needed for parsing signatures and attack remedies.

As expected,
\pltl proves to be highly memory efficient. In addition to this, DFA proves to require
a significantly higher number of bits than Mealy Machine because DFA keeps a state
machine per attack. The discrepancy between NAS and RRC for DFA is due to the fact that
the DFAs for NAS have 36 different states in total,
while the RRC DFAs have a total of 511 states.

\begin{table}[h!]
\centering

\begin{tabular}{c|l|l|l|}
\hline
\multicolumn{1}{|c|}{\textbf{Layer}} & \multicolumn{1}{c|}{\textbf{PLTL}} & \multicolumn{1}{c|}{\textbf{Mealy Machine}} & \multicolumn{1}{c|}{\textbf{DFA}} \\ \hline
\multicolumn{1}{|c|}{\textbf{NAS}} & 90 & 1186 & 8146 \\ \hline
\multicolumn{1}{|c|}{\textbf{RRC}} & 104 & 629 & 166886 \\ \hline
\multicolumn{1}{l|}{\textbf{Total}} & 194 & 1815 & 175033 \\ \cline{2-4}
\end{tabular}
\caption{Minimum bits required for internal monitor structure.}
\label{tab:memory_consumption_bits}
\end{table}

\end{document}